\documentclass[pdflatex,sn-nature]{sn-jnl}

\makeatletter
\def\arx@doc{document}

\renewcommand{\documentclass}[2][]{}
\let\arx@begin\begin
\def\begin#1{\def\arx@env{#1}%
  \ifx\arx@env\arx@doc\expandafter\@firstoftwo\else\expandafter\@secondoftwo\fi
  {\endinput}{\arx@begin{#1}}}
\@namedef{ver@algorithm.sty}{}  %
\@namedef{ver@subcaption.sty}{} %

\title[How language model agents reshape price competition]{How language model agents reshape price competition}

\author*[1]{\fnm{Martin} \sur{Bichler}}\email{m.bichler@tum.de}
\author[1]{\fnm{Philip} \sur{Hierhager}}
\author[1]{\fnm{Kassian} \sur{Köck}}
\author[1]{\fnm{Fabian} \sur{Pieroth}}
\affil*[1]{\orgdiv{School of Computation, Information and Technology}, \orgname{Technical University of Munich}, \country{Germany}}

\abstract{
Repeated Bertrand competition is a stringent benchmark for algorithmic pricing: in homogeneous-goods markets, standard learning procedures typically select the stage-game Nash outcome with prices near marginal cost. 
We show that LLM pricing agents can escape this benchmark, depending on deployment decisions. 
LLMs outperform standard online learners through faster adaptation, but this advantage largely disappears against full-information learners.
An extracted surrogate algorithm illustrates how LLMs in specific deployments profit from regularly undercutting slower learners. 
Agents that post one price every period from raw market histories that include rival prices can sustain supracompetitive regimes in self-play. 
Agents that periodically emit probability distributions from summarized histories of their own profits instead move toward the competitive region in self-play. 
Markets mixing these interfaces can generate dynamic regimes in which LLMs establish a high-price anchor and regularly undercut it to capture demand.
These regimes can keep both competitors' prices above marginal cost. Deployment choices shape these outcomes, shifting the focus from stationary equilibria to dynamic pricing regimes.

}


\maketitle

\tableofcontents

\section{Introduction}

Large language models (LLMs) are moving from passive text generation to autonomous
decision-making in environments where payoffs depend on repeated interaction. Pricing
is a particularly consequential domain. Algorithmic sellers repeatedly observe market
feedback, update prices and interact with other automated agents. Their behavior affects
profits, consumer welfare and market efficiency. The practical relevance of this setting
is immediate: LLM-based pricing systems have been proposed for e-commerce
\cite{wang2025llp}, and generative-AI vendors now provide commerce-agent blueprints
in which merchant agents connect to catalogue, inventory, pricing and campaign systems
and can stage price or promotion changes for operator approval
\cite{AnthropicCommerceAgents2026}. More broadly, LLM agents are increasingly
evaluated in strategic or feedback-driven environments
\cite{Horton2023LLMSimulatedEconomicAgents,ChenEtAl2023EconomicRationalityGPT,
AkataEtAl2023RepeatedGamesLLMs,CompeteAI2023CompetitionDynamicsLLMAgents,
ParkLiuOzdaglarZhang2025LLMAgentsRegret}. If such systems select high-price regimes,
the resulting private profits may come at the expense of consumer welfare. Yet we know
little about the long-run regimes selected when LLM pricing agents enter competition.

Classical repeated-game theory explains why many long-run outcomes can be sustained.
If agents maximize discounted cumulative profit, history-dependent strategies can support
outcomes close to the stage-game Nash equilibrium as well as outcomes close to monopoly
pricing \cite{FudenbergLevine1998LearningGames}. This theory is an existence theory,
not a selection theory: it characterizes which outcomes can be sustained, but it does not
predict which outcome a concrete adaptive process will reach.

Learning in games addresses this selection question. A learning rule is not only a
computational device for choosing actions; it is also a mechanism that selects a long-run
outcome, cycle or attractor of the repeated interaction. For many standard online-learning
algorithms, Bertrand competition provides a sharp equilibrium-learning benchmark. In the
stage game of a repeated Bertrand competition with homogeneous goods, the Nash
equilibrium price equals marginal cost, while in the heterogeneous-goods model, prices are
higher than marginal cost. We refer to the Nash equilibrium prices of the stage game as
the \emph{competitive outcome}. In Bayesian Bertrand
competition with homogeneous goods and private costs, standard Minty-type stability
arguments fail near equilibrium, but global asymptotic stability can be certified
numerically, and stochastic mirror-descent learners then converge to equilibrium in
all instances studied \cite{bichler2026convergencestochasticfirstorderalgorithms}. Most online-learning algorithms move toward the Nash
prediction in repeated Bertrand competition with homogeneous and heterogeneous goods
\cite{durmann2026online}, and in the discrete homogeneous game no-swap-regret learners
provably drive average profits down to the margin of a single grid step \cite{MaitiEtAl2026BertrandParadox}.

\subsection{Standard learners widely converge to Nash in all-or-nothing Bertrand competition}

We analyzed many online and reinforcement-learning algorithms but focus on a few
prominent examples in the main paper. Exp3 is an adversarial bandit algorithm for
external-regret minimization \cite{AuerCesaBianchiFreundSchapire2002Exp3};
BM-RM is a bandit implementation of the Blum--Mansour reduction from external to
internal or swap regret \cite{BlumMansour2007FromExternalToInternalRegret};
UCB represents optimism-based stochastic bandit learning
\cite{AuerCesaBianchiFischer2002UCB}. Exp3 has vanishing regret in the adversarial
model, while UCB has vanishing regret in the stochastic model of online learning.
Such online-learning algorithms are sequential learning procedures that update action
probabilities or indices from realized payoff feedback and aim at good long-run
performance, for example through regret minimization. They do not, however, explicitly
evaluate a current price through a continuation model of how this price will affect the
opponent's future learning dynamics.

Reinforcement-learning algorithms \citep{sutton1998reinforcement} such as Q-learning
and PPO \citep{SchulmanEtAl2017PPO} are more explicitly cumulative-reward oriented:
Q-learning estimates continuation values through a Bellman recursion, and PPO optimizes
cumulative reward from sampled trajectories. Symmetric Q-learning agents have been
shown to sustain supracompetitive prices in a heterogeneous-goods Bertrand model, i.e.,
prices higher than the stage-game Nash equilibrium
\cite{CalvanoEtAl2020AlgorithmicPricing,Klein2021AutonomousCollusion,
MeylahnDenBoer2022LearningCollude}, but subsequent work also shows that such outcomes
are sensitive to exploration, hyperparameters, market structure and algorithmic symmetry
\cite{AbadaLambin2023AICollusion,AbadaEtAl2024CollusionMistake,
DenBoerMeylahnSchinkel2026ArtificialCollusion}. Interestingly, in the homogeneous- and
heterogeneous-goods Bertrand models we study, PPO moves toward the Nash equilibrium of
the stage game rather than sustaining high-price regimes. Overall, there is significant
evidence that standard learning algorithms in Bertrand competition with homogeneous
goods often converge to the efficient Nash equilibrium, which is good news for consumer
welfare.

\subsection{LLMs adapt quickly against standard learners}

LLMs differ from standard online and reinforcement-learning algorithms. Their objective
is not encoded as a formal payoff function, Bellman recursion or regret criterion. It is
specified through a prompt and combined with language-mediated memory and an interface
that exposes market histories. Recent experiments show that LLM pricing agents can
sustain supracompetitive prices in differentiated Bertrand markets under point-price
deployments with rich histories and rival-price information
\cite{FishGonczarowskiShorrer2026AlgorithmicCollusionLLMs}. In these deployments the
model emits a single price every period from a raw history of recent periods. Recent
work also shows that different deployment decisions can reduce or break
LLM-mediated collusion \cite{KeppoEtAl2026Fragility}.

Computing a price for each interaction using an LLM is expensive. We therefore also
analyze deployments in which the LLM periodically emits a probability distribution over a
finite price grid. We denote recurring LLM interfaces by an output--history--interval
code: \textsf{P} and \textsf{D} denote point-price and distributional output,
\textsf{R} and \textsf{S} denote raw and summarized history, and the number denotes the
intervention interval in periods. These codes are shorthand interface identifiers
rather than complete deployment specifications; the remaining components, such as the
prompt, the persistent state and the action space, are reported separately (Methods,
Table~\ref{tab:deployments}). The Fish-style deployment, relating to the work of \cite{FishGonczarowskiShorrer2026AlgorithmicCollusionLLMs}, is \PRone. In our main
distributional deployment, \DStf,
the model receives summarized market histories and its previous
\texttt{plans\_and\_insights}, updates only at blockwise intervention times, and prices
in the following block are sampled from the emitted distribution. The
\texttt{plans\_and\_insights} field acts as a language-mediated memory of previous
pricing decisions. This deployment strategy significantly reduces pricing costs, which
is important for low-value goods. It constitutes a substantially different, but reasonable,
deployment strategy. For \DStf{}, we use neutral prompts that do not explicitly instruct LLMs to coordinate on high prices.

Prior literature mostly analyzes symmetric deployments in which LLM agents using the
same deployment strategy compete against each other. Real algorithmic markets are
likely to be more heterogeneous. Sellers may use standard online-learning algorithms,
LLM pricing agents with different deployment interfaces, fixed pricing rules or other adaptive systems.
We therefore first ask how LLM pricing deployments perform against standard learning
algorithms. This comparison is economically important because sellers have an incentive
to adopt LLM-based pricing systems only if such systems perform well against standard
algorithmic pricing rules. It is also practically important because an LLM query is more
expensive than an iteration of a standard online-learning algorithm.

To benchmark LLM behaviour against established online-learning rules, we report
experiments in which LLM pricing agents compete against standard learning algorithms
such as Exp3, UCB and PPO. These algorithms are model-free and do not need assumptions
on the number of competitors or the demand model, which is often unavailable or changing
in real-world markets. They operate on minimal information about their own payoff
feedback. For visible-rival-price treatments in a known Bertrand environment, however,
the feedback is richer: observing the rival price allows a firm to reconstruct the
counterfactual payoff of every price on the grid. Such settings call for full-information
benchmarks in addition to bandit learners. Throughout, we use \emph{standard learners}
to denote non-LLM pricing algorithms whose action probabilities or policies are updated
from realized market feedback over time. Static monopoly pricing and simple deterministic
rules such as short-term undercutting are treated as additional controls.

Standard learners also adapt to opponents because opponent behaviour determines the
payoff feedback they observe. However, this adaptation is mediated through a fixed
statistical update rule, such as regret accumulation, confidence-bound updates,
multiplicative weights or policy-gradient learning. LLM agents differ because their
deployed interface allows them to represent opponent behaviour in language and condition
current prices on qualitative hypotheses about future opponent responses. Thus, the
distinction is not whether agents adapt, but whether adaptation is restricted to a fixed
payoff-update rule or also involves language-mediated opponent-response reasoning.
Neither class best-responds myopically to the rival's current price: standard learners
respond to accumulated payoff statistics, and LLM agents may forgo a currently profitable
undercut because they expect it to lower the rival's future prices.

In the environments we study, LLM deployments create private profit advantages, but
through different mechanisms. Standard bandit learners explore substantially in the
initial periods, and convergence to the competitive region is slow under limited
feedback: in markets of standard learners alone, bandit learners need \(10^5\) to
\(10^7\) periods to settle in the competitive region (Fig.~\ref{fig:standard-learners-benchmark}b). 
Both \PRone{} and \DStf{} can exploit this learning phase and achieve higher
finite-horizon profits than standard learners. \DStf{} earns higher finite-horizon profits against adaptive learners by charging lower prices and undercutting rivals that remain above the competitive floor. To study the mechanism
behind \DStf{} competing against standard learning algorithms, we introduce
downward-biased belief search (DBBS), a surrogate of the LLM's distributional policy
updates. DBBS shows that their behaviour can be explained by a simple
ordered-search mechanism with inertia: the agent keeps much of its previous
distribution, gives more weight to recently profitable prices, avoids prices above its
current focal region and probes only a few prices upward. This mechanism explains both
competitive \DStf{} self-play and profitable undercutting of slower learners.

\subsection{Competing LLMs aim to escape the Bertrand paradox}

These findings motivate the next market-level question: \textit{if LLM pricing systems
are attractive to sellers, which price regimes emerge when multiple sellers deploy LLM
agents, possibly with different deployment strategies?} A central finding is that LLM
pricing outcomes are not determined only by the base model and the prompt. They are
selected by the interaction of the stage game, the deployment strategy and the opponent's
pricing behaviour. Fish-style \PRone{} agents sustain high prices not only in differentiated-goods
Bertrand competition, but also in homogeneous winner-takes-all competition. By contrast,
symmetric \DStf{} and \DSff{} agents, which output price distributions from summarized histories at blockwise
intervention times, move toward the finite-grid Nash region in homogeneous self-play and
do not sustain high-price regimes against standard online learners.

The difference between \PRone{} and \DStf{} or \DSff{} is not simply point versus distributional
output. The \PRone{} deployment exposes raw rival-price paths and asks for a single price,
making high-price conventions easy to represent and maintain. The \DStf{} and \DSff{} interfaces with their summarized history turn
pricing into an ordered payoff-feedback search. Failed high-price attempts shift
probability mass downward. In symmetric self-play this search tends to the finite-grid
Nash region, whereas \PRone{} agents can coordinate on supracompetitive prices. The resulting
price regimes need not correspond to stationary equilibria and can be highly dynamic.

Mixed LLM deployments provide further insights. Observable decision traces suggest that
LLM agents often identify high prices as profitable, but different deployments make
coordination on high prices more or less stable. When a \PRone{} agent competes with a \DSone{} agent with hidden rival prices, the
\PRone{} agent can set a high-price anchor, while the \DSone{} agent learns a profitable threshold below
that anchor. The \PRone{} agent may then periodically reset to the price ceiling or lower its price
briefly to harvest demand and make a profit. We refer to this recurrent pattern as
anchor-and-harvest. The selected regime depends on the specific deployment: action
representation, history representation, language-mediated memory, rival-price visibility
and opponent type.

The design space for LLM deployments is large. There are numerous deployment strategies
for LLMs, numerous standard learning algorithms, different foundation models and multiple
versions thereof. We tested different prompts and foundation models, and also deployment
strategies beyond the ones emphasized in the main paper. A fully combinatorial design is
infeasible. However, the main insights add value to the overall discussion on LLMs as
pricing agents. Even in repeated homogeneous-goods Bertrand competition, where standard
learning algorithms often converge to the efficient stage-game Nash equilibrium so consistently, LLMs
can sustain supracompetitive prices under some deployments.

The analysis of such multi-agent systems with LLMs poses new challenges. Studying these systems requires identifying systematic behavioural patterns, explaining how they depend on deployment choices and interactions with other agents, and developing models that account for the resulting trajectories and outcomes. The high adaptivity of LLM agents is beyond the online-learning rules analyzed in the past. It reinforces recent ideas saying that game theory needs to go beyond equilibrium finding and study the dynamics of a game \citep{papadimitriouGameDynamicsMeaning2019}. Importantly, resulting strategy profiles do not need to be stationary, but they can be dynamic over long time horizons. 

For competition authorities, our findings suggest that the tendency of standard pricing algorithms to converge toward the stage-game Nash outcome in homogeneous Bertrand markets need not persist when LLM-based agents enter the market. Depending on their deployment and the competitors they face, LLM agents can easily sustain supracompetitive price regimes. The homogeneous-goods Bertrand competition is a particularly insightful example as, although stylized, standard learning algorithms typically converge to the efficient Nash equilibrium.
We avoid the term algorithmic collusion, as collusion carries specific legal connotations and implications in competition law \citep{hartline2024regulationalgorithmiccollusion}. Nevertheless, if the widespread adoption of LLM-based pricing agents leads to prices substantially above the stage-game Nash equilibrium, this raises important concerns for competition authorities. Crucially, our findings suggest that auditing such agents requires more than examining their prompts and underlying foundation models: deployment decisions also play a critical role in shaping pricing outcomes.

\section{Results}

Next, we show that LLM-based pricing agents achieve high profits against standard learning algorithms, which exhibit substantial randomness in their pricing decisions due to extensive exploration in the initial rounds. When competing against simple pricing heuristics, such as short-term undercutting strategies, LLM agents quickly identify ways to avoid prices near marginal costs by establishing high price anchors to which the heuristics respond. Thus, despite not being specifically trained for sequential decision-making tasks, off-the-shelf LLMs demonstrate an ability to perform well in unfamiliar pricing environments.

We then turn to markets in which only LLM-based pricing agents compete. Here, outcomes vary substantially across deployment configurations: some lead to prices approaching the monopoly level, while others converge to the stage-game Nash equilibrium under symmetric play. The latter configurations (e.g., \DSff{}) rely on summarized price histories and rival price visibility, which appear to make it more difficult for agents to coordinate on high prices in the meta-game. In mixed-deployment settings, we again observe attempts by LLM agents to establish high price anchors that induce competitors to maintain elevated prices rather than converging toward marginal costs.

\subsection{Standard learners converge toward the competitive benchmark}

Before evaluating LLM pricing agents, we establish the benchmark against which
their behaviour should be interpreted. In homogeneous winner-takes-all Bertrand
competition, many standard learning procedures provide an equilibrium-learning
reference point. Their histories are not absent; rather, they are compressed into
statistical state variables such as weights, cumulative losses, empirical means,
confidence bonuses or policy parameters. In our experiments, Exp3, BM-RM, PPO,
UCB, full-information learners and simple pricing rules serve both as baselines
and as opponents for the LLM agents.

The non-LLM round robin confirms that homogeneous Bertrand competition is a
strong equilibrium-learning benchmark, but not a trivial one. Most standard
learners generate transaction prices close to the finite-grid Nash region and
high consumer surplus. This includes algorithms that learn from limited feedback,
such as Exp3 and BM-RM, the reinforcement learner PPO, and the full-information
learners FTRL (follow-the-regularized-leader with a Euclidean regularizer), full-information
regret matching and weighted fictitious play (Fig.~\ref{fig:standard-learners-benchmark}a). This complements recent results for Bayesian Bertrand
competition, where global stability of equilibrium learning is certified numerically
for all 243 systems studied although Minty-type arguments fail~\cite{bichler2026convergencestochasticfirstorderalgorithms}.
For our complete-information grid market, the regret notion matters: in every
correlated equilibrium of the discrete winner-takes-all game each firm earns at most
the margin of a single grid step, so two no-swap-regret learners such as BM-RM or
full-information regret matching are driven to the competitive region, whereas
no-external-regret learners admit coarse correlated equilibria with supracompetitive
prices \cite{MaitiEtAl2026BertrandParadox}.
Convergence is, however, not guaranteed for all algorithms. 
UCB is the most visible exception: in UCB self-play prices stay near the
ceiling (price index \(0.93\)), and against PPO they remain above the
competitive region (\(0.25\)) (Fig.~\ref{fig:standard-learners-benchmark}a).
Empirical NashConv identifies the same departures. Only three markets exceed
\(5\%\) of one-firm monopoly profit: UCB self-play (\(217\) profit units),
PPO against weighted fictitious play (\(87\)) and UCB against PPO (\(47\)); all
others are at most \(7.1\) (Supplementary Note~3). After \(10^7\) periods, \(20\) of
the \(21\) markets without UCB end in the finite-grid competitive region, compared
with \(5\) of the \(7\) markets with UCB; the only exception without UCB is PPO
against weighted fictitious play (\(0.46\)). Thus, the homogeneous Bertrand
environment disciplines many learning procedures, but the selected regime still
depends on the update rule (Fig.~\ref{fig:standard-learners-benchmark}a). This still makes it an important benchmark, because convergence 
is not as consistent for other types of games. 

The distinction between bandit and full-information feedback matters. If a firm
observes the rival price in the homogeneous Bertrand market and the payoff
function is known, it can reconstruct the counterfactual payoff of every price on
the grid. Such visible-rival-price treatments are full-information learning
problems, not pure bandit problems. In real-world markets, the demand model and the 
level of competition might not be known, which is why bandit algorithms are 
important. We therefore include bandit but also full-information algorithms: FTRL, full-information regret matching and weighted fictitious play, which best-responds to a discounted belief over the rival's price; it is belief-based, but like FTRL and full-information regret matching it uses the rival's observed price, so we group it with the full-information learners. The full-information learners provide a demanding comparison class: markets of two full-information learners settle in the competitive region within the first \(10^4\) periods, markets with one of them after a median of \(5\times10^4\) periods, whereas markets of bandit or reinforcement learners alone need a median of \(6\times10^5\) periods and up to several million (Fig.~\ref{fig:standard-learners-benchmark}b). Because the benchmark concerns where learners converge, the round robin runs for \(10^7\) periods with decaying, theoretically tuned exploration (Supplementary Note~3). 

\begin{figure}[t]
\centering
\includegraphics[width=\linewidth]{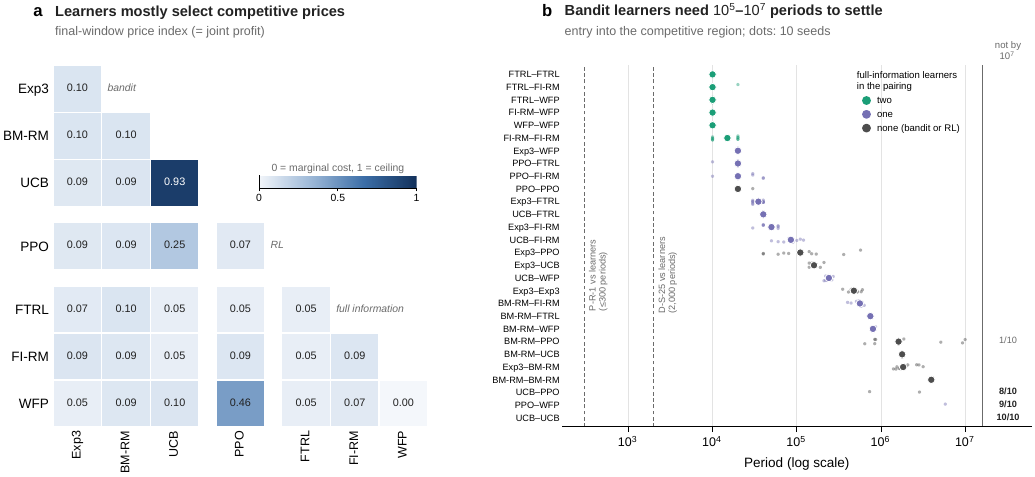}
\caption{\textbf{Standard learners mostly select competitive outcomes in homogeneous Bertrand competition.}
Round robin of seven standard learners (all unordered pairs including self-play; ten seeds per pair;
final window = last 250 periods), each with one frozen configuration. All firms observe the rival's
price, which FTRL, FI-RM and WFP require; Exp3, BM-RM and UCB learn from their own profit only.
\textbf{a}, Market-clearing price index, 0 = marginal cost, 1 = price ceiling; with winner-takes-all
demand it equals joint normalized profit. \textbf{b}, Period after which each market's price stays in the competitive region
(block means of \(10^4\) periods at most half a grid step above \(0.126\)); rows are markets sorted by the
median over seeds (large markers), small dots are seeds, colour gives how many of the two learners use full-information feedback;
\(10^4\) means competitive from the first block. Right column, number of seeds still outside the region
after \(10^7\) periods. FTRL, follow-the-regularized-leader with a Euclidean regularizer; BM-RM,
Blum--Mansour bandit regret matching; FI-RM, full-information regret matching; WFP, weighted fictitious
play (best response to a discounted belief over the rival's price, discount 0.9; grouped with the
full-information learners because it uses the rival's price). Horizon \(10^7\) periods; dashed lines in
\textbf{b}, the horizons of the \PRone{} (at most 300 periods) and \DStf{} (2,000 periods) runs against
standard learners.}
\label{fig:standard-learners-benchmark}
\end{figure}

This benchmark is especially important for interpreting PPO. PPO is not a myopic
price setter: it explicitly optimizes cumulative reward from sampled
trajectories. Nevertheless, in the homogeneous Bertrand environments we study,
PPO does not generally sustain high-price regimes. Instead, it moves toward
the competitive region in five of its seven markets; the exceptions are its
interactions with UCB and with weighted fictitious play. Thus, supracompetitive prices are not explained by cumulative-profit
optimization alone. The relevant distinction is between explicit algorithmic
update rules and deployment-induced LLM learning rules.

\subsection{LLM pricing agents earn private profit advantages against standard learners}

We next ask how LLM pricing deployments perform against existing algorithmic
pricing rules. This comparison is economically important because sellers have an
incentive to adopt LLM-based pricing systems only if such systems perform well
against standard learners. It is also practically important because an LLM query
is more expensive than an iteration of a standard online-learning algorithm. We
do not claim that every LLM deployment creates such an adoption incentive. Our
replicated quantitative evidence in this section concerns the \DStf{}
deployment; we complement it with an illustrative \PRone{} interaction
against Exp3 and use a simple adaptive pricing rule to illustrate how quickly a
\PRone{} agent can adapt to predictable opponent behavior.

The \DStf{} agent creates a robust finite-horizon profit advantage against the
standard learners considered here. Against Exp3, BM-RM, PPO and UCB, the \DStf{} agent
earns between 71 and 72 percent of cumulative industry profit on average over
the 2,000-period horizon, with ten independent seeds per matchup
(Fig.~\ref{fig:llm-adoption-incentive}a). The advantage does not arise from
sustaining high prices. Instead, the \DStf{} agent more often posts a lower price than its
opponent. In the final 10 percent of periods, the \DStf{} agent prices below the standard
learner in 57--60 percent of periods across the four matchups, while it prices
above the learner in only 28--38 percent
(Fig.~\ref{fig:llm-adoption-incentive}b). Standard learners often remain
temporarily above the competitive floor, because over the 2,000 periods of the \DStf{}
runs the bandit learners are still far from settling (Fig.~\ref{fig:standard-learners-benchmark}b), and the
\DStf{} agent identifies profitable prices below them. The resulting private adoption incentive therefore comes
from faster adaptation and undercutting rather than from maintaining a
supracompetitive price regime.

The \PRone{} agent can also generate a private profit advantage against a standard
learner. Figure~\ref{fig:llm-adoption-incentive}c shows an illustrative
300-period interaction between a \PRone{} agent and Exp3. Exp3 continues to explore
broadly over the price grid, whereas the \PRone{} agent adapts its price over time and
spends extended periods in comparatively persistent price regions above the
finite-grid competitive region. In this run, the \PRone{} agent earns 63 percent of
cumulative industry profit. This trajectory is illustrative rather than a
replicated treatment estimate, but it shows that a \PRone{} agent need not sustain a
monopoly-price regime to obtain a private profit advantage against an
exploratory standard learner. Against full-information learners the advantage
disappears, as discussed below.

The speed of \PRone{} adaptation is particularly transparent against a simple
win-up--lose-down pricing rule (Fig.~\ref{fig:llm-adoption-incentive}d). The
rule raises its price by one grid step after receiving positive demand and
lowers it by one grid step after receiving no demand. From period 3 onward, the
\PRone{} agent posts exactly the same price as the heuristic. A tie splits demand between
the firms and therefore counts as a sale for the heuristic, causing it to raise
its price by one grid step in the following period. The \PRone{} agent follows this price
increase, producing a sequence of matched upward moves. Both agents reach the
price ceiling in period 21 and remain there for the rest of the experiment.
Because they continue to set the same price, cumulative industry profit is split
approximately equally. This treatment therefore does not provide an additional
adoption-incentive result. Instead, it isolates the \PRone{} agent's ability to rapidly
match a simple opponent-response pattern and thereby move the interaction from
prices near the competitive region to a persistent high-price regime.

\begin{figure}[t]
\centering
\includegraphics[width=\linewidth]{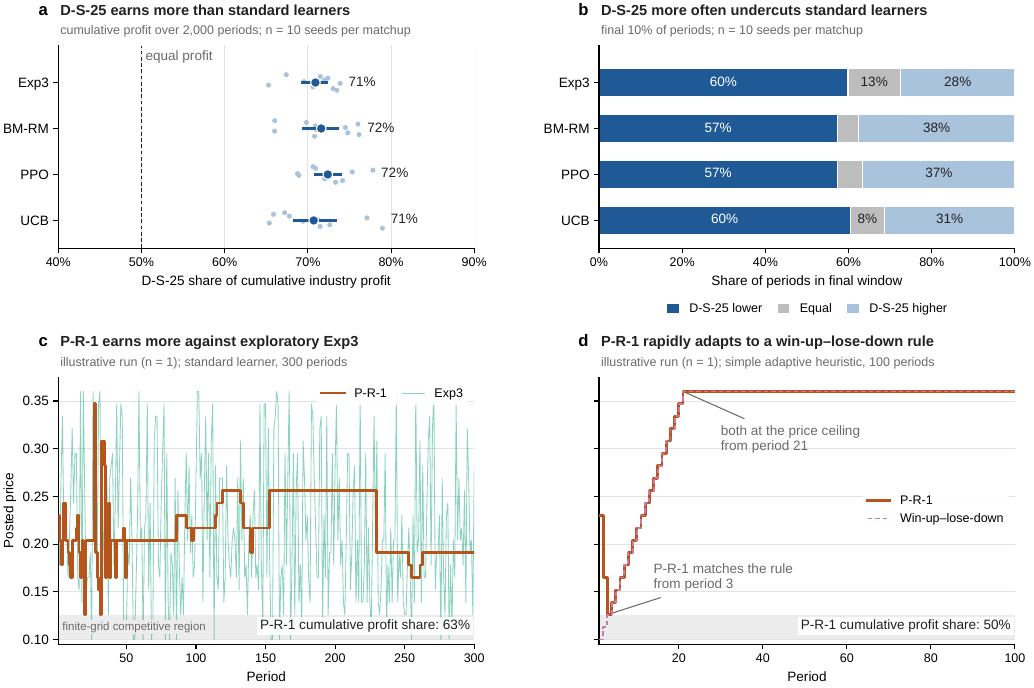}
\caption{\textbf{LLM pricing agents can earn private profit advantages against
standard learners and rapidly adapt to opponent behavior.}
\textbf{a}, \DStf{} earns more than Exp3, BM-RM, PPO and UCB over
the full 2,000-period horizon. Small points denote independent seeds, large
points denote matchup means, and the dashed line marks equal cumulative industry
profit. Each matchup contains ten independent seeds.
\textbf{b}, In the final 10 percent of periods, \DStf{} agents post lower prices than
their opponents more often than they post higher prices. Percentages may not sum
to 100 because of rounding.
\textbf{c}, Illustrative \PRone--Exp3 trajectory. Exp3 explores broadly over the
price range, while the \PRone{} agent moves through more persistent price regions and
earns 63 percent of cumulative industry profit.
\textbf{d}, Illustrative \PRone{} interaction with a win-up--lose-down heuristic.
From period 3 onward, the \PRone{} agent matches the heuristic's price. Because each tie
generates positive demand, the heuristic raises its price by one grid step in
the following period and the \PRone{} agent follows. Both agents reach the price ceiling
in period 21 and remain there, splitting cumulative industry profit
approximately equally. The shaded areas in panels \textbf{c} and \textbf{d}
denote the finite-grid competitive region. Panels \textbf{c} and \textbf{d}
show single illustrative runs (\(n=1\)); panels \textbf{a} and \textbf{b} use
ten independent seeds per matchup.}
\label{fig:llm-adoption-incentive}
\end{figure}

Additional experiments with simple pricing heuristics provide complementary
evidence that \PRone{} agents respond flexibly to predictable opponent behavior. Against
a randomized win-up--lose-down rule, high-price interaction is less stable and
the \PRone{} agent earns a larger share of industry profit. Against a deterministic
undercut-last heuristic, the \PRone{} agent responds to the predictable pricing cycle and
captures most of industry profit. These heuristic treatments are useful for
probing \PRone{} adaptation, but they are conceptually distinct from the replicated
comparison against standard learners and are therefore reported in detail in the
Supplementary Information.

The \DStf{} advantage is finite-horizon and path-conditioned. Realized external
regret is computed against the best fixed grid price on the observed opponent
path; it is not policy regret against an opponent that would have reacted
differently under counterfactual LLM actions. The experiments therefore do not
show that the \DStf{} agent dominates standard learners asymptotically or against
arbitrary opponents. They show that, on the structured price paths generated by
the standard learners considered here, the \DStf{} agent identifies profitable regions
of the ordered price grid quickly enough to obtain a substantial finite-horizon
profit advantage.

Taken together, these results establish the economic motivation for the
subsequent analysis. LLM pricing systems can be privately attractive relative
to standard learners even when the resulting market does not settle into a
high-price regime. The mechanism, however, depends on the deployment. The \DStf{}
advantage is associated with payoff-feedback-driven undercutting, whereas the
\PRone{} experiments illustrate rapid adaptation to opponent behavior: against
exploratory Exp3 this adaptation produces a private profit advantage, while
against the predictable win-up--lose-down rule the \PRone{} agent quickly matches the
opponent's response pattern and follows it into a persistent high-price regime.
We next use the agents' observable decision traces to distinguish these forms of
adaptation more systematically.

The \PRone{} advantage has limits, and they follow the opponent's feedback. The
strongest counterexample is coin betting, a parameter-free full-information learner:
it stays one or two grid steps below the \PRone{} agent's elevated prices and earns
71 percent of cumulative industry profit over 300 periods; after about 200 periods
the \PRone{} agent abandons its anchor and both prices fall to about \(0.19\).
Against weighted fictitious play, which best-responds to a discounted belief over
the rival's price, the \PRone{} agent earns 46 percent in an illustrative
100-period run, falling to about 35 percent in the last 40 periods, and against
full-information regret matching its larger share is earned at near-competitive
prices (Supplementary Note~4). The full-information learners therefore mark the
boundary of the \PRone{} evidence, in contrast to the slower-adapting bandit
learners such as Exp3.

\subsection{Decision traces reveal opponent-facing anchoring and payoff-feedback search}

At decision times, LLM agents emit not only prices or distributions but also
\texttt{plans\_and\_insights}. We treat these as behavioural traces, not as direct
access to hidden computation. Elicited explanations can be incomplete or
unfaithful and must be validated against actions and outcomes
\cite{YaminTangCortesGomezSharmaHorvitzWilder2026ElicitedBeliefs}. This caution
is especially important for pricing agents: recent work on reasoning models
reports that collusive outcomes need not be visible in the agents' own
chain-of-thought or prompt-level intent \cite{RiemerEtAl2026Certification}.

With this caveat, the traces suggest two distinct forms of adaptation. \PRone{} agents
more often use opponent-facing language, referring to anchoring, avoiding price
wars, resetting after low-price phases, preserving high-margin regimes and
harvesting demand: at least one such motif appears in \(94\%\) of \PRone{} traces
but in \(34\%\) of \DStf{} traces (Supplementary Note~5). In high-price \PRone{} regimes, the agent often reasons that a
locally profitable undercut could induce the opponent to lower future prices. It
therefore sometimes rejects a price that would be myopically profitable against
the opponent's current price.

\DStf{} agents more often use payoff-region language. Their traces refer to successful
and failed price regions, local probes and demand thresholds, and they express
caution through low-probability probes (payoff-region motifs in \(98\%\) of
\DStf{} traces against \(21\%\) of \PRone{} traces). In heterogeneous competition against online learners, the \DStf{} agent
often treats lower prices as profitable undercuts of opponents that remain above
the competitive floor. In hidden-rival-price \PRone--\DSone{} cells, the \DSone{} agent can move
upward from its own quantity and profit feedback, without explicitly observing
the rival price.

The protocol ablations indicate where this difference comes from. With raw
histories and visible rival prices, \(96\)--\(99\%\) of traces are
opponent-facing whether the LLM emits a point price or a distribution; with
summarized histories and hidden rival prices, \(3\)--\(6\%\) are. As for prices
(Fig.~\ref{fig:deployment-ablations}c), the information interface rather than the
action representation determines whether the LLM reasons about its rival or
about its own payoff regions.

This distinction is central. The point is not that standard learners have no
history: their histories are compressed into weights, cumulative losses,
empirical means, confidence bonuses, gradient states or policy parameters.
Rather, they do not evaluate the current action through a language-mediated
account of how it may change the opponent's future behaviour. The \PRone{} traces
suggest response-sensitive anchoring. The \DStf{} traces suggest payoff-feedback
ordered search.

\subsection{DBBS explains payoff-feedback search under the \DStf{} interface}
We construct downward-biased belief search to test whether \DStf{} updates have a
systematic low-dimensional structure that we can analyze. DBBS is intended as a surrogate for
the updates of \DStf, not as a model of \PRone{} behaviour. The
\PRone{} regimes documented above likely involve additional response-sensitive
reasoning and opponent-facing focal cues that are not captured by this ordered
payoff-feedback mechanism.

DBBS is not an explicit opponent-modelling algorithm. It does not estimate the
opponent's policy. Rather, it updates from recent payoffs, failures and the
ordered structure of prices. Opponent behaviour enters indirectly because it
determines realized payoffs.

DBBS maintains a recency-weighted payoff score \(v_k\) and failure score
\(\phi_k\) for each price tick \(k\). At each intervention epoch, it identifies
the modal recently played tick \(m\) and scores each tick as
\[
    S_k=v_k-\beta_{\mathrm{fail}}\phi_k-\beta_{\mathrm{loc}}(k-m)_+.
\]
The first term rewards recently profitable prices, the second penalizes failures,
and the third imposes one-sided caution for prices above the current focal
region. A softmax over scores is mixed with local upward exploration. The surrogate
then mixes this search policy with the agent's own recent price distribution, with an
inertia weight \(w\) that is cross-fitted on training seeds. This captures inertia,
recency weighting, failure sensitivity, ordered-price reasoning and local asymmetric
search. The fitted failure penalty is close to zero; the downward drift comes from the
one-sided locality term (Supplementary Note~6).

Teacher-forced validation fits DBBS to held-out LLM-emitted distributions, not
merely sampled realized prices. One parameter vector, fitted jointly on the seven
\DStf{} treatments of Section~1.2 with leave-one-seed-out cross-fitting, predicts how the
\DStf{} agent changes its distribution better than every alternative: on the 22\% of
blocks in which its mean price moves, the search component is more accurate than
persistence, win-up--lose-down and weighted fictitious play in all seven treatments, and
than the other baselines in six (Fig.~\ref{fig:dbbs-validation}a). In free-running
simulations on the same held-out environment seeds, DBBS reproduces the \DStf{} agent's
final price level more closely than every other mechanism, including those that need
full information (mean error \(0.020\) against \(0.033\)--\(0.074\); closer in at
least 48 of 70 games against each; Fig.~\ref{fig:dbbs-validation}b). It reproduces the
competitive self-play regime and the undercutting of adaptive learners and orders the
treatments by the agent's final price (Fig.~\ref{fig:dbbs-validation}c). Inertia is
essential for this: without it, DBBS undercuts the learners in 81\% of late periods,
against the agent's 59\%, and prices about \(0.04\) too low. We added the inertia
component after observing this bias, but its weight is estimated only from one-step
predictions on training seeds, so the free-running comparison remains held-out.

\begin{figure*}[t]
\centering
\includegraphics[width=\linewidth]{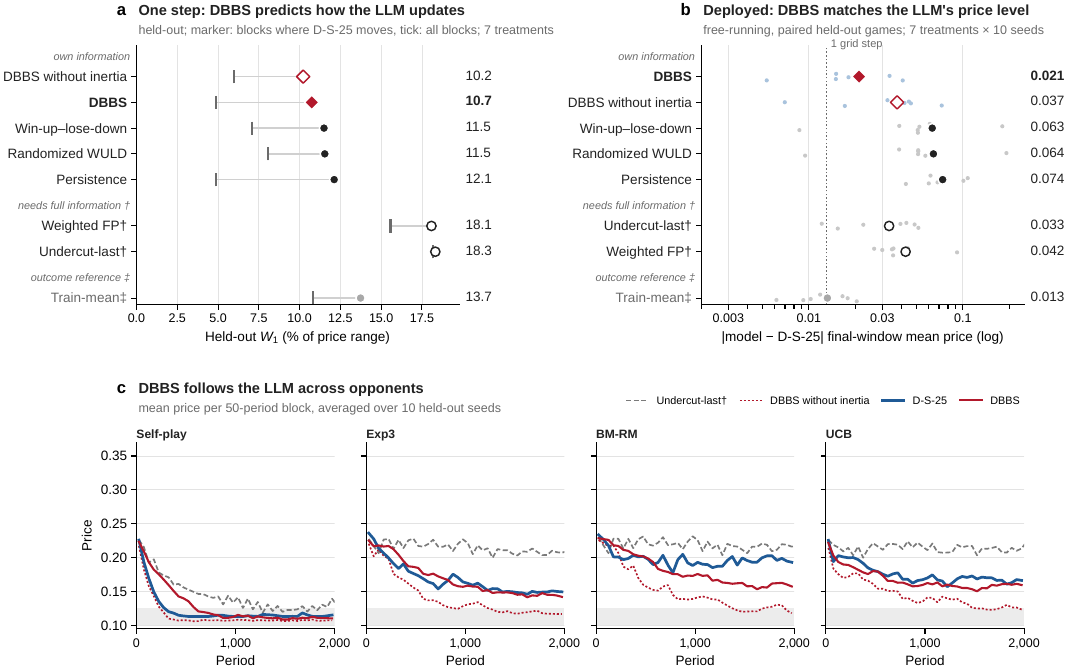}
\caption{\textbf{DBBS reproduces how \DStf{} updates and where it prices.}
DBBS uses one parameter vector fitted jointly on the seven \DStf{} treatments (ten environment seeds each,
leave-one-seed-out cross-fitting); every alternative is selected under the same folds and criterion, so all
results are held-out. DBBS, search policy mixed with the agent's own recent price distribution (inertia weight
cross-fitted on training seeds); DBBS without inertia, the search policy alone. Own-information mechanisms use
only what the agent observed (its own prices, sales and profits). \(^\dagger\)Undercut-last and weighted
fictitious play need full information---the rival's prices and the demand model and costs---which the agent
never observed. \(^\ddagger\)Train-mean emits the agent's own average distribution for that treatment and
encodes the outcome rather than a mechanism.
\textbf{a}, One-step prediction of the next emitted distribution: held-out Wasserstein-1 distance on all
blocks (ticks) and on the 22\% of blocks in which the agent's mean price moves by at least one grid step
(markers). Persistence is most accurate overall because emitted distributions are highly persistent; on
changing blocks the DBBS search policy is the most accurate model.
\textbf{b}, Free-running reproduction: each model replaces the agent against the same opponent on the same
held-out environment seed (7 treatments \(\times\) 10 seeds); absolute difference between the model's and the
agent's final-window mean price.
\textbf{c}, Mean price per 50-period block, averaged over the ten held-out seeds, for four treatments. Without
inertia DBBS undercuts the learners too far; with inertia it follows the agent. Grey band, finite-grid
competitive region.}
\label{fig:dbbs-validation}
\end{figure*}

DBBS also clarifies the limits of the \DStf{} behaviour. It is not a no-regret
algorithm. Deterministic opponent-price sequences can induce linear external
regret, and one-step prediction errors can compound under endogenous deployment.
This is informative: the \DStf{} agent can perform well on structured market paths without
satisfying worst-case online-learning guarantees. DBBS explains why \DStf{}
moves toward Nash in symmetric self-play and undercuts slow learners, and its
ordered payoff-feedback search is consistent with \DSone{} moving upward when a \PRone{}
agent creates a high-price anchor, a setting in which DBBS is not fitted. It does not explain the \PRone{} agent's anchoring
and anchor-and-harvest behaviour, which appears to depend on response-sensitive
reasoning and focal-price conventions.

Taken together, the analysis of \PRone{} and \DStf{} shows why LLM pricing may be attractive to sellers, but also why the mechanism depends on deployment. \PRone{} agents can earn private profit advantages by exploiting slow learners or simple pricing heuristics and
can sustain high-price regimes when the opponent makes such regimes viable. \DStf{}
can earn more against standard learners by adapting faster and charging lower
prices. These private incentives motivate the next market-level question: what
happens once multiple sellers deploy LLM pricing systems, possibly under
different interfaces?

\subsection{Deployment interfaces shift LLM competition between high-price coordination and undercutting}

Once sellers have incentives to adopt LLM pricing agents, the relevant
market-level question is which price regimes emerge when multiple sellers deploy
such systems. We find that LLM--LLM competition does not select a single
characteristic outcome. The same class of Bertrand games can produce
supracompetitive high-price regimes or prices close to the finite-grid competitive
region depending on how the LLM is deployed. \PRone{} and \DStf{}/\DSff{} select
sharply different self-play regimes (Fig.~\ref{fig:deployment-ablations}a). The
protocol ablations show that this contrast is not explained by output representation
alone (Fig.~\ref{fig:deployment-ablations}b,c).

We begin with \PRone, the Fish-style point-price deployment, as a positive control. In this
protocol, the LLM emits a single continuous price at each decision point,
observes raw market histories including rival prices, uses a long memory window
and receives profit-oriented prompt language adapted from prior LLM-pricing work
\cite{FishGonczarowskiShorrer2026AlgorithmicCollusionLLMs}. In the
differentiated-product logit market calibrated as in the algorithmic-pricing
literature \cite{CalvanoEtAl2020AlgorithmicPricing}, \PRone{} self-play reproduces
the supracompetitive pricing documented in prior LLM-pricing experiments
\cite{FishGonczarowskiShorrer2026AlgorithmicCollusionLLMs}. Across independent
control runs, market-average price elevation is close to the joint-profit
benchmark (Fig.~\ref{fig:deployment-ablations}b, control).

The high-price regime is not confined to differentiated demand. Holding the
\PRone{} deployment fixed while replacing logit demand by homogeneous
winner-takes-all Bertrand competition reduces market-average price elevation but
does not eliminate it. Thus, stronger price competition in the stage game
weakens the high-price outcome but does not by itself force the deployed LLM
toward the competitive benchmark (Fig.~\ref{fig:deployment-ablations}a).

Distributional self-play selects a qualitatively different regime in the
same homogeneous winner-takes-all market. Under \DStf{} and \DSff, prices
fall rapidly from elevated initial levels and remain near the lowest grid prices,
which form the finite-grid Nash region because of tie splitting and the discrete
price grid. This contrast is not an unconditional property of the foundation
model or of LLM-based decision making. \PRone{} sustains
high prices in the same market, whereas \DStf{} and \DSff{} move
toward the competitive region. The selected outcome is therefore a property of
the deployed adaptive system.

This comparison also sharpens the role of cumulative reward. PPO is a
cumulative-reward learner but does not generally sustain high-price regimes in
this environment. \PRone{} agents can sustain high prices without implementing a known
Bellman recursion, while \DStf{} and \DSff{} agents, using the same broad class of foundation models
but a different deployment interface, move toward the finite-grid Nash region.
High prices therefore do not follow from long-run optimization alone. They arise
under particular deployed adaptive systems that determine what information the
LLM receives, how past interaction is represented and how prices are selected.

A natural explanation for the difference between \PRone{} and \DStf{} or \DSff{} behaviour is the
change in output representation. \PRone{} directly emits a price, whereas the
\DS{} interfaces emit a probability distribution over a finite price grid. Direct protocol
ablations reject the view that this distinction alone explains the change in
regime. Starting from the Fish-style point-price control in the differentiated
market, replacing continuous point-price output by a point action on the
21-point grid leaves price elevation high. Replacing continuous output by a
probability distribution over the same grid also leaves elevation high
(Fig.~\ref{fig:deployment-ablations}b).
Changing the action representation alone therefore does not reproduce the
transition from the Fish-style high-price regime to the competitive behaviour
observed under the full \DSff{} bundle.

By contrast, changes to the information interface produce substantially larger
movements (Fig.~\ref{fig:deployment-ablations}c). Hiding the rival's price and replacing raw period-level histories
with summarized histories are the largest single-factor movements away from the
Fish-style control in our ablation design. These treatments suggest that what
the LLM observes, and how past market interaction is represented to it, is more
consequential for regime selection than the distinction between point and
distributional output alone.

\begin{figure}[t]
    \centering
\includegraphics[width=\linewidth]{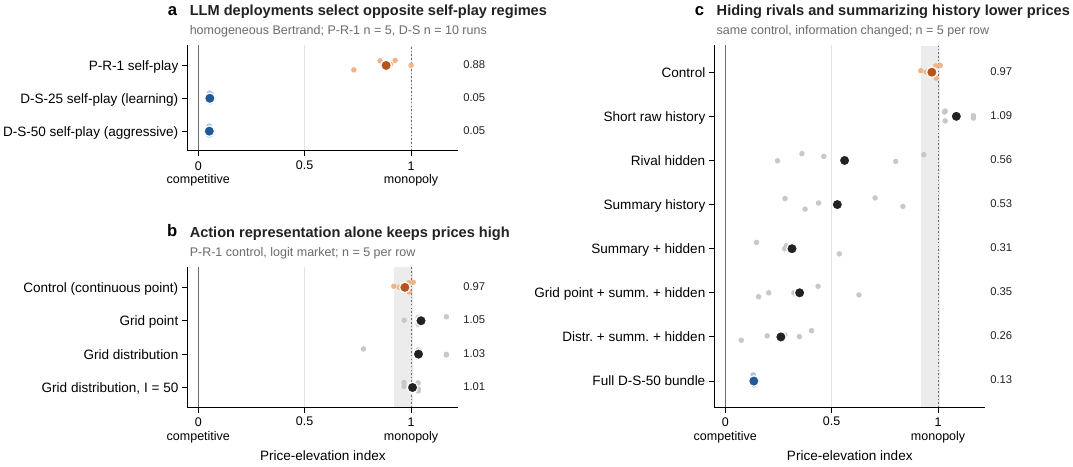}
\caption{\textbf{The information interface, not the action representation, separates high-price \PRone{}
play from competitive \DStf{} and \DSff{} play.} Values are the price-elevation indices of Methods, \(\Delta\) in the logit market (0 = continuous
one-shot Nash price, 1 = joint-profit price) and \(\Delta^{H}\) in the homogeneous market (0 = marginal
cost, 1 = the price ceiling, which is the joint-monopoly price), from the mean posted price over the final third of each run;
small points are independent runs, large points their means. Indices above 1 occur in the logit market,
where firms can price above the joint-profit price.
\textbf{a}, Self-play in the same homogeneous winner-takes-all market: \PRone{} (continuous price
every period, raw history, rival price visible; \(n=5\) runs of 300 periods) and the distributional deployments \DStf{} (learning) and
\DSff{} (aggressive) (\(n=10\) runs of 2,000 periods each).
\textbf{b}, \textbf{c}, Starting from the \PRone{} control of the protocol ablations in the differentiated logit market,
which differs from the Fish-style positive control only in using one merged scratchpad file
(Supplementary Note~1)
(orange; grey band, range across its runs), the action representation (\textbf{b}) or the information
interface (\textbf{c}) is changed. Grid-point or distributional output leaves prices at the monopoly
level. Hiding the rival's price or summarizing the history roughly halves the index, their combination
lowers it further, and the full \DSff{} bundle (distribution, summary, hidden rival price, 50-period
interval; blue) reaches 0.13. A short raw history does not lower prices. Each row \(n=5\) runs.}
\label{fig:deployment-ablations}
\end{figure}

The combined treatments reinforce this conclusion. Grid-based output remains
compatible with high-price play when raw histories and visible rival prices are
retained. In contrast, configurations combining summarized histories with hidden
rival prices produce substantially lower price elevation. Adding distributional
output and blockwise updating to this information interface moves play further
toward the competitive benchmark in the treatments examined. Every ablation
row is replicated over five runs (Fig.~\ref{fig:deployment-ablations}b,c); because
single runs of the information treatments vary widely, we interpret their
magnitudes as mechanism evidence rather than precise treatment-effect
estimates.

These results imply that the distinction between \PRone{} and \DStf{} or
\DSff{} should not be interpreted as an effect of output representation
alone. The configurations differ in a bundle of deployment choices---output
representation, history representation, language-mediated memory and
intervention interval---and, in the comparisons above, also in their information
setting, because \PRone{} observes the rival's price and \DStf{} and \DSff{} do
not. The ablation results identify what the LLM observes---rival-price
visibility and the representation of market histories---as the important
determinants of the selected regime.

This conclusion should not be read as showing that hiding rival prices or
summarizing histories is unconditionally pro-competitive. These changes alter
the information channel through which the LLM adapts rather than mechanically
lowering prices. Their effect can therefore depend on the opponent faced by the
agent. In heterogeneous LLM markets, for example, a \DSone{} agent that does not
observe rival prices can still infer profitable high-price regions from its own
quantity and profit feedback when another pricing agent provides a high-price
anchor. We examine these cross-deployment interactions next.

\subsection{Heterogeneous LLM deployments generate anchor-and-harvest dynamics}

The symmetric self-play experiments provide two useful endpoints. \PRone{}
can sustain supracompetitive price regions, whereas \DStf{} and \DSff{}
move toward the competitive region. Markets
with differently deployed LLM agents do not simply interpolate between these
outcomes. Instead, they generate asymmetric and often recurrent price dynamics.
A useful organizing principle is an \emph{anchor-and-harvest} tension. An agent
that remains at a high price can create a price umbrella under which another
agent earns positive margins by pricing below it. Harvesting this opportunity,
however, can destabilize the elevated price region that created it. The
resulting regime therefore depends on whether agents preserve the anchor,
harvest underneath it, or repeatedly move between these two behaviors.

Figure~\ref{fig:heterogeneous-regime-map} summarizes the regimes observed across
the heterogeneous and multi-agent LLM experiments, reporting for each run the
market-clearing price (Fig.~\ref{fig:heterogeneous-regime-map}a) and the
allocation of industry profit (Fig.~\ref{fig:heterogeneous-regime-map}b). We treat visible rival prices
as the primary information environment: posted prices are directly observable
in many online markets, and a pricing agent can therefore condition on rivals'
recent prices. Hidden-rival-price treatments provide a complementary mechanism
intervention. They remove this direct response channel while leaving agents able
to learn from their own quantities and profits. Across these treatments, neither
more information nor a particular action representation maps mechanically into
high or low prices. Instead, the interface changes how quickly harvesting
opportunities are identified and how strongly they destabilize a price anchor.
Every run in this section is a single 300-period run, so the reported prices and
profit shares describe representative paths rather than treatment effects.

\begin{figure}[t]
    \centering
    \includegraphics[width=\linewidth]{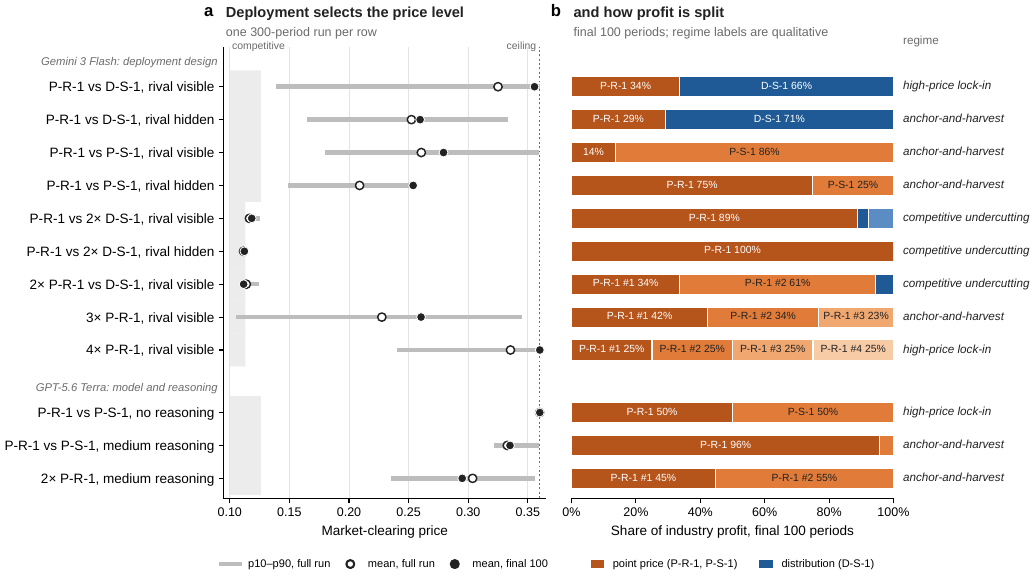}
    \caption{\textbf{Heterogeneous and multi-agent LLM deployments exhibit distinct price regimes.}
    Homogeneous winner-takes-all Bertrand market; each row is one representative 300-period run
    (\(n=1\)). Deployments \PRone, \PSone{} and \DSone{} as in Table~\ref{tab:deployments}; ``rival
    visible/hidden'' refers to the information setting, whether the LLMs observe competitors' prices. Upper block: Gemini 3 Flash, varying the deployment; lower block:
    GPT-5.6 Terra, varying reasoning. \textbf{a}, Market-clearing price: grey bar, 10th--90th percentile over
    the run; open marker, full-run mean; filled marker, mean over the final 100 periods; grey band,
    grid-based competitive benchmark (three lowest grid prices with two firms, two lowest with three or
    four); dotted line, price ceiling. \textbf{b}, Share of industry profit per
    firm in the final 100 periods (\PRone{} and \PSone{} agents orange, \DSone{} agents blue). Regime labels
    describe each run's price path qualitatively (Methods; Supplementary Note~8). Single-run rows are
    representative paths, not treatment-effect estimates.}
    \label{fig:heterogeneous-regime-map}
\end{figure}

The \PRone--\DSone{} experiments illustrate the tension most directly. When both
agents observe rival prices, the interaction settles quickly into a high-price
regime that the \DSone{} agent harvests (Fig.~\ref{fig:anchor-harvest-trajectories}a).
After about 45 periods of low and volatile prices, both agents move to the top of
the grid. In the final 100 periods the \PRone{} agent posts the \(0.36\) ceiling in every
period. The \DSone{} agent matches it in two thirds of the periods and posts one grid step
below it in the remaining third, capturing the whole market in those periods. The
market-clearing price averages \(0.356\), joint profit reaches \(98\%\) of the
monopoly benchmark, and the \DSone{} agent earns \(66\%\) of industry profit. The \PRone{} agent
thus provides a stable anchor, but does not respond to the \DSone{} agent's occasional
undercuts by abandoning it. With direct price observability, harvesting in this
run takes the form of small, repeated undercuts beneath a persistent anchor
rather than an unravelling of the anchor itself.

Removing rival-price information from the \DSone{} agent changes this interaction
substantially. The \DSone{} agent can no longer react directly to the \PRone{} agent's current
price, but it can infer from its own sales and profits which price regions remain
profitable. When the \PRone{} agent spends extended periods near the upper end of the
price range, the \DSone{} agent continues to sell and gradually moves upward. The \PRone{} agent
thereby creates a price umbrella from which the \DSone{} agent benefits. In the hidden-rival run the \DSone{} agent prices below the \PRone{} agent in about three quarters
of the final 100 periods and earns \(71\%\) of industry profit, even though the
\PRone{} agent has richer information and finer price control
(Fig.~\ref{fig:anchor-harvest-trajectories}b).
The \PRone{} agent periodically moves below the \DSone{} agent to regain demand and then returns to
the high-price region. We refer to this recurrent sequence as
\emph{anchor-and-harvest}: an elevated anchor is maintained or repeatedly
restored, but is intermittently interrupted by a lower price that captures the
market.

This comparison shows why rival-price visibility should not be interpreted as
a monotone pro- or anti-competitive treatment. With visible rival prices,
the \DSone{} agent can position itself exactly one grid step below a ceiling anchor, and
the market locks in near the monopoly price while the \DSone{} agent harvests. With hidden
rival prices, such precise undercutting is not possible; the \DSone{} agent instead learns
from its own payoff feedback that elevated prices remain profitable and follows
the \PRone{} agent up and down, producing recurrent cycles at lower average prices
(\(0.26\) rather than \(0.36\)). In both cases the \DSone{} agent, not the
better-informed \PRone{} agent, captures most of the profit. Reduced information can therefore
change the \emph{form} of adaptation without eliminating supracompetitive
regimes.

Other point-price deployments show that elevated prices need not be shared.
When a \PRone{} agent competes with a \PSone{} agent, the same point-price agent with a
summarized history, both
treatments end in anchor-and-harvest, but the harvester changes with the
information setting (Fig.~\ref{fig:heterogeneous-regime-map}b). With rival
prices visible to both, the \PRone{} agent holds the ceiling in most of the
final 100 periods while the \PSone{} agent prices below it and earns
\(86\%\) of industry profit. When the \PSone{} agent does not observe the
rival price, the \PRone{} agent undercuts it in most late periods and earns
\(75\%\). In both runs the market-clearing price stays well above the
competitive region (\(0.28\) and \(0.25\) in the final 100 periods;
Fig.~\ref{fig:heterogeneous-regime-map}a), so similar
price levels can coexist with very different allocations of the resulting
profit.

Adding firms does not by itself lower prices (Fig.~\ref{fig:heterogeneous-regime-map}a).
Low prices occur in the three-firm markets that combine \PRone{} and \DSone{} agents,
which end at or near the finite-grid Nash region: the final-100
market-clearing price is \(0.112\)--\(0.118\), against \(0.113\) at the top of the region. With one \PRone{} agent and two \DSone{} agents,
the continuous-price \PRone{} agent posts just below the \DSone{} agents' lowest grid prices and
earns \(89\%\) of industry profit with rival prices visible and all of it with rival prices hidden.
With two \PRone{} agents and one \DSone{} agent, the point agents undercut each other and earn
\(61\%\) and \(34\%\). Markets of \PRone{} agents alone, by contrast, stay elevated: the final-100
market price is \(0.26\) with three agents and the \(0.36\) ceiling with four.

\begin{figure*}[t]
    \centering
    \includegraphics[width=\linewidth]{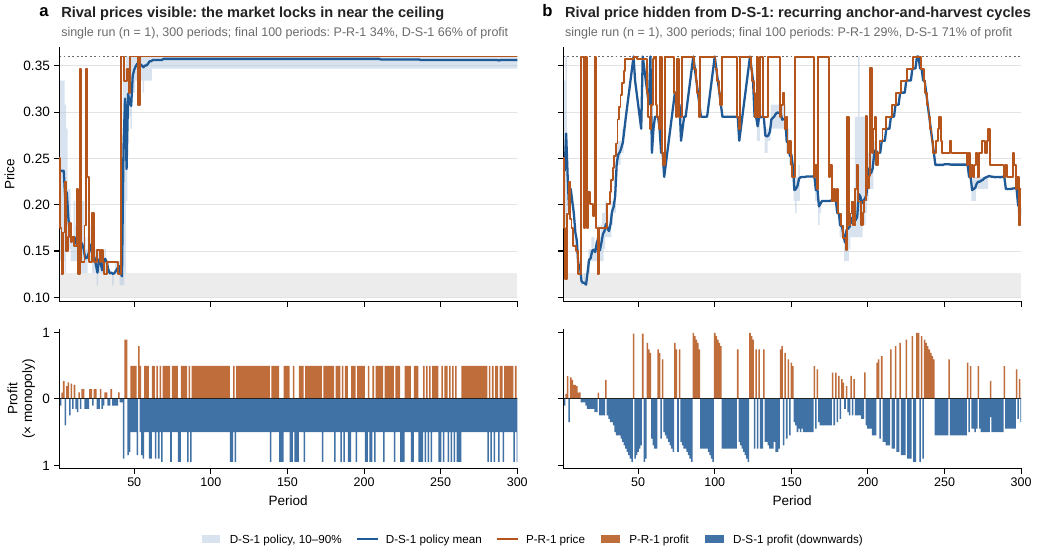}
    \caption{\textbf{Rival-price visibility shifts the balance between anchoring and harvesting.}
    One \PRone{} agent against one \DSone{} agent in the homogeneous Bertrand market; single run per panel (\(n=1\)), 300
    periods. Top: \PRone{} posted price (orange) and the \DSone{} agent's emitted policy (blue line, mean; shading,
    10th--90th percentile); grey band, grid-based competitive benchmark; dotted line, price ceiling. Bottom:
    per-period profit as a share of one-firm monopoly profit, \PRone{} agent upwards and \DSone{} agent downwards.
    \textbf{a}, With rival prices visible both firms settle near the ceiling after about 45 periods; the
    \DSone{} agent earns 66\% of profit in the final 100 periods. \textbf{b}, With the rival price hidden from \DSone{} the \PRone{} agent
    repeatedly re-anchors high while the \DSone{} agent follows below it, producing recurring cycles in which the
    \DSone{} agent earns 71\% of profit in the final 100 periods.}
    \label{fig:anchor-harvest-trajectories}
\end{figure*}

The two point-price-only runs show that the selected regime is path-dependent. In the
three-\PRone{} run the market price repeatedly falls toward the competitive region and resets
upward (10th--90th percentile \(0.11\)--\(0.35\) over the run): each reset restores a high price
that the other firms then undercut, and the firms earn \(42\%\), \(34\%\) and \(23\%\) of final-window
profit. In the four-\PRone{} run, high-price attempts also fail early, but from period 74 all
four agents hold the ceiling and split profit equally (Supplementary Note~8 shows both paths).
These are single runs, not estimates of the effect of market size: the same point-price
deployment produced recurrent anchor-and-harvest with three firms and a stable ceiling with four.

The lower block of Fig.~\ref{fig:heterogeneous-regime-map} repeats the
visible-rival-price \PRone--\PSone{} interaction with a second model, GPT-5.6 Terra,
and varies its reasoning setting (temperature 0.7 without reasoning; with reasoning
enabled the API does not accept a temperature, so the provider default applies). Without reasoning, the \PRone{} and
\PSone{} agents post the \(0.36\) ceiling in almost every period and
split industry profit equally. With medium reasoning, the market price remains
elevated (\(0.34\) in the final 100 periods, about \(90\%\) of the monopoly
joint profit), but the allocation becomes asymmetric: the \PRone{} agent
earns \(96\%\) of industry profit, the reverse of the Gemini run in which the
\PSone{} agent harvests. Two \PRone{} agents with medium reasoning
alternate between restoring and undercutting a high price; the final-window market price is
\(0.29\) and profit is split almost evenly (Fig.~\ref{fig:heterogeneous-regime-map}a,b). As in the Gemini rows, these are
single runs; they show that elevated prices recur across models while the
identity of the harvester is not fixed by the information interface alone.

Taken together, the heterogeneous and multi-agent experiments suggest a common
interpretation. High-price regimes persist when agents expect the continuation
value of preserving an anchor to outweigh the immediate gain from harvesting
it. When harvesting remains attractive but does not permanently destroy the
anchor, recurrent anchor-and-harvest patterns emerge. When several agents
aggressively pursue the same harvesting opportunity, repeated undercutting can
drive prices toward the competitive region. The relevant object is therefore
not simply whether LLMs ``coordinate'' on a high price, but how the deployed
agents resolve the dynamic trade-off between maintaining a profitable price
region and capturing demand underneath it.

These patterns are descriptive regimes of the coupled adaptive process and
should not be interpreted as equilibrium claims. A persistent high-price
plateau can contain profitable one-period deviations, and a recurrent
anchor-and-harvest cycle need not be an approximate dynamic equilibrium.
Establishing a dynamic-equilibrium interpretation would require
counterfactual deviation analysis against the induced opponent policies:
one would need to show that an agent cannot substantially improve its
discounted payoff by following an alternative policy while its competitors
continue to respond according to their deployed policies. We therefore use
the term \emph{price-regime selection} rather than equilibrium selection for
these interactions.

\section{Discussion}

Our results recast LLM pricing as a shift from equilibrium learning to
price-regime selection. Homogeneous Bertrand competition provides a particularly
sharp benchmark: in the stage game, the competitive outcome is marginal-cost
pricing, and standard learning procedures typically move play toward this prediction.
UCB is an exception, and it can produce higher-price and high-NashConv
outcomes in self-play. Full-information learners such as FTRL and full-information regret matching
converge to the Nash prediction more rapidly. 

Against this benchmark, LLM pricing agents select qualitatively different
regimes. The difference is not that LLMs simply have longer histories or optimize
a longer-run objective. Standard learners also carry history, but their histories
are compressed into statistical state variables such as weights, empirical means,
confidence bounds, regret sums or policy parameters. PPO even optimizes
cumulative reward from sampled trajectories, yet in our homogeneous Bertrand
benchmark, it does not sustain high-price regimes. LLM agents instead
carry language-mediated memory and may evaluate current prices through their
perceived effect on future opponent behavior. The same foundation model can
therefore produce high-price \PRone{} self-play, Nash-like \DStf{} and \DSff{} self-play,
finite-horizon undercutting against online learners and recurrent
anchor-and-harvest patterns in cross-deployment trajectories.

This distinction changes the economic interpretation of LLM pricing. The first
question is not only whether LLMs collude in self-play, but whether LLM pricing
systems create adoption incentives relative to existing pricing rules.
Our asymmetric experiments show that they can, but through different mechanisms.
\PRone{} can earn higher cumulative profit against Exp3 and can exploit
simple pricing heuristics such as deterministic undercut-last rules. They can
also coordinate with simple win-up--lose-down rules on high prices. \DStf{}
creates a different profit advantage: it can earn higher finite-horizon
profits against standard online learners by charging lower prices, undercutting
rivals that remain above the competitive floor, and capturing more demand. Thus,
seller incentives to adopt LLM pricing agents need not always come from raising
prices. They can also arise from faster adaptation to structured opponent price
paths.

Once such adoption incentives exist, the market-level question becomes
what happens when multiple LLM pricing systems compete. Our results show that
there is no single answer. \PRone{} agents with raw rival-price histories,
point-price actions and profit-oriented language can sustain high-price regimes,
including in homogeneous winner-takes-all competition. \DStf{} and \DSff{}, with summarized
histories, distributional output and blockwise updating, move toward the
finite-grid competitive region in homogeneous self-play and do not sustain
high-price regimes against standard online learners. Mixed \PRone--\DSone{} markets
can generate recurrent dynamic regimes such as anchoring, threshold learning and
anchor-and-harvest cycles. The stage game matters, but it is not sufficient. The
deployment protocol matters because it induces the learning rule. 

For learning in games, the key implication is that LLM deployments are induced
learning rules. Classical analyses usually fix a game and an algorithmic update
such as regret matching, mirror descent, projected gradient ascent, exponential
weights, Q-learning, or UCB. LLM agents do not come with such a compact analytic
update. The update is partly created by prompt, memory, history representation,
observation policy, action representation, and query schedule. Consequently,
changing from raw histories to summaries, from point prices to distributions, or
from every-period decisions to blockwise interventions is not merely an
implementation detail. It changes the adaptive system whose long-run regimes we
observe. These results point to a new empirical and theoretical agenda for
learning in games: studying foundation-model agents as deployment-induced
adaptive systems whose regimes depend on market structure, opponent ecology, and
interface design.

For game theory, the selected object is often a price regime rather than a
stationary equilibrium. Repeated-game theory already tells us that many
history-dependent outcomes can be sustained in principle. The empirical question
is which outcome a deployed adaptive process selects. Some outcomes in our
experiments resemble stationary Nash-like or high-price regimes. Others are
recurrent patterns, such as anchoring, intermittent undercutting and
anchor-and-harvest cycles. These are better understood as attractors of a coupled adaptive
system than as equilibria of the one-shot Bertrand game.

A recurrent price regime is not automatically a dynamic equilibrium. For a state
representation \(s\), deployed policies \((\psi_i,\psi_{-i})\) and discounted
value \(V_i\), an \(\varepsilon\)-dynamic equilibrium test would ask whether
\[
    V_i^{\psi_i,\psi_{-i}}(s)
    \ge
    \sup_{\psi_i'} V_i^{\psi_i',\psi_{-i}}(s)-\varepsilon
\]
for relevant states on the recurrent path. Our experiments document such
recurrent regimes, but we do not claim that all of them satisfy this
counterfactual deviation condition. Establishing this is a natural next step for
the theory of LLM agents in games. It will require methods that combine
trajectory classification, policy reconstruction, counterfactual deviation tests
and perturbation analysis.

The role of cumulative reward should therefore be interpreted carefully. A
long-run objective makes many history-dependent strategies feasible, but it does
not by itself generate high prices. PPO is an explicit cumulative-reward learner
in our evaluated configurations, yet it does not generally sustain high-price
regimes in homogeneous Bertrand competition. Conversely, \PRone{} agents can sustain high
prices without implementing a known Bellman recursion or explicitly specified
dynamic program, while \DStf{} and \DSff{} do not. What matters is the adaptive process that maps histories into future actions. For LLM agents, that process is induced by the prompt, memory, observation stream, action interface, and update cadence.

Our behavioural surrogate, DBBS, should be interpreted in this light. DBBS
captures \DStf{} as a payoff-feedback-driven ordered search rule with
inertia: the agent keeps much of its previous distribution, gives more weight to
recently profitable prices, avoids prices above its current focal region and probes
only a few prices upward. This explains why \DStf{} moves toward the competitive region in
self-play and undercuts slower online learners on structured paths, and it is
consistent with \DSone{} moving upward when a \PRone{} agent creates a high-price anchor. DBBS is not, however, a model of \PRone{}
behaviour. The \PRone{} regimes documented above likely involve additional
response-sensitive reasoning and focal cues created by raw rival-price histories
and point-price recommendations. A complete behavioural theory of LLM pricing
will therefore need to model both payoff-feedback search and richer
response-sensitive anchoring.

The mixed-deployment experiments refine the interpretation of rival-price
visibility. Reducing rival-price information does not mechanically make pricing
competitive. It changes the information channel. In some ablations, summarized
histories and hidden rival prices weaken simple high-price coordination. In
\PRone--\DSone{} interaction, however, a \DSone{} agent that does not observe rival prices can
still infer from its own quantities and profits that high prices remain
profitable when a \PRone{} agent acts as a high-price anchor. It can then learn a threshold
below that anchor and earn high profits. The \PRone{} agent may periodically reset to the
price ceiling or briefly lower price to harvest demand. This anchor-and-harvest
pattern illustrates why deployment components must be evaluated jointly rather
than one at a time.

The welfare implications are regime-dependent. A high-price \PRone{} regime raises
the familiar concern of supracompetitive pricing and lower consumer welfare. A
\DStf{} agent that undercuts standard learners keeps prices low but takes \(72\)--\(74\%\)
of industry profit in the final 10\% of periods. In the homogeneous market, total welfare is fixed and consumer
surplus is determined by the average market-clearing price, so average prices
measure consumer harm there. They do not identify the mechanism: the same average
can come from a stable plateau, from anchor-and-harvest or from one seller
undercutting another. Regulators and platforms therefore need transaction-level
price paths to detect anchoring, intermittent harvesting and systematic undercutting.

Supplementary Note~9 reports the welfare decomposition of the main regimes. In
the homogeneous market total welfare is fixed, and the regimes differ in its
division: consumers keep \(95\%\) of the surplus under \DStf{} and \DSff{} self-play and
\(72\)--\(84\%\) when a \DStf{} agent undercuts standard learners, but \(15\%\) under
\PRone{} self-play. In the logit market, where high prices also destroy surplus, the
Fish-style control loses \(12.9\%\) of the Nash welfare, and the full \DSff{}
deployment interface reduces this loss to \(0.8\%\).

The policy implication is not that this paper provides a complete audit or
certification protocol. Rather, our results identify why any behavioral
evaluation of LLM pricing agents must go beyond an evaluation of the prompts. The
relevant object is the deployed pricing system: base model, prompt, action space,
memory structure, history representation, rival-price visibility, update
frequency, sampling procedure, and opponent ecology. The same base model may
produce competitive prices under one deployment and supracompetitive prices under
another. Testing a model under one prompt, one market, or one opponent is therefore
insufficient. This complements recent calls for certification of AI reasoning
agents before market deployment \cite{RiemerEtAl2026Certification}: our
experiments specify dimensions along which such certification would need to vary.

These findings suggest that regulating LLM pricing requires system-level
governance rather than prompt-level intent checks. A pricing system should be
evaluated across market structures, opponent ecologies and interface designs.
Certification tests should vary rival-price visibility, raw versus summarized
histories, point-price versus distributional output, update cadence and memory
length. Logging requirements should record the model version, prompt,
observation stream, memory state, price recommendation, sampling procedure and
human overrides. Platform operators could also constrain the information
interface of pricing agents by providing lagged, aggregated or summary-based
market feedback rather than raw rival-price histories. Such constraints would
not guarantee competitive outcomes, but our experiments show that they can alter
the likelihood and form of supracompetitive regimes.

A second regulatory challenge is design: how should LLM pricing systems be
parameterized so that consumer welfare is not harmed? Our evidence suggests that
interface design can become a policy lever, but not a simple one. Summary
histories, distributional output, and blockwise updating can make simple high-price conventions harder to sustain. But these choices must be evaluated jointly with market
structure and opponent ecology. 

Several limitations remain. Observable decision traces are behavioural artefacts,
not direct observations of hidden computation. Our markets are stylized Bertrand
environments, chosen because Nash and joint-profit benchmarks are transparent.
Real pricing systems involve inventory, search, consumer heterogeneity,
multi-product portfolios, platforms and asynchronous updates. The current
\PRone-versus-learner evidence documents profit advantages against
Exp3 and simple heuristics, but a more systematic profit comparison against a
broader set of online learners would strengthen the
adoption-incentive analysis. DBBS is a behavioural surrogate for the updates of \DStf{}, not
a mechanistic theory of LLM cognition and not a no-regret guarantee. These
limitations point to a broader agenda: studying foundation-model agents as
deployment-induced adaptive systems whose regimes depend on market structure,
opponent ecology and interface design, and developing new tools to characterize
and govern the dynamic strategies that result.

\section{Methods}
\label{sec:methods}

\subsection{Market environments}

We study repeated Bertrand price competition. In each period, firms simultaneously
choose prices and then observe market feedback. The market accepts any price in the
feasible interval \([\underline p,\overline p]\); individual pricing agents may operate
on continuous or discrete action spaces within this interval. The standard learners
and the \DS{} interfaces use the grid of \(K=21\) prices
\[
    \mathcal P=\{p^0,\ldots,p^{K-1}\},\qquad p^0=c=\underline p,\quad p^{K-1}=\overline p,
\]
whereas \PRone{} and \PSone{} post continuous prices unless they are restricted to the grid
for a matched comparison. The grid has evenly spaced prices \(p^k=c+k\delta\) and grid step
\[
    \delta=\frac{p^{K-1}-c}{K-1}.
\]

In the homogeneous winner-takes-all market, firms sell identical goods and demand
is allocated to the firm with the lowest price. If multiple firms set the lowest
price, demand is split equally among them. For firm \(f\), let \(p_{ft}\) denote
its price in period \(t\), and define
\[
    L_t=\{f:p_{ft}=\min_g p_{gt}\}.
\]
Firm \(f\)'s quantity is
\[
q_f(p_t)=
\begin{cases}
D/|L_t|, & \text{if } f\in L_t,\\
0, & \text{otherwise,}
\end{cases}
\]
and profit is
\[
    \pi_f(p_t)=(p_{ft}-c)q_f(p_t).
\]
Unless otherwise stated, the homogeneous experiments use \(c=0.10\), \(D=1000\),
\(K=21\), \(\overline p=0.36\), and horizon \(T=2000\). The upper bound \(\overline p\) is the
consumers' reservation value: demand is \(D\) up to this price and zero above it, so
\(\overline p\) is also the monopoly price. When all \(n\) firms post the
same grid price \(p^k=c+k\delta\), each earns \(k\delta D/n\); undercutting by one grid step
earns \((k-1)\delta D\), and any higher price earns zero. The symmetric profile is therefore
a Nash equilibrium if and only if \(k\le n/(n-1)\). The finite-grid Nash region is
\[
    \{0.100,0.113,0.126\}\ \text{for } n=2,
    \qquad
    \{0.100,0.113\}\ \text{for } n\ge3.
\]
We mark this region in price-path figures rather than treating the competitive
benchmark as a single grid point. In markets that pair a continuous-price agent with
grid agents, the same prices serve as a grid-based competitive benchmark rather
than as the Nash region of the asymmetric game.

In the differentiated-product market, firm \(i\)'s demand follows the
multinomial-logit system
\[
q_i(p_i,p_{-i})
=
\frac{\exp((a_i-p_i)/\mu)}
{\sum_{j=1}^n \exp((a_j-p_j)/\mu)+\exp(a_0/\mu)},
\]
and profit is
\[
    \pi_i(p_i,p_{-i})=(p_i-c_i)q_i(p_i,p_{-i}).
\]
Unless otherwise stated, the differentiated experiments use \(n=2\),
\(a_1=a_2=2\), \(a_0=0\), \(\mu=0.25\), and \(c_1=c_2=1\). The continuous
one-shot Nash price is approximately \(p^{\mathrm{Nash}}=1.473\), and the
joint-profit price is approximately \(p^{\mathrm{monopoly}}=1.925\). We report
normalized price elevation in the differentiated market as
\[
    \Delta(p)
    =
    \frac{p-p^{\mathrm{Nash}}}
    {p^{\mathrm{monopoly}}-p^{\mathrm{Nash}}}.
\]

In the homogeneous market, we primarily report prices in levels and mark the
finite-grid Nash region. When a normalized homogeneous price index is used, we
define
\[
    \Delta^{H}(p)=\frac{p-c}{p^{K-1}-c},
\]
so that \(\Delta^{H}=0\) corresponds to marginal cost and
\(\Delta^{H}=1\) corresponds to the upper grid price. Marginal cost is the lowest
symmetric equilibrium price for every \(n\), so \(\Delta^{H}\) has the same anchor for
any number of firms; the finite-grid Nash region is \(\Delta^{H}\le0.10\) for two firms
and \(\Delta^{H}\le0.05\) for three or more.

\subsection{Feedback structures}

The information available to a pricing agent depends on the deployment. In
visible-rival-price treatments, firm \(i\) observes the rival price \(p_{-i,t}\)
after period \(t\). In the homogeneous Bertrand market with known cost, demand
and tie-breaking rule, this is full-information feedback. In a two-firm market,
observing \(p_{-i,t}\) allows firm \(i\) to reconstruct the counterfactual payoff
of every price \(p\in\mathcal P\):
\[
\pi_i(p,p_{-i,t})=
\begin{cases}
(p-c)D, & p<p_{-i,t},\\
(p-c)D/2, & p=p_{-i,t},\\
0, & p>p_{-i,t}.
\end{cases}
\]
Thus, visible-rival-price treatments are benchmarked not only against bandit
learners but also against full-information learners such as FTRL,
coin betting and full-information regret matching.

In hidden-rival-price treatments, the agent observes only its own price, quantity
and profit. Because the quantity follows from the own price and profit,
\(q=\pi/(p-c)\), this is bandit feedback: the payoff of the chosen price only.
The feedback is nevertheless structured. In the winner-takes-all market a sale of
the whole demand at \(p\) implies the same sale at every lower price, so the
payoffs of all lower prices follow; no sale at \(p\) implies no sale at any higher
price; and a split market reveals that the rival posted the same price. An agent
that treats prices as ordered can therefore infer part of the counterfactual
payoff vector from the same observations that a standard bandit learner, which
treats prices as unrelated arms, uses only for the price it played. LLM agents
receive the same own-market feedback as the bandit learners, together with the
price values and their marginal cost; they are not told the demand model.

Rival-price observability is part of the information setting, not of the
deployment, and is specified per firm: in the heterogeneous experiments one firm
can observe its rival's price while the other does not.

\subsection{Pricing agents and LLM deployments}

We distinguish three kinds of pricing agents: standard non-LLM learners, LLM agents
and simple pricing controls. Each firm \(i\) is described by its decision engine
(a learning algorithm, DBBS or a foundation model), its information setting
\(\iota_i\) (whether it observes the rival's price) and, for LLM agents, its
deployment.

\paragraph{Standard learners.}
The non-LLM baselines include bandit-feedback learners, full-information learners
and reinforcement-learning baselines. Bandit-feedback learners include Exp3,
BM-RM, UCB and related stochastic or adversarial bandit algorithms. Exp3 is an
adversarial-bandit external-regret algorithm; BM-RM is a bandit implementation of
the Blum--Mansour reduction from external to internal or swap regret; UCB is an
optimism-based stochastic bandit learner. Full-information learners include
FTRL, coin betting, full-information regret matching \cite{HartMasColell2000} and weighted fictitious play. These
algorithms are used when the rival price is visible and the counterfactual payoff
vector can be reconstructed. Reinforcement-learning baselines include PPO and
additional variants reported in the Supplementary Information.

\paragraph{LLM deployments.}
A deployment is the interface through which a foundation model prices,
\[
D=(\mathcal H,\mathcal A,\mathcal O,\Delta,\mathcal M,\mathcal P),
\]
where \(\mathcal H\) is the history representation and memory window, \(\mathcal A\)
the action space, \(\mathcal O\) the output representation, \(\Delta\) the
intervention interval, \(\mathcal M\) the persistent state carried between
decisions and \(\mathcal P\) the prompt and invocation configuration (objective
framing, price presentation, response template, temperature and reasoning
setting). We name deployments by the three components that the protocol
ablations vary: output (\textsf{P}, a point price; \textsf{D}, a probability
distribution over the price grid), history (\textsf{R}, raw period-level
observations; \textsf{S}, a per-price summary) and intervention interval.
Table~\ref{tab:deployments} lists the five deployments used in the paper; the
Supplementary Information gives every component.

\PRone{} reimplements the pricing loop of Fish et al.\ and serves as the positive
control: one price every period from the raw history of the last 100 periods, the
profit-oriented framing \prompt{1} and a persistent scratchpad that the model rewrites at
each decision. Against standard learners its prices are restricted to the
21-point grid, so that both firms share one action set and exact ties can occur;
elsewhere it posts continuous prices. \PSone{} is the same agent with a
summarized history. The distribution--summary deployments \DS{} emit a
probability distribution over the 21-point grid from a per-price summary that
reports the number of observations, mean quantity, mean profit, profit dispersion,
normalized-profit statistics and a trend; prices are sampled from this
distribution during the following intervention block. \DStf{} and \DSff{} are the
configurations frozen by the learning and aggressive selection rules of the
configuration screen; they return a JSON object containing the distribution and an
updated \texttt{plans\_and\_insights} field and also see their previous
distribution. \DSone{} applies the summary-and-distribution interface every
period inside the Fish harness, so that it can be paired with \PRone{} under the
same template and cadence. The persistent text is fed back at the next decision
and acts as language-mediated memory; we treat it as an observable behavioural
trace, not as hidden chain-of-thought.

\begin{table}[t]
\centering
\caption{LLM deployments. Interval, periods between LLM decisions; window,
periods summarized or listed in the history. All deployments use Gemini~3 Flash
at temperature 0.7 unless stated otherwise; rival-price observability is set by
the information setting and varies within deployments.}
\label{tab:deployments}
\small
\begin{tabularx}{\linewidth}{@{}lllccX@{}}
\toprule
\textbf{Interface code} & \textbf{Output} & \textbf{History} & \textbf{Interval} & \textbf{Prompt} & \textbf{Used in} \\
\midrule
\PRone & point & raw, 100 & 1 & 1 & Figs.~2c,d, 4, 5, 6 \\
\PSone & point & summary, 100 & 1 & 1 & Fig.~5 \\
\DSone & distribution & summary, 100 & 1 & 1 & Figs.~5, 6 \\
\DStf & distribution & summary, 75 & 25 & 0 & Figs.~2a,b, 3, 4a \\
\DSff & distribution & summary, 150 & 50 & 1 & Fig.~4a,c \\
\bottomrule
\end{tabularx}
\par\smallskip\footnotesize The
full-bundle arm of the protocol ablations (Fig.~4c) runs \DSff{} inside the Fish harness with a
250-period window. \PRone{} keeps the two-file scratchpad of Fish et al.\ in the Fish-style positive
control and the logit-market matchups; the protocol ablations and all homogeneous-market runs use
one merged file (Supplementary Note~1).
\end{table}

\paragraph{Simple pricing controls.}
We also use static and deterministic pricing controls, including static monopoly
pricing, static competitive pricing, win-up--lose-down rules, randomized
win-up--lose-down rules and deterministic undercut-last heuristics. These
controls are used to test whether LLM agents can identify and exploit simple
pricing regularities.

\subsection{Configuration screening}

Candidate configurations are screened only in homogeneous self-play before
cross-agent evaluation. The configuration screen evaluates LLM and non-LLM
candidates over five seeds with \(K=21\) and \(T=2000\). For LLMs, the screen
varies model, temperature, intervention interval, history representation and
objective framing. For non-LLM agents, it tunes Exp3, BM-RM, PPO and UCB. The
learning score ranks candidates by external regret, swap/internal regret,
NashConv and price volatility. The aggressive score first removes candidates in
the worst quartile of any learning-quality metric and then ranks remaining
candidates by profit and price elevation.

The frozen learning configuration, \DStf, uses Gemini 3 Flash Preview, temperature \(0.7\),
neutral \prompt{0}, summary history and intervention interval 25. The frozen
aggressive configuration, \DSff, uses Gemini 3 Flash Preview, temperature \(0.7\), \prompt{1},
summary history and intervention interval 50. The frozen learning baselines are
Exp3 with \(\gamma=0.30\), BM-RM with \(\gamma=0.50\) and \(\eta=\gamma/21\),
PPO with rollout \(R=50\) and learning rate \(0.005\), and UCB with
\(\alpha=1.7\). Aggressive baselines use Exp3 with \(\gamma=0.125\), PPO with
learning rate \(0.010\), and UCB with \(\alpha=1.0\); BM-RM is unchanged.
Additional full-information learners and heuristic controls are specified in the
Supplementary Information.

\subsection{Experiment families}

The experiments are organized into seven families.

\paragraph{Standard learner round robin.}
Seven standard learners (Exp3, BM-RM, UCB, PPO, FTRL with a Euclidean regularizer,
full-information regret matching and weighted fictitious play) compete in every unordered pair, including
self-play, in the homogeneous Bertrand market for \(10^7\) periods with ten seeds
and full rival-price information. Each learner uses one frozen configuration with
decaying exploration; Exp3 and every BM-RM row learner use the anytime tuning of
\cite{AuerCesaBianchiFreundSchapire2002Exp3}. This establishes the
equilibrium-learning benchmark and identifies exceptions such as UCB-driven
high-price cells. The Supplementary Information reports the full
round-robin matrices for prices, profits, consumer surplus, total welfare,
NashConv and regret diagnostics.

\paragraph{\PRone{} versus standard learners and heuristics.}
\PRone, restricted to the price grid, is matched against Exp3, PPO, coin betting,
full-information regret matching, weighted fictitious play and simple pricing
rules in the homogeneous market. These cells test whether it can generate private profit
advantages against slow learners or exploitable heuristics.

\paragraph{\DStf{} versus standard learners.}
\DStf{} is matched against Exp3, BM-RM, PPO and UCB in the
homogeneous market. These cells test whether it earns profits by undercutting
learners that remain above the competitive floor.

\paragraph{LLM self-play.}
We evaluate \PRone{} self-play and \DStf{} and \DSff{} self-play in homogeneous and differentiated
Bertrand environments. These cells test whether the deployed LLM system selects
a high-price regime, a finite-grid Nash-like regime or another recurrent pattern.

\paragraph{Deployment ablations.}
Starting from the \PRone{} control, we vary the output representation, history
representation and intervention interval of the deployment and the rival-price
visibility of the information setting. These ablations test
whether high-price regimes are driven by point actions alone or by the broader
deployment interface.

\paragraph{Mixed LLM deployments.}
We evaluate \PRone--\DSone{} and \PRone--\PSone{} interactions, and markets with
three or four LLM agents, under visible and hidden rival prices. These cells test whether heterogeneous LLM deployments generate
asymmetric or dynamic regimes, including anchor-and-harvest cycles.

\paragraph{Robustness families.}
Additional experiments vary prompts, foundation models, opponent types, demand
systems and evaluation windows. Detailed results are reported in the online
supplement.

\subsection{Held-out evaluation}

The held-out homogeneous evaluation uses new trajectories not used in the
configuration screen. Unless otherwise stated, homogeneous \DStf{} and \DSff{} evaluation uses
ten independent seeds per cell and horizon \(T=2000\). Main estimates use the
final \(10\%\) window,
\[
    W=\{1801,\ldots,2000\}.
\]
Evaluation-window robustness varies the final window over \(5\%\), \(10\%\),
\(20\%\), \(50\%\) and the full horizon.

Point-price experiments use either the differentiated logit market or the
homogeneous market, as stated in the corresponding figure or table. Deployments
with an LLM call in every period (\PRone, \PSone, \DSone) use shorter horizons;
blockwise deployments use longer horizons. Exact horizons, call budgets,
replication counts, prompt variants, model versions and random seeds are reported
in the Supplementary Information.

\subsection{Outcome measures and uncertainty}

For each seed \(s\) and evaluation window \(W\), we compute average price,
transaction price, quantity, profit, cumulative profit and profit share. For
asymmetric cells, we compute LLM-minus-opponent gaps in price, profit and
quantity:
\[
    G^{Y}_s(W)
    =
    \frac{1}{|W|}
    \sum_{t\in W}
    \left(Y_{s,\mathrm{LLM},t}-Y_{s,\mathrm{opp},t}\right),
    \qquad
    Y\in\{p,\pi,q\}.
\]
The undercut share is
\[
    U_s(W)
    =
    \frac{1}{|W|}
    \sum_{t\in W}
    \mathbf 1\{p_{s,\mathrm{LLM},t}<p_{s,\mathrm{opp},t}\}.
\]

Realized external regret is computed relative to the best fixed grid price on
the observed opponent path:
\[
    ER_{s,f}(W)
    =
    \max_{p\in\mathcal P}
    \sum_{t\in W}\pi_f(p,p_{s,-f,t})
    -
    \sum_{t\in W}\pi_f(p_{s,f,t},p_{s,-f,t}).
\]
This is a finite-horizon, path-conditioned diagnostic and not policy regret
against a strategically responding opponent. We report it for the held-out runs in
Supplementary Note~4.

For empirical mixed profiles \(\sigma\), we compute NashConv as
\[
\mathrm{NashConv}(\sigma)
=
\sum_f
\left[
\max_{p_f\in\mathcal P}
\mathbb E_{p_{-f}\sim\sigma_{-f}}\pi_f(p_f,p_{-f})
-
\mathbb E_{p\sim\sigma}\pi_f(p)
\right].
\]

For welfare analyses, we report consumer surplus, producer surplus and total
welfare using the market primitives of the corresponding environment. In the
homogeneous market, welfare calculations use the demand and value primitives
reported in the Supplementary Information. In the differentiated logit market, welfare
calculations use the logit demand system. The Supplementary Information provides exact
formulas and all welfare tables.

All main estimates are computed at the seed level. We report means and bootstrap
confidence intervals obtained by resampling independent seed-level trajectories.

\subsection{Price regimes}

We use the term price regime descriptively for persistent or recurrent patterns
in the coupled price path of deployed agents. The labels in
Fig.~\ref{fig:heterogeneous-regime-map} are qualitative descriptions of each run's
price path, not the output of a classification rule. \emph{Competitive undercutting}:
the market-clearing price (the lowest posted price) stays in or near the finite-grid
Nash region while agents undercut each other. \emph{High-price lock-in}:
the market-clearing price stays at or one grid step below the ceiling. \emph{Anchor-and-harvest}:
one agent holds or repeatedly restores a high price, and another prices below it and
captures demand.

In the \PRone--\DSone{} runs, anchor-and-harvest has three features: (i) a high-price anchor phase in
which the \PRone{} agent remains near the upper part of the price range, (ii) a threshold-learning phase
in which the \DSone{} agent prices below the anchor and earns positive profits, and (iii) intermittent
harvest moves in which the \PRone{} agent lowers its price briefly to regain demand. All heterogeneous
runs are single runs; the Supplementary Information shows the price path of every run.

We do not infer from persistence alone that a dynamic regime is an equilibrium.
A dynamic-equilibrium interpretation would require counterfactual deviation
analysis against the induced opponent policy. The present experiments document
the regimes selected by deployed agents and report path-level diagnostics;
formal equilibrium certification of recurrent regimes is left to follow-up work.

\subsection{Reasoning traces and DBBS}

We analyse \texttt{plans\_and\_insights} outputs as observable behavioural traces.
They are not treated as hidden chain-of-thought or direct observations of model
cognition. For motif extraction, complete LLM call records were analysed using a
fixed coding prompt and then checked by the authors. Each record contained the
market history available to the pricing agent, the previous pricing strategy, the
newly emitted price or distribution and the updated \texttt{plans\_and\_insights}
field.

We distinguish two trace-level patterns. \PRone{} traces often use opponent-facing
language, referring to anchoring, avoiding price wars, resetting and harvesting.
\DStf{} and \DSff{} traces more often use payoff-region language, referring to successful and
failed price regions, local probes and thresholds. Motif frequencies are measured
by coding every trace against a fixed keyword lexicon, with precision checked on
randomly drawn matches (Supplementary Note~5). DBBS is constructed only as a
surrogate for \DStf{} updates.

DBBS treats prices as an ordered grid. It maintains a recency-weighted payoff
belief \(v_k\) and a failure belief \(\phi_k\) for each tick \(k\). At each
intervention epoch, it observes realized prices and normalized profits and
identifies the modal played tick \(m\) of the block. For each period of the block,
in order, with played tick \(k\) and normalized profit \(r_k\), it updates
\[
    v_k\leftarrow \lambda r_k+(1-\lambda)v_k,
    \qquad
    \phi_k\leftarrow \lambda y_k+(1-\lambda)\phi_k,
    \qquad
    y_k=\mathbf 1\{k>0\}\,\mathbf 1\{r_k\le0\}.
\]
Ticks not played keep their beliefs. The failure signal \(y_k\) is the zero-sales
indicator: in the winner-takes-all market a price above marginal cost earns zero
profit exactly when it does not sell, and marginal cost (\(k=0\)) is never a failure.
DBBS is fitted and deployed only in the homogeneous market. DBBS scores ticks by
\[
    S_k=v_k-\beta_{\mathrm{fail}}\phi_k-\beta_{\mathrm{loc}}(k-m)_+.
\]
The first term rewards recently profitable prices, the second penalizes failures,
and the third imposes one-sided caution for prices above the current focal
region. A softmax with temperature \(\tau\) maps scores into a base distribution.
The base distribution is mixed with local upward exploration: with weight \(\rho\),
probability mass is placed on ticks up to \(G_{\mathrm{up}}\) steps above the
current focal tick \(m\). The emitted distribution mixes this search policy with a
persistence component, a recency-weighted histogram of the agent's own recent prices:
\(\sigma_{n+1}=w\,\sigma^{\mathrm{pers}}_{n+1}+(1-w)\,\sigma^{\mathrm{search}}_{n+1}\), with the inertia
weight \(w\in\{0,0.05,\ldots,1\}\) selected for each fold on the training seeds by one-step loss; \(w=0\)
is DBBS without inertia.

DBBS is evaluated under teacher forcing and free-running deployment. Under
teacher forcing, DBBS follows recorded LLM histories and predicts the next
LLM-emitted distribution. Under free-running deployment, the fitted DBBS policy
replaces the LLM and acts endogenously. Parameter selection uses leave-one-seed-out
cross-validation over environment seeds and minimizes a normalized squared-CDF
loss. Evaluation reports first Wasserstein distance normalized by the price
range. The validation fits one parameter vector per fold jointly on the seven
\DStf{} treatments (ten environment seeds, ten folds). Parameters are
selected from 20,000 Latin-hypercube candidate vectors over
\[
\lambda\in[0.05,0.99],\quad
\beta_{\mathrm{fail}}\in[0,2],\quad
\beta_{\mathrm{loc}}\in[0,0.30],\quad
\tau\in[0.01,0.20],\quad
\rho\in[0,0.40],\quad
G_{\mathrm{up}}\in\{0,\ldots,6\}.
\]
The Supplementary Information reports the full DBBS construction,
heuristic baselines, validation results and theoretical statements.

\subsection{Generative AI use}

GPT-5.6 Sol was used for qualitative motif extraction from LLM call logs and for
assistance with language editing. All extracted motifs, generated text, code
outputs, statistical estimates and mathematical claims were checked by the
authors. Market outcomes and numerical results were computed from the
experimental code and were not generated by the language model.

\subsection{Supplementary Information}

The Supplementary Information provides the complete experimental specification and the
supporting evidence for each main-text figure: the markets, deployments, standard
learners and metrics (Supplementary Note~1); the configuration screen and frozen
configurations (Note~2); the full round-robin matrices and convergence paths
(Note~3, Fig.~1); held-out price paths, evaluation-window robustness and all \PRone{}
runs against standard learners and pricing rules (Note~4, Fig.~2); the trace-motif
extraction and motif frequencies (Note~5); the DBBS construction, fitting, baselines, free-running
controls and proofs (Note~6, Fig.~3); the
deployment-ablation design and diagnostics (Note~7, Fig.~4); logit-market \PRone{}
matchups and the trajectories of every heterogeneous-deployment run (Note~8,
Figs.~5--6); the welfare decomposition of the main regimes (Note~9); and all
prompt templates (Note~10).

\section{Data availability}
All data needed to reproduce the figures and tables will be made available in a public repository upon publication, including simulation outputs, prompt templates, processed analysis tables and LLM call records permitted for release.

\section{Code availability}
Code for the market environments, agent implementations, statistical analysis and figure generation will be made available in a public repository upon publication.

\bibliography{references}

\let\arx@newcommand\newcommand \let\newcommand\providecommand
\let\arx@newtheorem\newtheorem \def\newtheorem#1#2{}
\let\arx@figurename\figurename \let\arx@tablename\tablename

\title[Supplementary Information]{Supplementary Information for ``How language model agents reshape price competition''}

\author*[1]{\fnm{Martin} \sur{Bichler}}
\author[1]{\fnm{Philip} \sur{Hierhager}}
\author[1]{\fnm{Kassian} \sur{Köck}}
\author[1]{\fnm{Fabian} \sur{Pieroth}}

\affil*[1]{\orgdiv{School of Computation, Information and Technology}, \orgname{Technical University of Munich}, \country{Germany}}

\maketitle

\tableofcontents

\suppnote{Experimental setup}{supp:setup}

This note defines the markets, the two LLM deployments, the standard learners and the evaluation
metrics used throughout the paper. The configuration screen that fixed the hyperparameters of the main
homogeneous experiments is documented in Supplementary Note~\ref{supp:protocol}.

\subsection{Markets}

\paragraph{Homogeneous winner-takes-all Bertrand market.}
Two (or more) firms with marginal cost \(c=0.10\) post prices on the 21-point grid
\(\Pgrid=\{0.100,0.113,\ldots,0.360\}\). The market demand of \(1{,}000\) units goes to the firm with the
lowest price and is split equally among tied firms; demand is inelastic up to the reservation value
\(0.36\), which is also the grid ceiling. Normalized profit divides profit by the one-firm monopoly profit
\((0.36-0.10)\times1{,}000=260\).

Because of the discrete grid and tie splitting, the symmetric pure Nash profiles depend on the number of
firms \(n\). If all firms post \(c+k\delta\), each earns \(k\delta D/n\), and undercutting by one step earns
\((k-1)\delta D\); the profile is an equilibrium if and only if \(k\le n/(n-1)\). With two firms the symmetric
equilibrium prices are \(0.100\), \(0.113\) and \(0.126\); with three or more firms they are \(0.100\) and
\(0.113\). Figures display this finite-grid competitive region rather than treating the stage-game benchmark
as a single price.

\paragraph{Differentiated logit Bertrand market.}
The differentiated market follows the symmetric multinomial-logit Bertrand model. Firm \(i\)'s demand is
\[
q_i(p_i,p_{-i})=
\frac{\exp((a_i-p_i)/\mu)}
{\sum_{j=1}^n \exp((a_j-p_j)/\mu)+\exp(a_0/\mu)} ,
\]
and profit is \(\pi_i(p_i,p_{-i})=(p_i-c_i)q_i(p_i,p_{-i})\). Unless otherwise stated, we use \(n=2\), \(a_1=a_2=2\), \(a_0=0\), \(\mu=0.25\), and \(c_1=c_2=1\). The continuous one-shot Nash price is approximately \(p^{\mathrm{Nash}}=1.473\), and the joint-profit price is approximately \(p^{\mathrm{monopoly}}=1.925\). We report
\[
\Delta(p)=
\frac{p-p^{\mathrm{Nash}}}{p^{\mathrm{monopoly}}-p^{\mathrm{Nash}}}.
\]

\subsection{Deployment parametrization}
\label{supp:deployments}

We separate four layers of an experiment. The \emph{market} fixes demand (homogeneous winner-takes-all or
differentiated logit), the number of firms, costs, valuations, the feasible price interval and the
tie-breaking rule. The \emph{information setting} \(\iota_i\) of firm \(i\) states whether it observes its
rivals' prices; own-market feedback (own price, quantity and profit) is always observed. The \emph{decision
engine} is a learning algorithm, DBBS or a foundation model. For LLM agents, the \emph{deployment}
\[
D=(\mathcal H,\mathcal A,\mathcal O,\Delta,\mathcal M,\mathcal P)
\]
is the interface through which the model prices: \(\mathcal H\), the history representation (raw or
summary) and its window; \(\mathcal A\), the action space (continuous interval or a grid of \(K\) prices);
\(\mathcal O\), the output representation (a point price or a probability distribution over the grid);
\(\Delta\), the intervention interval; \(\mathcal M\), the persistent state carried between decisions (the
free-text plans, in one or two files, and for distributional outputs the previous distribution); and
\(\mathcal P\), the prompt and invocation configuration (objective framing, price presentation, response
template and parsing, temperature and reasoning setting). Recurring deployments are denoted by an
output--history--interval interface code, for example \DStf{} for a distribution from a summarized
history every 25 periods. The codes are shorthand identifiers rather than complete specifications.
Supplementary Table~\ref{tab:supp-deployments} gives every component of the five deployments.

\begin{table*}[t]
\centering
\caption{Components of the LLM deployments. Windows in periods. Fish template: the free-text reply
template of \cite{FishGonczarowskiShorrer2026AlgorithmicCollusionLLMs} with observations, scratchpad and
price fields; structured: a JSON response validated against a schema. Rival-price observability is part of
the information setting and is reported per experiment.}
\label{tab:supp-deployments}
\small
\begin{tabularx}{\textwidth}{@{}lXXXcXX@{}}
\toprule
\textbf{Interface code} & \(\mathcal H\) & \(\mathcal A\) & \(\mathcal O\) & \(\Delta\) & \(\mathcal M\) & \(\mathcal P\) \\
\midrule
\PRone & raw, 100 & continuous; 21-point grid against standard learners & point & 1 &
scratchpad, one file (two in the logit positive control) & \prompt{1}, Fish template; prices \(\times 20\) in the
homogeneous market \\
\PSone & summary, 100 & continuous & point & 1 & scratchpad, one file & \prompt{1}, Fish template, \(\times 20\) \\
\DSone & summary, 100 & 21-point grid & distribution & 1 & scratchpad, one file & \prompt{1}, Fish template,
\(\times 20\) \\
\DStf & summary, 75 (\(3\Delta\)) & 21-point grid & distribution & 25 & \texttt{plans\_and\_insights},
previous distribution & \prompt{0}, structured, \(\times 20\) \\
\DSff & summary, 150 (\(3\Delta\)) & 21-point grid & distribution & 50 & \texttt{plans\_and\_insights},
previous distribution & \prompt{1}, structured, \(\times 20\) \\
\bottomrule
\end{tabularx}
\par\smallskip\footnotesize All deployments use \texttt{gemini-3-flash-preview} at temperature 0.7, except in
the GPT-5.6 Terra runs of main-text Fig.~5. The protocol ablations of Supplementary Note~\ref{supp:ablations} vary these components one at a time
inside the Fish harness, starting from \PRone; their full-bundle arm is \DSff{} with a 250-period window.
\end{table*}

\subsection{Distribution--summary deployments \DStf{} and \DSff}

The configuration screen for the distributional interfaces varies five dimensions:
\begin{enumerate}[label=(\roman*)]
    \item language model,
    \item sampling temperature,
    \item intervention interval,
    \item history representation, and
    \item objective framing.
\end{enumerate}

We use a one-factor-at-a-time design rather than a full factorial: each dimension is varied in turn relative to a common baseline, and the baseline level is included as a candidate within every varied dimension. The baseline is \texttt{gemini-3-flash-preview} at temperature \(0.7\), intervention interval \(\Delta=50\), the summary history over \(3\Delta\), and the neutral framing \prompt{0}. This yields the \(10\) distinct configurations and \(14\) screened cells reported above.

For model capability we compare a weaker variant, \texttt{gemini-3.1-flash-lite}, with the stronger \texttt{gemini-3-flash-preview}; note that these differ in generation as well as in size. We evaluate sampling temperatures \(0\), \(0.7\), and \(1.0\), and intervention intervals of \(25\), \(50\), and \(100\) periods.

Each \DStf{} and \DSff{} decision uses a wrapper prompt that provides the objective and task, a public product and cost block stating the firm's own marginal cost, the observed history, the available pricing actions together with the admissible price range, the agent's previous \texttt{plans\_and\_insights}, and its previous mixed strategy. The agent is told its own marginal cost but never the rival's cost, price, or profit. The response schema contains a probability distribution over the price grid, validated to be non-negative and to sum to one, and a new free-text \texttt{plans\_and\_insights} field. The full wrapper is reproduced in Listing~\ref{lst:supp-base-wrapper}.

\paragraph{History representations.}
We consider three representations, which jointly vary the rendering and the depth of the observation window. The window is specified as a multiple of the intervention interval rather than in absolute periods, so a \(3\Delta\) window covers the last \(3\Delta\) periods and the same label corresponds to \(75\) periods at \(\Delta=25\) but \(150\) periods at \(\Delta=50\).

The summary representation aggregates the window into one row per price tick, preceded by a legend, and reports: price in euros, number of observations, mean quantity, mean profit, profit standard deviation, mean normalized profit, normalized-profit standard deviation, best normalized profit, most recent normalized profit, the trend in normalized profit, the most recent period in which the price was used, the mean observed competitor price, and a flag indicating whether the pricing strategy changed before any observation at that price. The trend is the slope of a least-squares fit of normalized profit on period for that tick, set to zero when its absolute value falls below \(5\times10^{-4}\). Under the local information scope used in the screen the competitor-price column is always \texttt{NA}; the column and its legend entry are nonetheless present in the prompt.

The two raw representations instead list one row per period, reporting the period, the agent's own price, the competitor price, the quantity sold, and the profit. As in the summary representation, the competitor-price column is rendered as \texttt{NA} throughout the screen, since the rival's price is not in the agent's information set under the local scope; the column is populated only in the full-information cells described below. The two raw variants differ only in window depth, covering either \(1\Delta\) or \(3\Delta\).

\paragraph{Objective framings.}
The neutral framing \prompt{0} asks the agent, in the first person, to choose prices that perform well for its own product using only the information provided. Prompts~1 and~2 instead adopt the profit-oriented objective of \cite{FishGonczarowskiShorrer2026AlgorithmicCollusionLLMs}, which casts the agent as assisting a user and makes long-run profit its top priority. Prompts~0 and~1 therefore differ in their entire base framing and not only in an appended instruction; Prompts~1 and~2 share that base and differ only in the appended exploration instruction. \prompt{1} adds profit-protective exploration. \prompt{2} permits risky or aggressive exploration for data-gathering purposes and additionally states that pricing below the competitor typically increases units sold, so it supplies a directional hint about undercutting that Prompts~0 and~1 do not. The exact text of each framing is reproduced in Supplementary Note~\ref{supp:prompts}: \prompt{0} in Listing~\ref{lst:supp-p0}, the shared profit-oriented base of Prompts~1 and~2 in Listing~\ref{lst:supp-fish-base}, and their respective addenda in Listings~\ref{lst:supp-p1} and~\ref{lst:supp-p2}.

\paragraph{Price scale.}
All monetary values in the prompt, including marginal cost, the price grid,
the history statistics, and profits, are converted to euros and multiplied by
a factor of \(20\). The scaling affects only the presentation: demand,
profits, and all metrics are computed in the market's internal units. In the
homogeneous market, for example, the marginal cost of \(0.10\) appears as
\texteuro\(2.000\) and the price grid runs from \texteuro\(2.000\) to
\texteuro\(7.200\) in steps of \texteuro\(0.260\).

\subsection{Deployments in the Fish harness: \PRone, \PSone{} and \DSone}
\label{app:point-llm-configuration}

\PRone{} is a standalone reimplementation of the pricing loop of
\cite{FishGonczarowskiShorrer2026AlgorithmicCollusionLLMs}. Unlike \DStf{} and
\DSff, its configuration is not screened. It is held fixed at the
Fish-style control configuration. \PSone{} and \DSone{} run in the same harness and
differ from \PRone{} only in the components stated in Supplementary
Table~\ref{tab:supp-deployments}: \PSone{} replaces the raw history by the per-price
summary, and \DSone{} additionally emits a distribution over the 21-point grid.
The default configuration consists of
\begin{enumerate}[label=(\roman*)]
    \item Gemini 3 Flash Preview at sampling temperature \(0.7\),
    \item one decision in every period, that is, an intervention interval of \(1\),
    \item a raw history of the previous \(100\) periods,
    \item the rival's price shown in the history, and
    \item the profit-oriented framing \prompt{1}.
\end{enumerate}

Each \PRone{} decision uses a single free-text prompt in the format of
\cite{FishGonczarowskiShorrer2026AlgorithmicCollusionLLMs} which is detailed in \ref{supp:prompts}. It states the
objective and task, the firm's marginal cost and the admissible prices, the
current content of a persistent scratchpad file written by the agent in the
previous decision, and the market data. The agent answers by filling in a
fixed template whose fields are its observations and thoughts, the new
scratchpad content, and the chosen price. The new scratchpad content replaces
the old one. \cite{FishGonczarowskiShorrer2026AlgorithmicCollusionLLMs} use
two scratchpad files, one for plans and one for insights, and the reply template
has a separate slot for each. The scratchpad layout is a configurable component
rather than a fixed choice, and the experiments reported here do not all use the
same setting. The Fish-style positive control and the logit-market \PRone{} matchups of
Supplementary Note~\ref{supp:pllm-matchups} retain Fish's original two-file layout, so that they reproduce
their reference as closely as possible. The direct protocol ablations of
Supplementary Note~\ref{supp:ablations} and the homogeneous \PRone{} experiments (Supplementary
Notes~\ref{supp:llm-vs-learners} and~\ref{supp:pllm-matchups}) instead use a single merged
\texttt{PLANS\_AND\_INSIGHTS} file with one template slot, which matches the
single persistent-text field of \DStf{} and \DSff. The control of the
direct protocol ablations therefore differs in this one respect from the
separate Fish-style positive control, and the two should not be read as the same
configuration. If no valid price can be parsed from the
response, or the parsed price lies outside the admissible range, the agent keeps
its previous price.

\paragraph{Action representation.}
In the continuous variant the agent may choose any positive price, and the
prompt states the price above which no customer would buy
(Listing~\ref{lst:supp-point-continuous}). In the grid variant
the prompt lists the \(21\) prices of the action grid
(Listing~\ref{lst:supp-point-grid}), and the reply is mapped to
the nearest grid price. The matchups against non-LLM learners in the
homogeneous market use the grid variant, so that the LLM and its opponent choose
from the same action set and exact price ties remain possible.

\paragraph{History representation.}
The raw history lists one entry per period, most recent first. Each entry
contains the agent's own price, the rival's price, the quantity sold, and the
realized profit, rounded to two decimals. Protocol ablations that use the
summary representation replace this block with the per-price statistics of the
\DStf{} and \DSff{} described above.

\paragraph{Objective framing and price scale.}
\PRone{} uses \prompt{1}, the collusion-prone framing of
\cite{FishGonczarowskiShorrer2026AlgorithmicCollusionLLMs}. It asks the agent to
maximize profit in the long run, to explore many pricing strategies, and to
avoid actions that undermine profitability. In the logit market, prices and
profits are shown on the original scale of
\cite{FishGonczarowskiShorrer2026AlgorithmicCollusionLLMs}. In the homogeneous
market, they are shown in euros scaled by a factor of \(20\), as for the
\DStf{} and \DSff.

\paragraph{Deviations from the reference implementation.}
Relative to \cite{FishGonczarowskiShorrer2026AlgorithmicCollusionLLMs}, the configurations used here differ
in five respects, each described above: (i) the homogeneous-market experiments present prices and profits
in euros scaled by \(20\), while the logit market keeps the original scale; (ii) the homogeneous matchups
use the 21-point price grid instead of continuous prices, with replies mapped to the nearest grid price;
(iii) all experiments except the Fish-style positive control and the logit matchups merge the two
scratchpad files into a single \texttt{PLANS\_AND\_INSIGHTS} file; (iv) the summary history replaces the raw
history only in the protocol ablations that state it; and (v) the prompt keeps the \prompt{1} framing but adds the
market and grid information required by each variant (Supplementary Note~\ref{supp:prompts}). The
Fish-style positive control retains the continuous action, raw history, original scale and two-file
scratchpad.

\subsection{Standard learners}

\DStf{} and \DSff{} are matched against four standard learners, spanning distinct sequential-learning paradigms:
\begin{itemize}
    \item Exp3: adversarial bandit external-regret minimization.
    \item BM-RM: the bandit form of the Blum--Mansour reduction, which combines \(K\) external-regret learners into a swap-regret minimizer.
    \item UCB: optimism-based stochastic bandit learning.
    \item PPO: policy-gradient reinforcement learning.
\end{itemize}

All baselines use the same price grid and run under the same local information scope as the LLM, so none of them observes the rival's price. They receive bandit feedback from realized actions: after choosing a price, an agent observes the profit generated by that price but not the counterfactual profit of unchosen prices. The baselines nevertheless differ in how much of their own past they retain. Exp3, BM-RM, and UCB are stateless across the action set, updating only from the latest realized payoff, whereas PPO conditions on a fixed-length encoding of its own recent prices, quantities, and profits. PPO's state representation also contains slots for rival prices; under the local scope used here those slots are inactive.

Their hyperparameters were fixed by the configuration screen (Supplementary Note~\ref{supp:protocol}). The
standard-learner benchmark of main-text Fig.~1 additionally includes three full-information
learners and uses separate, convergence-oriented configurations under full information; it is described in
Supplementary Note~\ref{supp:round-robin}.

\subsection{Evaluation metrics}

For seed \(s\) and evaluation window \(W\), the seed-level LLM-minus-opponent gap for outcome \(Y\in\{p,\pi,q\}\) is
\[
\Delta Y_s(W)=
\frac{1}{|W|}
\sum_{t\in W}
\left(Y_{s,\mathrm{LLM},t}-Y_{s,\mathrm{opp},t}\right).
\]
The undercut share is
\[
U_s(W)=
\frac{1}{|W|}
\sum_{t\in W}
\1\{p_{s,\mathrm{LLM},t}<p_{s,\mathrm{opp},t}\}.
\]
Realized external regret on the observed opponent path is
\[
ER_{s,f}(W)=
\max_{p\in\Pgrid}\sum_{t\in W}\pi_f(p,p_{s,-f,t})
-
\sum_{t\in W}\pi_f(p_{s,f,t},p_{s,-f,t}).
\]
This is a finite-horizon, path-conditioned diagnostic. It is not policy regret against an opponent that would have reacted differently under counterfactual actions.

\suppnote{Configuration screening}{supp:protocol}

This note documents the configuration screening and the frozen agent specifications used in the main homogeneous-market experiments. 
The purpose of the screen is not to tune agents against their later evaluation opponents. Candidate configurations are selected exclusively in homogeneous self-play.
Across agent families, candidate configurations are evaluated with the same \(K=21\) price grid and horizon \(T=2000\) in the homogeneous-goods Bertrand market, with five seeds per candidate and both firms drawn from the same candidate configuration. The screen comprises \(67\) candidate cells: \(14\) distributional-LLM cells, \(11\) Exp3, \(11\) BM-RM, \(6\) PPO, \(22\) UCB, and \(3\) strategic-heuristic cells. Every distributional-LLM and learner cell runs under a local information scope, so the rival's price is not in the agent's information set and an agent sees only its own realized price, quantity, and profit. The three strategic-heuristic cells are the single exception and run under a full information scope, because the reactive rules they implement are defined in terms of the rival's last price. The scope is a property of the market environment rather than of an individual agent, so it is shared by both firms within a cell; consequently any later cell that pairs an agent against one of these reactive heuristics also reveals rival prices to that agent. No such cell enters the configuration screen for the LLM or learner families. Because the LLM screen varies one dimension at a time and re-runs the baseline level within each varied dimension, its \(14\) cells comprise \(10\) distinct configurations, with the baseline configuration appearing five times.

Because the screen is run at this horizon, the selected parameters do not necessarily carry over unchanged to the $300$ periods used with intervention interval $1$, but we treat them as a good approximation. Selection statistics are computed within each agent family. Each candidate's statistics average over its screening runs, both firms, and four recorded metric windows: the first \(10\%\) and the first \(40\%\) of the horizon, the final \(20\%\), and the full horizon. Screening statistics therefore summarize a candidate's whole trajectory, including its initial exploration phase, and are not comparable to the final-\(10\%\) estimates reported for the held-out evaluation in Supplementary Note~\ref{supp:llm-vs-learners}. The selected configuration aliases are then held fixed for subsequent experiments; no trajectory from the configuration screen is reused in held-out evaluation.

The distributional candidates are the one-factor variations described in Supplementary
Note~\ref{supp:setup}. The candidate grids of the standard learners are as follows.

\subsection{Exp3 candidates}

Exp3 has a single tuned parameter, the exploration rate \(\gamma\), which both sets the uniform exploration mass mixed into the played distribution and scales the importance-weighted payoff updates: the multiplicative weight update uses step size \(\eta=\gamma/K\). We evaluate ten constant values,
\[
\gamma \in \{0.01,0.03,0.05,0.075,0.10,0.125,0.15,0.175,0.20,0.30\}.
\]
The grid is anchored at the horizon-dependent reference of \cite{AuerCesaBianchiFreundSchapire2002Exp3}. For \(K\) actions, horizon \(T\), and rewards bounded in \([0,1]\),
\[
\gamma_T^{\mathrm{Exp3}}
=
\min\left\{1,\sqrt{\frac{K\log K}{(e-1)T}}\right\}
\]
yields an expected adversarial-regret bound of \(2\sqrt{(e-1)TK\log K}\). For \(K=21\) and \(T=2000\), this gives \(\gamma_T^{\mathrm{Exp3}}\approx0.136\). This value is bracketed by the candidates \(0.125\) and \(0.15\) rather than being itself a candidate, and the grid extends well below and above it.

An eleventh candidate decays exploration instead of holding it fixed, evaluating the same expression at the elapsed period \(t\) rather than at the horizon,
\[
\gamma_t
=
\sqrt{\frac{K\log K}{(e-1)t}},
\]
so that exploration falls as \(t^{-1/2}\) and the agent needs no horizon input. In the implementation the cap is the configured maximum exploration rate, which is set just below one rather than exactly one. This anytime reference is scored by the same rules as every other candidate and remains eligible under both selection objectives.

\subsection{BM-RM candidates}

We use the \emph{bandit} form of the \cite{BlumMansour2007FromExternalToInternalRegret} reduction, which turns \(K\) external-regret learners into an internal- and swap-regret minimizer. Concretely, the agent maintains one Exp3-style exponential-weights learner per source action. These row learners define a row-stochastic replacement matrix \(Q\), mixed with uniform exploration of mass \(\gamma\), and the played distribution \(p\) is the stationary distribution of that explored matrix, so \(p=pQ\). After sampling a price \(k\) from \(p\) and observing only the realized normalized profit \(b_k\), the observed gain is distributed across the row learners as \(g_{i,k}=p_i\,b_k\,Q_{i,k}/p_k\), and row learner \(i\) applies the importance-weighted estimate \(g_{i,k}/Q_{i,k}=p_i\,b_k/p_k\). No counterfactual profit for an unplayed price is ever used.

The instantiation is governed by the exploration rate \(\gamma\) and the row-learner step size \(\eta\). We do not tune these independently: every candidate sets \(\eta=\gamma/K\), the coupling under which the reduction's regret bound is stated. A candidate is therefore one exploration rate,
\[
\gamma \in \{0.10,0.20,0.30,0.40,0.50,0.60,0.70,0.80,0.875,0.90,0.95\},
\]
giving eleven candidates with step sizes \(\eta=\gamma/21\) between \(0.005\) and \(0.045\). The horizon-dependent theoretical anchor is
\[
\gamma_T^{\mathrm{BM}}
=
\left(\frac{K^2\log K}{T}\right)^{1/3}
\approx0.875,
\qquad
\eta_T^{\mathrm{BM}}\approx0.0417,
\]
for \(K=21\) and \(T=2000\); unlike the Exp3 anchor, this value is itself one of the candidates. The anchor is high because the bandit reduction converges slowly: at \(T=2000\) its guarantee is far from binding, and the schedule accordingly prescribes near-uniform exploration. We therefore also cover the range well below the anchor, since substantially lower exploration rates are what make the agent behave like a pricing competitor rather than a randomizing one at this horizon.

\subsection{PPO candidates}

For PPO, we vary rollout length \(R\), the number of periods collected before each policy update; the learning rate \(\eta\), which controls the update step size; and the entropy coefficient \(\lambda_H\), which weights the entropy bonus encouraging continued exploration. We first obtain a baseline by manual tuning and then conduct a staged local search around it: we double the rollout length, then reduce the learning rate while holding the longer rollout fixed, and finally add entropy regularization to the best lower-learning-rate configuration. All remaining PPO settings are held at their defaults. This yields six candidates, in which ``baseline'' denotes the starting point of this search rather than the repository's default configuration:
\[
\begin{array}{lll}
\text{baseline:}     & R=25, & \eta=0.020,\ \lambda_H=0,\\
\text{rollout-50:}   & R=50, & \eta=0.020,\ \lambda_H=0,\\
\text{low-lr:}       & R=50, & \eta=0.010,\ \lambda_H=0,\\
\text{very-low-lr:}  & R=50, & \eta=0.005,\ \lambda_H=0,\\
\text{low-entropy:}  & R=50, & \eta=0.010,\ \lambda_H=0.001,\\
\text{medium-entropy:} & R=50, & \eta=0.010,\ \lambda_H=0.005.
\end{array}
\]

\subsection{UCB candidates}

UCB has a single tuned parameter, the optimism coefficient \(\alpha\), which scales the confidence bonus added to each price's empirical mean profit. The index of price \(a\) after \(t\) updates is
\[
\hat\mu_a+\alpha\sqrt{\frac{\log t}{n_a}},
\]
where \(n_a\) is the number of times \(a\) has been played. We evaluate a fine grid spanning two orders of magnitude,
\[
\alpha \in \{0.025,0.05\}\cup\{0.1,0.2,\ldots,2.0\},
\]
for twenty-two candidates. Because the factor \(2\) of the classical bound is carried by \(\alpha\) in this parametrization, the UCB1 reference of \cite{AuerCesaBianchiFischer2002UCB} corresponds to \(\alpha=\sqrt{2}\approx1.414\); the grid brackets this value and extends to the near-greedy regime at the lower end. Each price is played once before the index is used, so the first \(K=21\) periods are forced exploration irrespective of \(\alpha\).

\subsection{Selection criteria}

No single metric identifies the best configuration for a pricing agent. We therefore select two configurations per family: a \emph{learning} configuration that balances several measures of learning quality, and an \emph{aggressive} configuration that favors profit and elevated prices subject to learning-quality requirements. ``Learning'' and ``aggressive'' name these two \emph{selection rules} and not prompt framings; both rules range over the same candidate screen. The rules defined in this subsection produce the scores in Supplementary Table~\ref{tab:supp-frozen-selection-scores} and the frozen configurations in Supplementary Tables~\ref{tab:supp-frozen-learning} and~\ref{tab:supp-frozen-aggressive}; Supplementary Table~\ref{tab:supp-selection-window-sensitivity} reports how those choices change when the rules are applied to a single metric window, and Supplementary Fig.~\ref{fig:supp-screening} visualizes where the selected LLM configurations sit within the candidate screen.

For each candidate \(c\) in the candidate set \(\mathcal C_A\) of agent family \(A\), let
\[
\overline{ER}_c,\quad \overline{SR}_c,\quad \overline{NC}_c,\quad \overline{PD}_c
\]
denote its mean external regret, swap regret, NashConv, and price dispersion, each averaged over screening runs, both firms, and the four metric windows defined above. Two conventions matter for reading these quantities. First, \(\overline{ER}_c\) and \(\overline{SR}_c\) are cumulative over each window rather than per-period averages, so their magnitudes scale with window length. Second, the fourth criterion is cross-firm price dispersion, the mean over periods of the within-period standard deviation of prices across firms; we use it as the available proxy for price stability. It penalizes persistent price gaps between the two firms rather than temporal fluctuation, and in symmetric self-play it is therefore small for a pair that moves in lockstep even if that pair moves a great deal over time.

For a candidate set \(\mathcal D\) and a metric to be minimized, let \(r_{\mathcal D}^{\uparrow}\) denote the rank within \(\mathcal D\) in ascending order, with rank \(1\) assigned to the most preferred, that is smallest, value. The descending rank \(r_{\mathcal D}^{\downarrow}\), used for metrics to be maximized, is defined analogously with rank \(1\) assigned to the largest value. Ranks are one-indexed, and ties share the lower rank. The learning-quality score assigns equal weight to the four criteria:
\[
S_c^{\mathrm{learn}}
=
\frac{1}{4}
\left[
r_{\mathcal C_A}^{\uparrow}(\overline{ER}_c)
+
r_{\mathcal C_A}^{\uparrow}(\overline{SR}_c)
+
r_{\mathcal C_A}^{\uparrow}(\overline{NC}_c)
+
r_{\mathcal C_A}^{\uparrow}(\overline{PD}_c)
\right],
\]
and we select \(c_A^{\mathrm{learn}}\in\arg\min_{c\in\mathcal C_A}S_c^{\mathrm{learn}}\). Ranking places metrics with different scales on a common ordinal scale and prevents any single metric from dominating because of its numerical magnitude.

The aggressive configuration is a stress test of profit-seeking and high-price behavior. To exclude candidates with clearly poor learning properties, we first remove the worst quartile according to each learning metric. Letting \(Q_A^{0.75}(M)\) denote the empirical \(75\)th percentile of metric \(M\) within family \(A\),
\[
\mathcal C_A^{\mathrm{aggr}}
=
\left\{
c\in\mathcal C_A:
\begin{array}{l}
\overline{ER}_c\le Q_A^{0.75}(\overline{ER}),\quad
\overline{SR}_c\le Q_A^{0.75}(\overline{SR}),\\[2pt]
\overline{NC}_c\le Q_A^{0.75}(\overline{NC}),\quad
\overline{PD}_c\le Q_A^{0.75}(\overline{PD})
\end{array}
\right\}.
\]
Among the eligible candidates we rank mean profit \(\overline{\Pi}_c\) and mean price elevation \(\overline{PE}_c\) in descending order,
\[
S_c^{\mathrm{aggr}}
=
\frac{1}{2}
\left[
r_{\mathcal C_A^{\mathrm{aggr}}}^{\downarrow}(\overline{\Pi}_c)
+
r_{\mathcal C_A^{\mathrm{aggr}}}^{\downarrow}(\overline{PE}_c)
\right],
\]
and select \(c_A^{\mathrm{aggr}}\in\arg\min_{c\in\mathcal C_A^{\mathrm{aggr}}}S_c^{\mathrm{aggr}}\). Note that the aggressive ranks are computed within the eligible subset \(\mathcal C_A^{\mathrm{aggr}}\), not within the full family, so the two rules do not in general rank the same candidate set. In the implementation, the eligibility filter falls back to the full candidate set if it would otherwise be empty; this fallback is not triggered for any family reported here.

All ranks and quantiles are computed separately within each agent family and are therefore not comparable across families. Remaining ties are resolved by a deterministic cascade: the learning rule breaks ties by lower price dispersion, then lower NashConv, then a parametrization-simplicity proxy; the aggressive rule breaks ties by higher profit, higher price elevation, lower price dispersion, lower NashConv, and then the same proxy. The proxy first prefers candidates identified as baseline, theory-anchored, neutral-prompt, or strong-model levels, then those with fewer explicit parameter overrides, and resolves any remainder by configuration identifier so that the outcome is fully deterministic. Because several families contain exact ties on the aggressive score, this cascade is load-bearing and we report it for completeness.

\subsection{Screening results and frozen configurations}

For each agent family the two selection rules each freeze one configuration from the same self-play candidate screen. Supplementary Table~\ref{tab:supp-frozen-selection-scores} reports the score each frozen configuration attains under the rule that selected it, and Supplementary Tables~\ref{tab:supp-frozen-learning} and~\ref{tab:supp-frozen-aggressive} list the resulting hyperparameters for the learning and aggressive rules respectively. For the non-LLM families the two rules freeze nearly the same configuration, differing only in the single tuned exploration or learning constant---Exp3's \(\gamma\), PPO's \(\eta\), and UCB's \(\alpha\)---and for BM-RM they coincide exactly at \(\gamma=0.50\). The LLM is the family in which the two rules diverge most: the learning rule freezes the neutral-framing configuration at intervention interval \(25\), and the aggressive rule freezes a distinct configuration with \prompt{1} at interval \(50\). These are genuinely different configurations rather than one configuration relabeled. Scored under the other rule, the learning-frozen LLM attains an aggressive score of \(7.50\) and the aggressive-frozen LLM a learning score of \(9.50\), so each is a poor choice by the other rule's criterion. Note also that the learning-frozen PPO and UCB configurations do not pass the aggressive eligibility filter at all, which is a further reason the two columns of Supplementary Table~\ref{tab:supp-frozen-selection-scores} refer to different configurations. Supplementary Fig.~\ref{fig:supp-screening} locates the two frozen LLM configurations within the self-play profit and price elevation of all candidates (Supplementary Fig.~\ref{fig:supp-screening}a) and within the two selection objectives (Supplementary Fig.~\ref{fig:supp-screening}b).

\begin{table}[t]
\centering
\caption{Selection scores of the frozen configurations, each evaluated under the rule that selected it. Lower is better. The two columns refer to \emph{different} configurations within a family, not to one configuration scored twice; the only exception is BM-RM, for which both rules select the same candidate. Scores are averages of one-indexed within-family ranks, so the attainable minimum is \(1.00\).}
\label{tab:supp-frozen-selection-scores}
\small
\begin{tabular}{@{}lcc@{}}
\toprule
\textbf{Class} & \(S^{\mathrm{learn}}\) & \(S^{\mathrm{aggr}}\) \\
\midrule
LLM       & 1.25 & 1.00 \\
Exp3      & 1.25 & 2.50 \\
BM-RM     & 1.00 & 1.00 \\
PPO       & 2.25 & 1.00 \\
UCB       & 6.25 & 1.00 \\
\bottomrule
\end{tabular}
\end{table}

\begin{table}[t]
\centering
\caption{Frozen learning-quality configurations. The history window \(3\Delta\) is relative to the intervention interval, so it resolves to \(75\) periods at \(\Delta=25\) here and to \(150\) periods for the aggressive configuration at \(\Delta=50\).}
\label{tab:supp-frozen-learning}
\small
\begin{tabularx}{\linewidth}{@{}lX@{}}
\toprule
\textbf{Class} & \textbf{Configuration} \\
\midrule
\DStf{} &
\prompt{0} (neutral), Gemini 3 Flash Preview, temperature \(0.7\), intervention interval \(25\), summary history \(3\Delta\) \\
Exp3 & Constant \(\gamma=0.30\) \\
BM-RM & Constant \(\gamma=0.50\), \(\eta=\gamma/21\) \\
PPO & \(R=50\), \(\eta=0.005\), \(\lambda_H=0\) \\
UCB & \(\alpha=1.7\) \\
\bottomrule
\end{tabularx}
\end{table}

\begin{table}[t]
\centering
\caption{Frozen aggressive configurations.}
\label{tab:supp-frozen-aggressive}
\small
\begin{tabularx}{\linewidth}{@{}lX@{}}
\toprule
\textbf{Class} & \textbf{Configuration} \\
\midrule
\DSff{} &
\prompt{1}, Gemini 3 Flash Preview, temperature \(0.7\), intervention interval \(50\), summary history \(3\Delta\) \\
Exp3 & Constant \(\gamma=0.125\) \\
BM-RM & Constant \(\gamma=0.50\), \(\eta=\gamma/21\) \\
PPO & \(R=50\), \(\eta=0.010\), \(\lambda_H=0\) \\
UCB & \(\alpha=1.0\) \\
\bottomrule
\end{tabularx}
\end{table}

\paragraph{Sensitivity to the metric window.}
The screening statistics defined above aggregate the four recorded metric windows. Because the selection rules act on ranks rather than levels, we report in Supplementary Table~\ref{tab:supp-selection-window-sensitivity} how far the frozen choices depend on that aggregation by applying the identical rules to a single window. BM-RM and PPO select the same configuration under every window, as does the aggressive rule for BM-RM, PPO, and UCB. The learning rule is more window-dependent: for the LLM it moves from the \(25\)-period intervention interval to the baseline interval of \(50\), and for UCB from \(\alpha=1.7\) to \(\alpha=1.3\) or \(\alpha=0.025\). Part of the movement in the LLM row reflects the structure of that screen, which contains five independent runs of the baseline configuration, one per varied dimension, so neighboring ranks can separate replicates of the same configuration; the two baseline entries in the LLM learning row are in fact won by the replicate that sits in the model dimension. We retain the aggregate statistics, because the scoring rule and its inputs were specified and frozen before any held-out trajectory was generated, and report the alternatives here so that the dependence can be read off directly.

\begin{table}[t]
\centering
\caption{Configuration selected by each rule when the identical selection rules are applied to a single metric window rather than to the aggregate. We report the final \(20\%\) and the full horizon, the two windows on which a selection could reasonably be based; the two early windows measure behavior before learning has settled and are not offered as alternatives. Every entry is one selected candidate. The LLM candidates are one-factor deviations from a common baseline (\texttt{gemini-3-flash-preview}, temperature \(0.7\), interval \(\Delta=50\), summary history, \prompt{0}), so an LLM entry names the dimension in which the selected candidate differs from that baseline, and the two rules may select along different dimensions. All PPO entries use \(R=50\) and \(\lambda_H=0\); all BM-RM entries use \(\eta=\gamma/21\).}
\label{tab:supp-selection-window-sensitivity}
\small
\begin{tabular}{@{}lllll@{}}
\toprule
\textbf{Class} & \textbf{Rule} & \textbf{Aggregate (frozen)} & \textbf{Final \(20\%\)} & \textbf{Full horizon} \\
\midrule
LLM & learning   & interval \(\Delta=25\) & baseline (\(\Delta=50\)) & baseline (\(\Delta=50\)) \\
    & aggressive & \prompt{1}              & \prompt{2}                & \prompt{1} \\
\addlinespace
Exp3 & learning   & \(\gamma=0.30\)  & \(\gamma=0.20\) & \(\gamma=0.30\) \\
     & aggressive & \(\gamma=0.125\) & \(\gamma=0.10\) & \(\gamma=0.125\) \\
\addlinespace
BM-RM & learning   & \(\gamma=0.50\) & \(\gamma=0.50\) & \(\gamma=0.50\) \\
      & aggressive & \(\gamma=0.50\) & \(\gamma=0.50\) & \(\gamma=0.50\) \\
\addlinespace
PPO & learning   & \(\eta=0.005\) & \(\eta=0.005\) & \(\eta=0.005\) \\
    & aggressive & \(\eta=0.010\) & \(\eta=0.010\) & \(\eta=0.010\) \\
\addlinespace
UCB & learning   & \(\alpha=1.7\) & \(\alpha=0.025\) & \(\alpha=1.3\) \\
    & aggressive & \(\alpha=1.0\) & \(\alpha=1.0\) & \(\alpha=1.0\) \\
\bottomrule
\end{tabular}
\end{table}

\begin{figure*}[t]
\centering
\suppfig[width=0.98\textwidth]{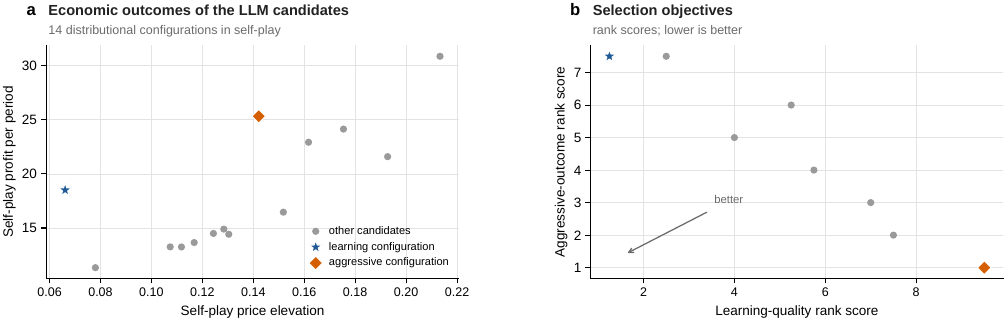}
\caption{\textbf{LLM configuration screen.} \textbf{a}, Self-play price elevation and profit per period of the 14 screened distributional cells (five seeds, 2,000 periods each). \textbf{b}, Learning-quality and aggressive-outcome rank scores (lower is better). The star and diamond mark the configurations frozen by the learning and aggressive rules.}
\label{fig:supp-screening}
\end{figure*}

\suppnote{Standard learners in homogeneous competition}{supp:round-robin}

The benchmark in main-text Fig.~1 is a round robin of seven standard learners in the homogeneous
winner-takes-all market: the bandit learners Exp3, BM-RM and UCB; the reinforcement learner PPO; and the
full-information learners FTRL, full-information regret matching (FI-RM) and weighted
fictitious play (WFP); WFP is belief-based but, like the other two, uses the rival's observed price. Every unordered pair, including self-play, is simulated for
\(T=10^7\) periods with ten seeds, using the seed derivation of the main simulator; the final window is
the last 250 periods. Unlike the LLM experiments, all firms observe the rival's
price, because FTRL, FI-RM and WFP require it. Exp3, BM-RM and UCB still learn only from their own
realized profit; PPO's state includes the rival price.

The round robin asks where learners \emph{converge}, so each learner uses one frozen configuration with
decaying exploration rather than the finite-horizon configurations selected for the 2,000-period LLM
experiments. Exp3 uses the anytime tuning
\(\gamma_t=\min\{1,\sqrt{K\log K/((e-1)t)}\}\) of \cite{AuerCesaBianchiFreundSchapire2002Exp3}. BM-RM uses
the same schedule for every row learner, with \(\eta_t=\gamma_t/K\), so that both bandit learners follow
the same theoretically motivated tuning. UCB uses \(\alpha=1\); PPO uses rollout length 25, learning rate
0.05 and entropy coefficient 0.03; FTRL is follow-the-regularized-leader with a Euclidean regularizer: it
plays \(x_{t+1}=\arg\max_{x\in\mathcal X}\{\eta_t\langle U_t,x\rangle-\tfrac12\|x\|_2^2\}=\operatorname{proj}_{\mathcal X}(\eta_t U_t)\),
the Euclidean projection onto the probability simplex \(\mathcal X\) over the 21 grid prices of the
cumulative full-information payoff vector \(U_t=\sum_{s\le t}u_s\) (payoffs normalized by one-firm monopoly
profit), with \(\eta_t=1/\sqrt t\). Because it projects the cumulative payoff once (lazy projection, or dual
averaging), it differs from projected gradient ascent \(x_{t+1}=\operatorname{proj}_{\mathcal X}(x_t+\eta u_t)\)
whenever the projection binds.
FI-RM uses the
\(\mu\)-damped procedure without exploration; WFP best-responds to a belief over the rival's price that
decays geometrically with factor 0.9, opens at marginal cost and keeps its current price when it remains
a best response. With these schedules all learners have vanishing exploration, so the final-window
outcomes describe the selected long-run regime rather than finite-horizon exploration.

Supplementary Fig.~\ref{fig:supp-round-robin-matrices} reports the full \(7\times7\) matrices of focal profit (a),
profit advantage (b), external (c) and internal (d) regret per period, consumer surplus (e) and own posted price (f), and
Supplementary Fig.~\ref{fig:supp-round-robin-paths} the convergence paths of every market. Total welfare equals the
one-firm monopoly profit in every cell because demand is inelastic up to the ceiling, and is therefore
not shown.

After \(10^7\) periods, 25 of the 28 markets end in the finite-grid competitive region (final-window
market price at most \(0.126\)), and the same 25 have a NashConv below 5\% of one-firm monopoly profit. The
three exceptions are UCB self-play, which stays near the ceiling (market price \(0.343\), NashConv \(217\)),
UCB against PPO (\(0.166\), \(47\)) and PPO against WFP (\(0.219\), \(87\)); Supplementary
Fig.~\ref{fig:supp-round-robin-nashconv} reports NashConv for every market. Main-text Fig.~1b measures
convergence speed as the end of the first \(10^4\)-period block after which every block mean of the market
price stays at most half a grid step above the competitive region (\(\le0.1325\)); the tolerance keeps markets
that settle exactly at \(0.126\), such as Exp3 self-play, from counting as leaving the region through block-mean
noise. Markets of two full-information learners settle within \(10^4\) periods, markets with
one of them after a median of \(5\times10^4\), and markets of bandit or reinforcement learners alone after a
median of \(6.4\times10^5\) periods, BM-RM self-play after about \(4\times10^6\)
(Supplementary Fig.~\ref{fig:supp-round-robin-paths}), which is why the
benchmark runs for \(10^7\) periods.

\begin{figure*}[t]
\centering
\suppfig[width=0.98\textwidth]{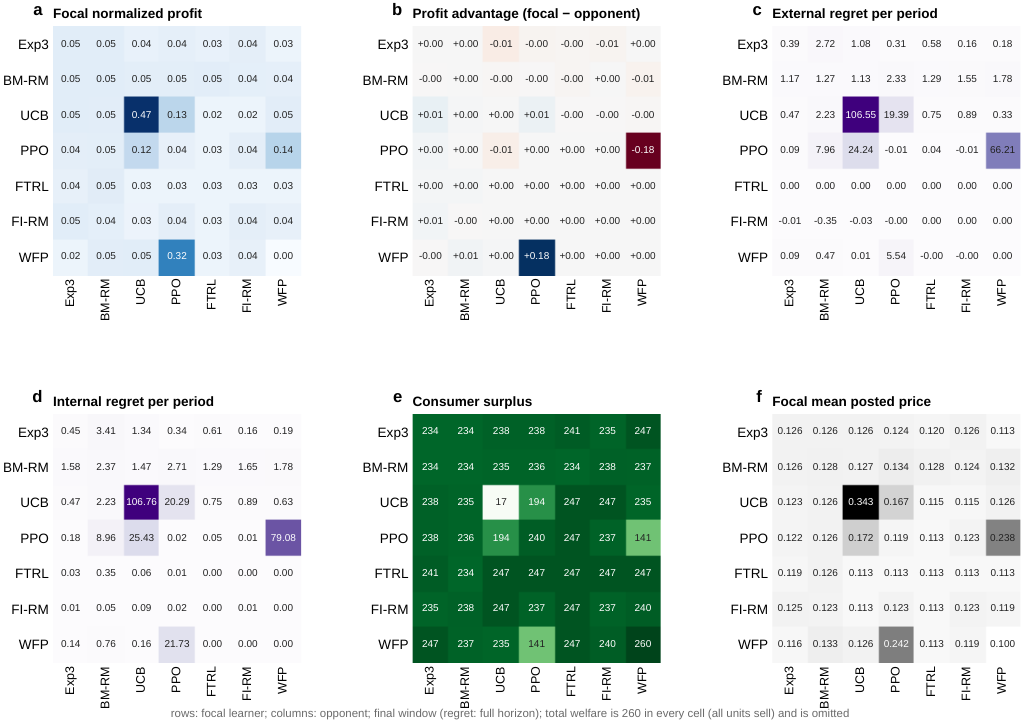}
\caption{Standard-learner round robin: full matrices. Rows are the focal learner, columns the opponent. \textbf{a}, Focal normalized profit. \textbf{b}, Profit advantage over the opponent. \textbf{c}, \textbf{d}, External and internal regret per period. \textbf{e}, Consumer surplus. \textbf{f}, Focal mean posted price.
Final-window means over ten seeds; regret is per period over the full horizon.}
\label{fig:supp-round-robin-matrices}
\end{figure*}

\begin{figure}[t]
\centering
\suppfig[width=0.75\linewidth]{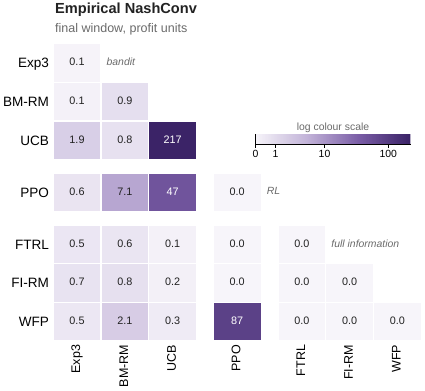}
\caption{Standard-learner round robin: empirical NashConv of the final-window product of empirical price
distributions (profit units; one-firm monopoly profit is 260). Colour uses \(\log(1+\mathrm{NashConv})\);
printed values are untransformed. Means over ten seeds.}
\label{fig:supp-round-robin-nashconv}
\end{figure}

\begin{figure*}[t]
\centering
\suppfig[width=0.98\textwidth]{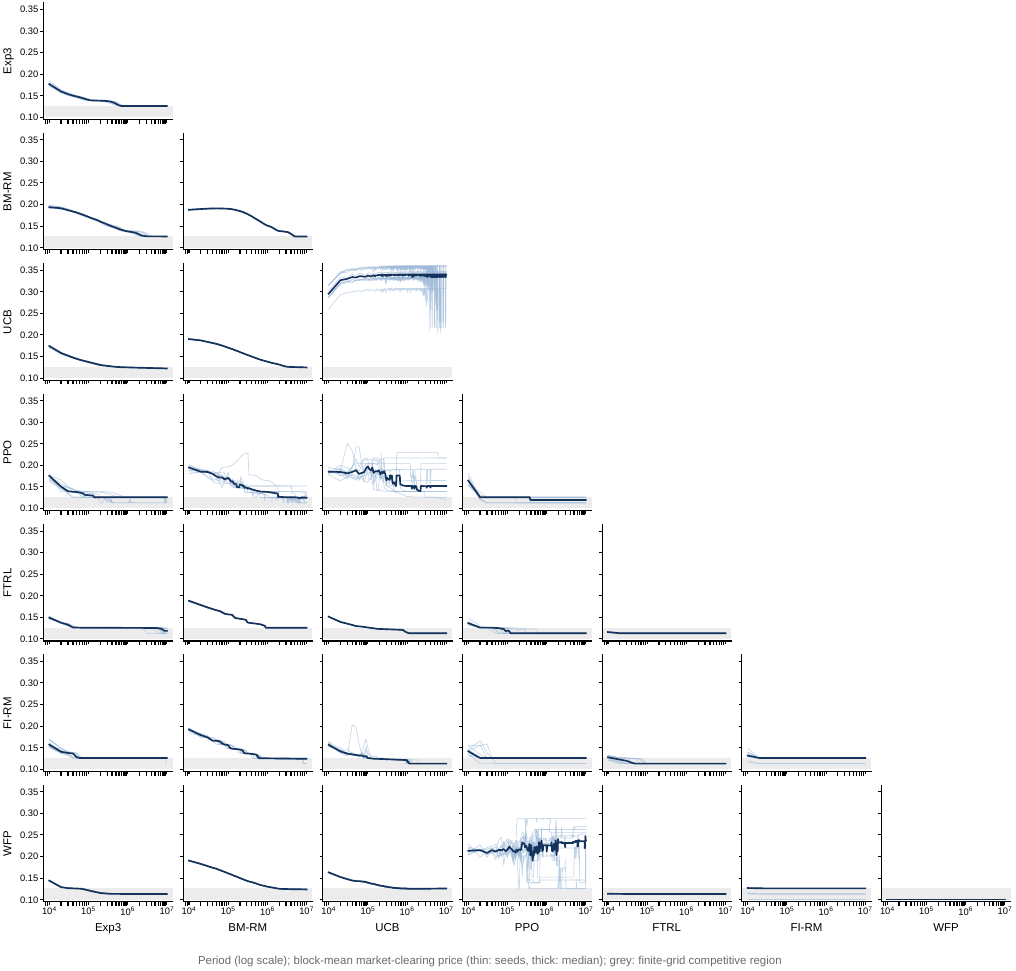}
\caption{Standard-learner round robin: market-clearing price over time for every pair (log time axis).
Thin lines are seeds, thick lines their median; grey band, finite-grid competitive region.}
\label{fig:supp-round-robin-paths}
\end{figure*}
\suppnote{LLMs against standard learners}{supp:llm-vs-learners}

This note supports main-text Fig.~2. The replicated cells pair \DStf{}
with Exp3, BM-RM, PPO and UCB (ten seeds, 2,000 periods each); two stress tests pair \DStf{} or
\DSff{} with the aggressive UCB configuration. The \PRone{} runs against standard learners and simple
pricing rules are single runs.

\subsection{Held-out price dynamics}

Supplementary Fig.~\ref{fig:supp-stage2-price-paths} shows the price paths behind main-text Fig.~2a,b. Against every
learner the \DStf{} agent's median price falls below the learner's within the first few hundred periods and stays
there, which is the undercutting summarized in main-text Fig.~2b; the aggressive-UCB stress cells show the
same ordering.

\begin{figure*}[t]
\centering
\suppfig[width=0.98\textwidth]{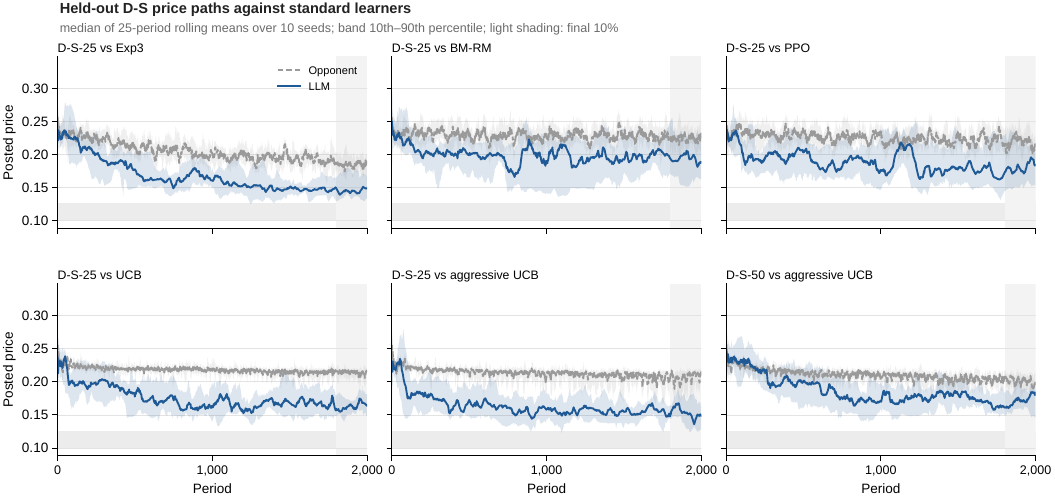}
\caption{\textbf{Held-out \DStf{} price paths against standard learners.} Median over ten seeds of 25-period rolling mean prices; bands span the 10th--90th percentiles across seeds. Grey band, finite-grid competitive region; light shading, final 10\% of periods used in main-text Fig.~2b.}
\label{fig:supp-stage2-price-paths}
\end{figure*}

\subsection{Evaluation-window robustness}

All main-text estimates use the final \(10\%\) of each run as the primary evaluation window. Supplementary Table~\ref{tab:supp-window-robustness} reports robustness over final \(5\%\), \(10\%\), \(20\%\), \(50\%\), and full-horizon windows, averaging cell-level means across the four learner matchups, and Supplementary Fig.~\ref{fig:supp-window-sensitivity} plots the dependence per learner for the profit gap (a), the price gap (b), the share of periods in which the \DStf{} agent posts the lower price (c) and the external-regret gap (d).

\begin{table}[t]
\centering
\caption{Robustness to evaluation-window length. Entries average cell-level means across the four learner matchups. The ER gap is cumulative over the window, not divided by its length.}
\label{tab:supp-window-robustness}
\small
\begin{tabular}{@{}lrrrrr@{}}
\toprule
\textbf{Metric} & \textbf{Final 5\%} & \textbf{10\%} & \textbf{20\%} & \textbf{50\%} & \textbf{Full} \\
\midrule
Profit gap & 24.64 & 24.92 & 25.69 & 25.99 & 27.70 \\
Price gap & -0.04 & -0.04 & -0.04 & -0.04 & -0.04 \\
ER gap (cumul.) & -2198.30 & -4007.74 & -7739.55 & -16364.73 & -33867.11 \\
\bottomrule
\end{tabular}
\end{table}
\begin{figure*}[t]
\centering
\suppfig[width=0.98\textwidth]{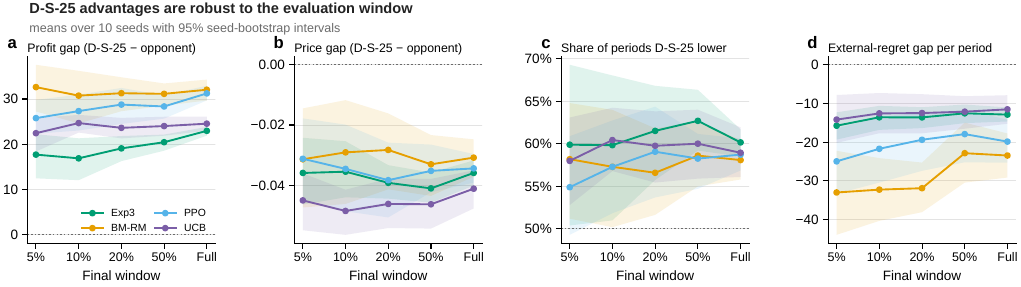}
\caption{\textbf{Robustness to the evaluation window.} \textbf{a}, Profit gap; \textbf{b}, price gap; \textbf{c}, share of periods in which the \DStf{} agent posts the lower price; \textbf{d}, external-regret gap per period; for final windows of 5--100\% of the 2,000 periods. Means over ten seeds with 95\% seed-bootstrap intervals.}
\label{fig:supp-window-sensitivity}
\end{figure*}

\subsection{\PRone{} against standard learners and pricing rules}
\label{supp:pllm-homogeneous}

Main-text Fig.~2c,d shows two single-run \PRone{} interactions in the homogeneous market. Supplementary Table~\ref{tab:supp-pllm-homogeneous}
lists the complete set of such runs. In all of them the \PRone{} agent (Gemini 3 Flash, temperature 0.7) emits a
grid price every period from a raw 100-period history with the rival's price visible; the blockwise Exp3
run instead uses the variant \dep{P-R-25}, which decides every 25 periods from a raw 75-period history. Each cell is one run (seed 301), so the
table documents illustrative behaviour rather than treatment effects.

The runs separate by the opponent's feedback. Against the bandit learners and the exploitable rules the
\PRone{} agent earns the larger share, and against win-up--lose-down it matches the rule up to the ceiling. Against
the full-information learners it gains no profitable advantage. Coin betting stays one or two grid steps
below the \PRone{} agent's elevated prices and earns \(71\%\) of industry profit; after about 200 periods the \PRone{} agent
abandons its anchor and both prices fall towards the competitive region, with joint profit dropping from
\(68\%\) to \(30\%\) of the monopoly benchmark between the second and the last 50-period block.
Against full-information regret matching the \PRone{} agent's \(75\%\) share is earned mainly at prices between
\(0.112\) and \(0.126\), where joint profit is only 5--10\% of the monopoly benchmark, so the share
overstates the advantage. Weighted fictitious play also ends ahead of the \PRone{} agent. The \PRone{} advantage is
therefore an advantage against slow, exploring learners and predictable rules, not against learners that
can evaluate every price from the rival's observed price.

\begin{table}[t]
\centering
\caption{\PRone{} against standard learners and simple pricing rules in the homogeneous market.
Cumulative \PRone{} share of industry profit and mean per-period profit of each firm as a share of the one-firm
monopoly profit (\PRone/opponent) over the run; single runs (\(n=1\)). Learners use the configurations frozen in the
configuration screen (Supplementary Note~\ref{supp:protocol}) and are far from converged at these horizons (Supplementary Note~\ref{supp:round-robin}).}
\label{tab:supp-pllm-homogeneous}
\footnotesize
\setlength{\tabcolsep}{4pt}
\begin{tabularx}{\linewidth}{@{}>{\raggedright\arraybackslash}p{2.7cm}ccc>{\raggedright\arraybackslash}X@{}}
\toprule
\textbf{Opponent} & \textbf{Periods} & \textbf{\PRone{} share} & \textbf{Profit} & \textbf{Behaviour} \\
\midrule
\multicolumn{5}{@{}l}{\emph{Bandit feedback}} \\
Exp3 & 300 & 62.7\% & 0.20/0.12 & \PRone{} agent adapts to Exp3's exploration (main-text Fig.~2c) \\
Exp3 (\dep{P-R-25}) & 2,000 & 62.1\% & 0.17/0.10 & advantage persists over a longer horizon with blockwise calls \\
PPO & 230\(^{*}\) & 66.1\% & 0.25/0.13 & \PRone{} agent prices just below PPO's exploratory prices \\
\multicolumn{5}{@{}l}{\emph{Full information}} \\
Coin betting & 300 & 28.9\% & 0.16/0.38 & learner undercuts the \PRone{} agent's anchor for 200 periods; prices then fall to about 0.18 \\
Full-information regret matching & 300 & 75.1\% & 0.12/0.04 & share won by undercutting at the grid floor; joint profit 16\% of monopoly \\
Weighted fictitious play & 100 & 46.4\% & 0.10/0.11 & share falls to about 35\% in the last 40 periods \\
\multicolumn{5}{@{}l}{\emph{Simple pricing rules}} \\
Win-up--lose-down & 100 & 50.0\% & 0.45/0.45 & \PRone{} agent matches the rule from period 3; both reach the ceiling at period 21 (main-text Fig.~2d) \\
Randomized win-up--lose-down & 100 & 62.1\% & 0.34/0.21 & supracompetitive but less stable coordination \\
Undercut-last & 86\(^{*}\) & 84.3\% & 0.58/0.11 & \PRone{} agent exploits the predictable undercutting rule \\
\bottomrule
\end{tabularx}
\par\smallskip\footnotesize \(^{*}\)Run stopped before the planned horizon (PPO 300, undercut-last 100 periods); values are computed over the completed periods.
\end{table}

Supplementary Fig.~\ref{fig:supp-pllm-learner-paths} shows the price paths behind the table, and they make
the different forms of adaptation visible. Against Exp3 the \PRone{} agent revises its price in \(30\%\) of the
first 150 periods and in \(4\%\) afterwards, settling on a few persistent prices while Exp3 keeps exploring
the whole grid; the blockwise \dep{P-R-25} variant drifts below Exp3's mean (\(0.16\) against \(0.19\) in the
final quarter). Against PPO it holds prices between \(0.19\) and \(0.27\). Against coin betting it alternates
between the ceiling and lower prices, changing its price in \(81\%\) of the first 150 periods, while the
learner stays just below; both then slide to about \(0.19\). Against full-information regret matching both
firms step down to the grid floor within 150 periods; the \PRone{} agent then jumps to the ceiling and stays
there for 45 periods without being followed, and only at the end do both move to about \(0.24\).
Against weighted fictitious play it oscillates between about \(0.14\) and \(0.23\), changing its price in most
periods, while the learner prices just below. Against the rules it settles into structured patterns: it
moves in lockstep with win-up--lose-down to the ceiling, posts a constant \(0.256\) from period 35 while
the randomized rule fluctuates around it, and against undercut-last it resets to the ceiling every five
periods and steps down ahead of the rule, posting the lower price in \(78\%\) of periods.

\begin{figure*}[t]
\centering
\suppfig[width=0.98\textwidth]{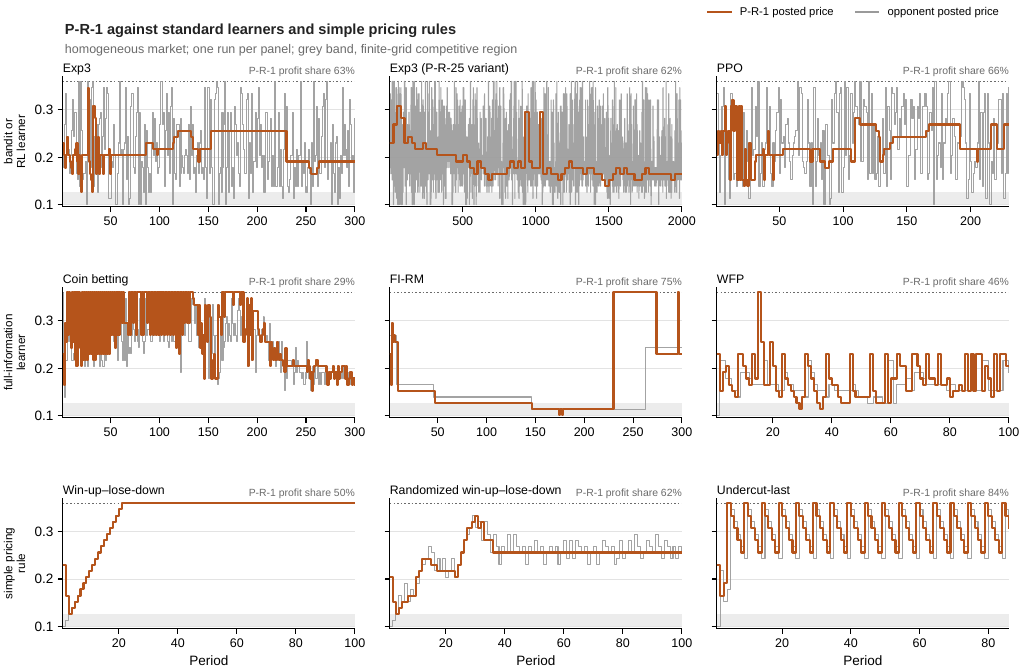}
\caption{\textbf{\PRone{} price paths against standard learners and simple pricing rules.} Homogeneous
market; one run per panel, the runs of Supplementary Table~\ref{tab:supp-pllm-homogeneous}. Orange,
\PRone{} posted price; grey, opponent posted price; grey band, finite-grid competitive region; dotted line,
price ceiling. Rows group the opponents by feedback: bandit or RL learners, full-information learners and
simple pricing rules. Each panel states the cumulative \PRone{} share of industry profit. The Exp3 and
win-up--lose-down panels repeat the runs of main-text Fig.~2c,d for comparison; the second panel is the
blockwise \dep{P-R-25} variant.}
\label{fig:supp-pllm-learner-paths}
\end{figure*}

\suppnote{Reasoning traces and motif extraction}{supp:traces}

At each intervention epoch, the LLM emits a \texttt{plans\_and\_insights} field in addition to a price or distribution. We treat this text as an observable behavioural trace, not as faithful access to hidden computation. The trace analysis is used only to identify candidate mechanisms; the resulting surrogate is validated separately against emitted distributions and realized price paths.

\subsection{Motif-extraction procedure}

Complete LLM call records were supplied to GPT-5.6 Sol using the fixed analysis prompt reproduced in Listing~\ref{lst:supp-motif-extraction}. Each record contained the market history available to the pricing agent, its previous distribution, its newly emitted distribution, and its \texttt{plans\_and\_insights} text. Supplementary Table~\ref{tab:supp-trace-motifs} summarizes the recurring motifs this procedure identified and maps each to the DBBS component that represents it.

\begin{lstlisting}[style=prompt,caption={Motif-extraction prompt.},label={lst:supp-motif-extraction}]
Analyze the attached JSON records of repeated pricing decisions.
Each record contains:
- the market history available to the pricing agent;
- its previous probability distribution;
- its new probability distribution;
- its plans_and_insights text.

Identify recurring behavioral rules that explain how the agent changes its
probability distribution after new feedback.

Focus on:
1. how successful and unsuccessful prices are evaluated;
2. whether recent outcomes receive more weight than older outcomes;
3. whether prices are treated as ordered or as independent actions;
4. how the focal pricing region moves;
5. whether exploration is local or global;
6. whether higher and lower neighboring prices are treated asymmetrically.

Do not propose a mathematical algorithm initially. First report recurring
patterns, supporting each with multiple examples from different runs and
treatments. Then provide the simplest qualitative decision rule that summarizes
the evidence.
\end{lstlisting}

\begin{table*}[t]
\centering
\caption{Recurring behavioural motifs extracted from LLM decision traces and their representation in DBBS.}
\label{tab:supp-trace-motifs}
\small
\begin{tabular}{@{}p{0.20\textwidth}p{0.40\textwidth}p{0.31\textwidth}@{}}
\toprule
\textbf{Motif} & \textbf{Trace-level pattern} & \textbf{DBBS component} \\
\midrule
Broad initialization
& Before receiving feedback, the LLM distributes probability broadly around the middle of the price grid.
& Cold-start distribution. \\
\addlinespace
Recency weighting
& Recent profitable outcomes receive more emphasis than older averages or historical peaks.
& Recency-weighted payoff score \(v_i\). \\
\addlinespace
Failure sensitivity
& Zero-profit or failed outcomes reduce probability on the corresponding price region.
& Failure score \(\phi_i\) and penalty \(\beta_{\mathrm{fail}}\phi_i\). \\
\addlinespace
Ordered-price reasoning
& Failures at relatively high prices are interpreted as evidence that the profitable region lies lower.
& One-sided locality penalty \(\beta_{\mathrm{loc}}(i-m)_+\). \\
\addlinespace
Local asymmetric search
& Probability concentrates around a focal price; higher neighbours serve as margin probes and lower neighbours as safer fallbacks.
& Search centre \(m\) with local upward probing \(\{m+1,\ldots,m+G_{\mathrm{up}}\}\). \\
\bottomrule
\end{tabular}
\end{table*}

\subsection{Motif frequencies}

To make the frequency statements of main-text Section~1.3 auditable without further model calls, every
persisted planning text is coded against a fixed keyword lexicon. A \emph{trace} is the planning text one
firm writes at one LLM call: the \texttt{plans\_and\_insights} field of \DStf{} and \DSff{} (calls that
fell back to a default action are excluded), and the concatenated \texttt{plans} and \texttt{insights}
scratchpad entries of the Fish-style harness. Supplementary Table~\ref{tab:supp-motif-frequencies} reports
the resulting shares.

\paragraph{Keyword lexicon.} The lexicon codes ten motifs in two families
(Supplementary Table~\ref{tab:supp-motif-lexicon}). Five motifs are opponent-facing: the trace
reasons about how the rival reacts, about a price war, about signalling or resetting a price level, about
keeping prices high, or about undercutting the rival and harvesting its demand. Five refer to payoff
regions of the firm's own price grid: a price region that worked, a price that failed or sold nothing, a
local probe of nearby prices, a demand threshold, or a gradual price step. Each motif is a list of
case-insensitive regular expressions. Most patterns are word stems (\emph{retaliat*} matches
\emph{retaliate}, \emph{retaliation} and \emph{retaliatory}) or short phrases in which a space and a hyphen
are interchangeable; a few require two terms in the same sentence within a fixed distance. Three
exclusions remove systematic false matches: \emph{undercut*} does not count as undercutting when preceded
by \emph{avoid}, \emph{avoiding} or \emph{not} (\emph{avoid undercutting} counts as preserving a high price
instead); \emph{signal*} does not count when it describes market feedback (\emph{signals a shift},
\emph{signals weakening demand}); and a local-probe term does not count when one of the three preceding
words is \emph{competitor}, \emph{rival}, \emph{their} or \emph{they}, because it then describes the
rival's probe rather than the firm's own. The exact regular expressions are in the replication
files.

\begin{table*}[t]
\centering
\caption{Keyword lexicon for the motif frequencies of Supplementary Table~\ref{tab:supp-motif-frequencies}.
Matching is case-insensitive. An asterisk marks a word stem that matches any continuation; slashes separate
alternatives; parentheses mark optional words; ``\(\ldots\)'' requires both terms in the same sentence
within the stated number of characters or words. A trace exhibits a motif if any listed pattern matches.}
\label{tab:supp-motif-lexicon}
\footnotesize
\begin{tabular}{@{}p{0.17\textwidth}p{0.79\textwidth}@{}}
\toprule
\textbf{Motif} & \textbf{Keywords and patterns} \\
\midrule
\multicolumn{2}{@{}l}{\emph{Opponent-facing}} \\
Rival response
& \emph{competitor(s)/rival(s)/opponent(s)/other firm(s)} \(\ldots\) (\(\le 80\) characters)
\emph{follow*/match*/respond*/react*/retaliat*/mirror*/copy*/punish*/reciprocat*};
the verbs \emph{follow*/match*/respond*/react*/retaliat*/mirror*/copy*} \(\ldots\) (\(\le 40\) characters)
\emph{competitor/rival/opponent/other firm}; \emph{tit-for-tat}, \emph{retaliat*}, \emph{forgiv*},
\emph{punish*}. \\
\addlinespace
Price war
& \emph{price war(s)}, \emph{race to the bottom}, \emph{downward spiral}, \emph{mutually destructive}. \\
\addlinespace
Signal or reset
& \emph{signal(s/ed/ing)}, unless followed by \emph{(a/the) need/shift/change/instability/potential/that/weak*/declin*/drop/deteriorat*/saturat*};
\emph{reset*}; \emph{price leader*}; \emph{lead a/the (price) increase/rise/move}. \\
\addlinespace
Preserving a high price
& \emph{cooperat*}, \emph{collu*}, \emph{tacit*}, \emph{truce}; \emph{avoid(ing) (one word) undercut*};
\emph{do not/don't/never undercut/price below}; \emph{maintain(ing) (the) high/ceiling/parity/price level};
\emph{preserv* (the) high/margin/price/ceiling}. \\
\addlinespace
Undercutting or harvesting
& \emph{undercut*}, unless preceded by \emph{avoid/avoiding/not}; \emph{harvest*};
\emph{capture (the) (entire/full/whole/all) market/demand/volume}. \\
\midrule
\multicolumn{2}{@{}l}{\emph{Payoff-region}} \\
Successful price region
& \emph{most/highly effective/profitable/successful/reliable (recent) price/action/region/range/point/level/option/zone/bracket/corridor};
\emph{best/top/strongest-perform*}; \emph{profit(able) peak/anchor/region/zone/range/plateau/window/corridor};
\emph{sweet spot}; \emph{highest/best (normalized/average/recent/last) profit/return}. \\
\addlinespace
Failed price or zero sales
& \emph{zero profit/sales/demand/units/volume/quantity}; \emph{no sales/demand/units sold}; \emph{fail*};
\emph{loss of volume/demand/market}; \emph{demand cliff}; \emph{unviable/non-viable};
\emph{lost all/the volume/demand/market/sales}. \\
\addlinespace
Local probe
& \emph{probe(s)}, \emph{probing}; \emph{neighbo*}; \emph{adjacent}; \emph{local(ly) search/refine*/explor*/optim*};
\emph{right/left shoulder}; \emph{upper/lower boundary/edge/tail}; \emph{fine-tun*}; not counted when
\emph{competitor/rival/their/they} occurs among the three preceding words. \\
\addlinespace
Demand threshold
& \emph{demand curve/cliff/ceiling/threshold/elasticity/boundary}; \emph{elasticit*}; \emph{price-sensitiv*};
\emph{willingness to pay}. \\
\addlinespace
Gradual price step
& \emph{gradual(ly)/incremental(ly)/cautious(ly)/step-by-step/slowly} followed within two words by
\emph{rais*/lower*/increas*/decreas*/mov*/shift*/adjust*/nudg*/reduc*/climb*};
\emph{rais*/lower*/increas*/decreas*/nudg*} followed within three words by
\emph{gradually/incrementally/cautiously/in small steps/step by step};
\emph{increase/raise/lower/decrease/nudge (the/our/my) (price) by one/a single (grid) step/action/tick/notch}. \\
\bottomrule
\end{tabular}
\end{table*}

\paragraph{From matches to shares.} For firm-run \(u\) with traces \(t=1,\ldots,T_u\) let
\(x^k_{ut}=1\) if trace \(t\) matches motif \(k\) and \(0\) otherwise. Consecutive traces of one firm share
a scratchpad and are strongly autocorrelated, so the unit of analysis is the firm-run, not the trace: we
compute its share \(s^k_u=T_u^{-1}\sum_t x^k_{ut}\) and report the mean of \(s^k_u\) over the firm-runs of a
group. Every firm-run therefore receives equal weight regardless of how many LLM calls it made. The two
``any'' rows set \(x_{ut}=1\) if trace \(t\) matches at least one motif of the family; because one trace can
match several motifs, they are not sums of the rows above them. The 95\% intervals are percentile
intervals from 2,000 bootstrap draws that resample games with replacement, because the two firms of one game
interact; the numbers of firm-runs and traces are given in the last row of each panel.

\paragraph{Precision audit.} Precision was checked by reading randomly drawn matches, each shown with
about 110 characters of context on either side, from \PRone{} and \PSone{} traces and from \DStf{} and
\DSff{} traces, and judging whether the passage expresses the motif. The Precision column of
Supplementary Table~\ref{tab:supp-motif-frequencies}a reports the true positives per motif (the audited
matches are in the replication files); \(58\) of \(66\) matches exhibit the motif. Seven of the eight false
positives add to the motif family that is rarer in that deployment---point-price descriptions of the rival's probes
coded as payoff-region motifs, and \DStf{} statements such as matching a rival at the price floor coded as
opponent-facing---so they work against the \PRone--\DStf{} contrast reported below. The lexicon measures wording, not
intent, and its recall is not estimated.

\begin{table*}[t]
\centering
\caption{Share of LLM decision traces exhibiting each motif. \textbf{a}, Deployed interfaces in the
homogeneous market (Gemini 3 Flash): \PRone, \PSone{} and \DSone, firms of the main-text Fig.~5 runs (\PRone{}
also the runs against standard learners); \DStf{} and \DSff, their replicated runs against standard learners and in self-play
(main-text Figs.~2a,b and 4a). Precision, true positives
among audited matches. \textbf{b}, Protocol ablations in the logit market at intervention interval 1 and
memory 100 (Supplementary Note~\ref{supp:ablations}); the control is \PRone{} with the rival's price visible,
and each other column changes the stated components. Values are means over firm-runs; intervals are in the
replication files.}
\label{tab:supp-motif-frequencies}
\small
\textbf{a}\quad Deployed interfaces\\[2pt]
\begin{tabular}{@{}lcccccc@{}}
\toprule
\textbf{Motif} & \textbf{Precision} & \PRone & \PSone & \DSone & \DStf & \DSff \\
\midrule
Rival response & 5/6 & 0.54 & 0.53 & 0.38 & 0.11 & 0.09 \\
Price war & 6/6 & 0.20 & 0.25 & 0.04 & 0.01 & 0.02 \\
Signal or reset & 5/6 & 0.60 & 0.75 & 0.52 & 0.11 & 0.07 \\
Preserving a high price & 6/6 & 0.21 & 0.13 & 0.25 & 0.02 & 0.02 \\
Undercutting or harvesting & 5/6 & 0.87 & 0.87 & 0.53 & 0.19 & 0.33 \\
\emph{Any opponent-facing motif} &  & 0.94 & 0.97 & 0.71 & 0.34 & 0.40 \\
\addlinespace
Successful price region & 7/8 & 0.01 & 0.01 & 0.09 & 0.31 & 0.30 \\
Failed price or zero sales & 6/6 & 0.16 & 0.30 & 0.64 & 0.90 & 0.90 \\
Local probe & 5/8 & 0.02 & 0.04 & 0.44 & 0.36 & 0.47 \\
Demand threshold & 6/6 & 0.03 & 0.03 & 0.11 & 0.23 & 0.25 \\
Gradual price step & 7/8 & 0.02 & 0.03 & 0.01 & 0.00 & 0.00 \\
\emph{Any payoff-region motif} &  & 0.21 & 0.37 & 0.75 & 0.98 & 0.97 \\
\midrule
Firm-runs/traces & & 23/6,016 & 2/600 & 7/2,100 & 90/7,200 & 30/1,200 \\
\bottomrule
\end{tabular}

\vspace{8pt}
\textbf{b}\quad Protocol ablations\\[2pt]
\begin{tabular}{@{}lcccc@{}}
\toprule
\textbf{Motif} & \makecell{Control\\(point, raw,\\visible)} & \makecell{Grid\\distribution} &
\makecell{Grid point,\\summary,\\hidden} & \makecell{Distribution,\\summary,\\hidden} \\
\midrule
Rival response & 0.82 & 0.78 & 0.00 & 0.00 \\
Price war & 0.26 & 0.40 & 0.00 & 0.00 \\
Signal or reset & 0.27 & 0.37 & 0.01 & 0.01 \\
Preserving a high price & 0.42 & 0.63 & 0.00 & 0.02 \\
Undercutting or harvesting & 0.62 & 0.67 & 0.05 & 0.01 \\
\emph{Any opponent-facing motif} & 0.98 & 0.96 & 0.06 & 0.03 \\
\addlinespace
Successful price region & 0.38 & 0.19 & 0.37 & 0.47 \\
Failed price or zero sales & 0.04 & 0.02 & 0.07 & 0.03 \\
Local probe & 0.06 & 0.00 & 0.36 & 0.40 \\
Demand threshold & 0.13 & 0.09 & 0.10 & 0.10 \\
Gradual price step & 0.02 & 0.03 & 0.00 & 0.00 \\
\emph{Any payoff-region motif} & 0.55 & 0.24 & 0.67 & 0.71 \\
\midrule
Firm-runs/traces & 10/3,000 & 10/3,000 & 10/3,000 & 10/3,000 \\
\bottomrule
\end{tabular}
\end{table*}

The deployed traces separate as main-text Section~1.3 describes (Supplementary Table~\ref{tab:supp-motif-frequencies}a). At least one opponent-facing motif
appears in \(94\%\) of \PRone{} traces (95\% interval \(89\)--\(98\%\)) but in \(34\%\) (\(30\)--\(39\%\))
of \DStf{} traces, and at least one payoff-region motif in \(21\%\) (\(17\)--\(26\%\)) of \PRone{} traces
but in \(98\%\) of \DStf{} traces. Rival response (\(0.54\) against \(0.11\)), price wars (\(0.20\) against
\(0.01\)) and preserving a high price (\(0.21\) against \(0.02\)) are \PRone{} motifs; failed prices
(\(0.90\) against \(0.16\)), local probes (\(0.36\) against \(0.02\)) and successful price regions
(\(0.31\) against \(0.01\)) are \DStf{} motifs. \DSff{} is close to \DStf. The opponent-facing traces of
\DStf{} and \DSff{} mostly infer that an unseen rival is undercutting a price that stopped selling.
\DSone{}, which observes the rival's price in some of its runs, lies between the two. Gradual-step wording
is rare in every group (\(\le 3\%\)); \DStf{} and \DSff{} express caution through low-probability probes
instead.

The protocol ablations (Supplementary Table~\ref{tab:supp-motif-frequencies}b) show that the contrast follows the information interface, not the action
representation. With raw histories and visible rival prices, \(96\)--\(99\%\) of traces are opponent-facing
whether the LLM emits a continuous price, a grid price or a grid distribution. With summarized histories and
hidden rival prices, the share falls to \(6\%\) for grid point prices and \(3\%\) for distributions, and
payoff-region motifs dominate in both. Hiding the rival's price alone lowers the share to \(45\%\);
summarizing the history alone lowers it to \(82\%\). The trace evidence therefore matches the price
evidence of main-text Fig.~4: what the LLM observes, more than how it emits a price, determines whether it
reasons about its rival or about its own payoff regions.

\suppnote{DBBS construction and validation}{supp:dbbs}

This note defines downward-biased belief search (\dbbs) and documents its validation under teacher forcing
and free-running deployment on \DStf, the learning configuration of main-text Section~1.2 (main-text
Fig.~3). All DBBS fits and deployments use
the homogeneous winner-takes-all market. In the differentiated logit market every firm keeps positive
demand, so the zero-sales failure signal below would never fire and \(\beta_{\mathrm{fail}}\) would be
inert; we therefore do not fit DBBS there.

\subsection{DBBS update}

DBBS treats prices as an ordered grid \(\Pgrid=\{p_0,\ldots,p_{K-1}\}\). It maintains a recency-weighted payoff score \(v_i\) and failure score \(\phi_i\) for each tick \(i\). At each intervention epoch it observes the realized block of played ticks and normalized profits and identifies the modal played tick \(m\). For each period of the block, in order, with played tick \(i\) and normalized profit \(r_i\), it updates
\[
v_i \leftarrow \lambda r_i+(1-\lambda)v_i,
\qquad
\phi_i \leftarrow \lambda y_i+(1-\lambda)\phi_i,
\qquad
y_i=\1\{i>0\}\,\1\{r_i\le0\}.
\]
The failure signal \(y_i\) is the zero-sales indicator: in the winner-takes-all market a price above
marginal cost earns zero profit exactly when it does not sell. Tick \(0\) is marginal cost, which earns zero
profit even when it sells, and is never a failure. Ticks not played keep their scores. The score assigned to
tick \(i\) is
\[
S_i=v_i-\beta_{\mathrm{fail}}\phi_i-\beta_{\mathrm{loc}}(i-m)_+.
\]
The base policy is the softmax
\[
\sigma^{\mathrm{base}}(i)
=
\frac{\exp(S_i/\tau)}
{\sum_{k=0}^{K-1}\exp(S_k/\tau)}.
\]
The final policy mixes this base policy with a local search distribution of weight \(\rho\), uniform on
the \(G_{\mathrm{up}}\) ticks above the focal tick \(m\). Before the first observation, DBBS emits a
cold-start distribution estimated per treatment from the training seeds of the fold.

\subsection{Inertia}

The LLM keeps much of its previous distribution at each update. The surrogate therefore mixes the search
policy with a persistence component,
\[
\sigma_{n+1}=w\,\sigma^{\mathrm{pers}}_{n+1}+(1-w)\,\sigma^{\mathrm{search}}_{n+1},
\]
where \(\sigma^{\mathrm{search}}\) is the policy above and \(\sigma^{\mathrm{pers}}\) is a recency-weighted
histogram of the agent's own recent prices (the persistence baseline below). Both components are updated
from the same feedback. The weight \(w\) is selected for each fold on the training seeds by one-step RPS over
\(w\in\{0,0.05,\ldots,1\}\); \(w=0\) recovers the search policy alone (``DBBS without inertia''). We added the
inertia component after observing that the search policy alone undercuts standard learners more than
\DStf{} does in free-running deployment; its weight is estimated only from one-step predictions on training
seeds, never from free-running outcomes, so the free-running evaluation remains held-out.

\subsection{Teacher forcing and free-running reproduction}

Under teacher forcing, DBBS follows recorded LLM trajectories. At each intervention epoch it receives the
same realized history as the LLM and predicts the LLM's next emitted distribution. The target is the full
probability distribution, not the realized sampled price. Under free-running reproduction, the fitted DBBS
policy replaces the LLM and acts endogenously. This tests whether one-step prediction fidelity translates
into similar market dynamics when prediction differences accumulate over time. Each recorded trajectory spans
\(T=2000\) periods, in which \DStf{} emits 80 distributions (79 prediction targets after the cold start).

\subsection{Fit set}

The fit set consists of the seven treatments in which the learning configuration was run with ten seeds:
self-play and play against Exp3, BM-RM, PPO, UCB, the aggressive UCB configuration and a static
marginal-cost anchor. A second run against the marginal-cost anchor with identical configuration and seeds
is counted once. Every treatment has the same ten environment
seeds, so leave-one-seed-out cross-fitting has ten folds.

One parameter vector per fold is fitted jointly on all seven treatments. This is the
most parsimonious specification and the strongest test: the opponent-specific price levels in the free
games must emerge from one mechanism rather than from per-treatment tuning.

\subsection{Prediction metrics and parameter selection}

Let \(\sigma_n\) denote a predicted distribution and \(\hat\sigma_n\) the LLM-emitted target distribution. Define the cumulative difference
\[
d_{n,k}
=
\sum_{i=0}^{k}\sigma_n(i)
-
\sum_{i=0}^{k}\hat\sigma_n(i),
\qquad k=0,\ldots,K-2.
\]
The training objective is the normalized ranked-probability score
\[
\operatorname{RPS}_n
=
\frac{1}{K-1}
\sum_{k=0}^{K-2}d_{n,k}^2.
\]
For evaluation, we report first Wasserstein distance
\[
W_{1,n}
=
\sum_{k=0}^{K-2}|d_{n,k}|(p_{k+1}-p_k),
\]
normalized by the price range. On the equally spaced 21-point grids, one tick corresponds to \(5\%\) of the price range.

In fold \(j\), environment seed \(j\) is held out for every treatment and parameters are fitted on the
remaining seeds. We draw \(20{,}000\) candidate parameter vectors by Latin-hypercube sampling over
\[
\lambda\in[0.05,0.99],\quad
\beta_{\mathrm{fail}}\in[0,2],\quad
\beta_{\mathrm{loc}}\in[0,0.30],\quad
\tau\in[0.01,0.20],\quad
\rho\in[0,0.40],\quad
G_{\mathrm{up}}\in\{0,\ldots,6\},
\]
with \(\rho=0\) when \(G_{\mathrm{up}}=0\). The box was widened once because, in a first fit with
\(\rho\le0.15\) and \(G_{\mathrm{up}}\le3\), the selected \(\rho\) and \(G_{\mathrm{up}}\) lay at their limits;
with the box above, every selected parameter lies inside its range except \(\beta_{\mathrm{fail}}\), which is
close to its natural lower bound of zero (Supplementary Table~\ref{tab:supp-dbbs-pooled-params}).

\begin{table}[t]
\centering
\caption{Selected DBBS parameters by fold. \(w\), inertia weight on the persistence component.}
\label{tab:supp-dbbs-pooled-params}
\small
\begin{tabular}{@{}lccccccc@{}}
\toprule
Folds & \(\lambda\) & \(\beta_{\mathrm{fail}}\) & \(\beta_{\mathrm{loc}}\) & \(\tau\) & \(\rho\) & \(G_{\mathrm{up}}\) & \(w\) \\
\midrule
all 10 & 0.822 & 0.018 & 0.096 & 0.049 & 0.176 & 4 & 0.55--0.60 \\
\bottomrule
\end{tabular}
\end{table}

The failure penalty is close to zero, so DBBS moves probability downward mainly through the one-sided
locality penalty, combined with wide upward probing.

\subsection{Heuristic baselines}

Every baseline is lifted to the block level, selected on the same folds and by the same criterion as
DBBS, and grouped by the information it requires.
\emph{Own information} (what the LLM observed): persistence, a recency-weighted histogram of the
LLM's own recent prices with per-period decay \(\lambda\in\{0.02,\ldots,0.98\}\); and win-up--lose-down,
deterministic or with random step sizes, applied to every period of the previous block (a sale moves the
next price up, no sale down) and pooled with recency weights.
\emph{Full information} (rival prices and the demand model, which the LLM did not observe):
undercut-last, one grid step below each rival price; and weighted fictitious play, a logit best response
to the expected winner-takes-all profit under a discounted belief over rival prices.
\emph{Outcome reference}: the training-set mean distribution for the same treatment.
Earlier versions also used win-stay--lose-shift and undercut-own-last-price rules; these remain in the
replication outputs. Given a selected centre \(c_{n+1}\), the kernel-based dynamic heuristics predict
\[
\sigma_{n+1}(k)
=
(1-\varepsilon)
\frac{\exp(-\kappa |k-c_{n+1}|)}
{\sum_{j=0}^{K-1}\exp(-\kappa |j-c_{n+1}|)}
+
\frac{\varepsilon}{K}.
\]
For dynamic heuristics we search
\[
s\in\{1,2\},\qquad
\kappa\in\{0.4,0.8,1.5,2.5,5\},\qquad
\varepsilon\in\{0,0.02,0.05,0.10\}.
\]

\subsection{Changes, persistence and free-running controls}

Emitted distributions are highly persistent: in 78\% of \DStf{} blocks the mean
price moves by less than one grid step. A no-change forecast is therefore the most accurate one-step
predictor overall, and one-step accuracy alone cannot separate persistence from payoff-feedback search.
Supplementary Table~\ref{tab:supp-dbbs-change} therefore also scores the blocks in which the LLM's
distribution changes. On those blocks DBBS without inertia is the most accurate model: it beats
persistence, win-up--lose-down and weighted fictitious play in all seven treatments, and randomized
win-up--lose-down and undercut-last in six (paired Wilcoxon \(p\le0.047\)). The cross-fitted inertia weight
lowers held-out RPS relative to persistence alone in 6 of 7 treatments (\(p=0.047\)), so the search component carries predictive information beyond
inertia.

\begin{table}[t]
\centering
\caption{Held-out one-step accuracy on \DStf{} on all blocks, on the 22\% of blocks in which the LLM's mean
price moves by at least one grid step, and on stable blocks (W\(_1\) in \% of the price range; means over the
seven treatments).}
\label{tab:supp-dbbs-change}
\small
\begin{tabular}{@{}lccc@{}}
\toprule
Model & All & Changing & Stable \\
\midrule
DBBS & 4.9 & 10.7 & 3.4 \\
DBBS without inertia & 6.0 & \textbf{10.2} & 4.8 \\
Persistence & \textbf{4.9} & 12.1 & \textbf{3.1} \\
Win-up--lose-down & 7.1 & 11.5 & 6.0 \\
Randomized win-up--lose-down & 8.1 & 11.5 & 7.3 \\
Weighted fictitious play\(^\dagger\) & 15.6 & 18.1 & 15.6 \\
Undercut-last\(^\dagger\) & 18.2 & 18.3 & 18.5 \\
Train-mean\(^\ddagger\) & 10.8 & 13.7 & 10.4 \\
\bottomrule
\end{tabular}
\par\smallskip\footnotesize \(^\dagger\)needs full information; \(^\ddagger\)outcome reference.
\end{table}

For the free-running controls, every model replaces the LLM exactly as DBBS does: it samples a price each
period from its current distribution and updates at the LLM's intervention interval, against the same
opponent with the same configuration and on the same held-out environment seed. Full-information baselines
receive the rival's realized prices from the simulation without changing the information available to the
opponent. Supplementary Table~\ref{tab:supp-dbbs-freegame} summarizes the 70 paired games (7 treatments
\(\times\) 10 seeds).

\begin{table}[t]
\centering
\caption{Free-running reproduction of final-window prices (last 25\% of 2,000 periods). Error, mean absolute
difference between the model's and the \DStf{} agent's mean price over 70 games; DBBS closer, games in which DBBS with inertia is
closer to the LLM than the model (paired Wilcoxon test).}
\label{tab:supp-dbbs-freegame}
\small
\begin{tabular}{@{}lcc@{}}
\toprule
Model & Error & DBBS closer \\
\midrule
DBBS & \textbf{0.020} & -- \\
DBBS without inertia & 0.030 & 59/70 (\(p=5\times10^{-7}\)) \\
Win-up--lose-down & 0.063 & 56/70 (\(p=4\times10^{-8}\)) \\
Randomized win-up--lose-down & 0.064 & 55/70 (\(p=3\times10^{-8}\)) \\
Persistence & 0.074 & 58/70 (\(p=3\times10^{-10}\)) \\
Weighted fictitious play\(^\dagger\) & 0.042 & 55/70 (\(p=1\times10^{-6}\)) \\
Undercut-last\(^\dagger\) & 0.033 & 48/70 (\(p=8\times10^{-5}\)) \\
Train-mean\(^\ddagger\) & 0.013 & 31/70 (\(p=0.03\)) \\
\bottomrule
\end{tabular}
\end{table}

For \DStf{}, inertia is essential in free running. Without it, DBBS undercuts the standard learners in 81\% of
late periods (the LLM in 59\%) and prices about 0.038 below the LLM; with it, the bias shrinks to 0.013 and
DBBS is closer to the LLM than every mechanism, including the full-information references. DBBS also orders
the treatments by the LLM's final price (Spearman correlation 0.89 across the seven treatments, against 0.21
for win-up--lose-down and 0.18 for persistence).

\subsection{Theoretical statements and proofs}
\label{supp:proofs}

This subsection gives formal statements clarifying what DBBS does and does not guarantee. Proposition~\ref{prop:supp-dbbs-linear-regret} shows that DBBS is not an adversarial no-regret algorithm: against a suitable deterministic opponent its regret grows linearly. Propositions~\ref{prop:supp-dbbs-best-response} and~\ref{prop:supp-dbbs-nash-bracket} then study an idealized full-information version to explain the direction of the empirical update, showing respectively that the search centre tracks the lowest best response to a fixed opponent and that idealized self-play enters and remains in the finite-grid Nash bracket.

\begin{definition}[No external adversarial regret]
Consider a learning algorithm that at each period \(t=1,2,\ldots\) selects a mixed strategy \(\sigma_t\in\Delta(\{0,\ldots,K-1\})\) and then draws an action \(i_t\sim\sigma_t\). Let \((j_t)_{t\ge1}\) be an arbitrary deterministic opponent price sequence. Let \(r(i,j)\) denote the normalized payoff from playing tick \(i\) against opponent tick \(j\). The expected external regret after \(T\) periods is
\[
R_T
=
\max_{a\in\{0,\ldots,K-1\}}
\sum_{t=1}^T r(a,j_t)
-
\E\left[\sum_{t=1}^T r(i_t,j_t)\right].
\]
The algorithm satisfies no external adversarial regret if, for every deterministic sequence \((j_t)_{t\ge1}\),
\[
\limsup_{T\to\infty}\frac{R_T}{T}\le0.
\]
\end{definition}

\begin{proposition}[DBBS can have linear adversarial regret]
\label{prop:supp-dbbs-linear-regret}
Consider DBBS in the winner-takes-all Bertrand market with normalized payoffs and \(K=21\). Let \(0<\lambda\le1\) and \(0\le\beta_{\mathrm{fail}},\beta_{\mathrm{loc}}<\infty\). Then DBBS does not satisfy no external adversarial regret. Specifically, the claim holds in either of the following cases:
\begin{enumerate}[label=(\roman*)]
    \item DBBS is run with fixed \(\tau>0\).
    \item DBBS is run with \(\rho=0\) and a decreasing temperature sequence \(\tau_n\downarrow0\) satisfying
    \[
    \sum_{n=1}^{\infty}\exp(-\gamma/\tau_n)<\infty
    \]
    for some \(\gamma\in(0,18/20)\).
\end{enumerate}
\end{proposition}

\begin{proof}
We construct an oblivious deterministic opponent sequence in each case.

\paragraph{Case 1: fixed \(\tau>0\).}
Because rewards and failure indicators lie in \([0,1]\) and \(i,m_n\in\{0,\ldots,20\}\), the DBBS score satisfies
\[
-\beta_{\mathrm{fail}}-20\beta_{\mathrm{loc}}
\le S_{i,n}\le1.
\]
Hence score differences are bounded by
\[
D=1+\beta_{\mathrm{fail}}+20\beta_{\mathrm{loc}}.
\]
Every tick receives base-policy probability at least \(\exp(-D/\tau)/21\), and after mixing with local search,
\[
\sigma_n(i)
\ge
(1-\rho)\frac{\exp(-D/\tau)}{21}.
\]
Against an opponent that always plays \(j_t=20\), any positive probability on sufficiently low suboptimal ticks generates a constant expected per-period regret. Thus \(R_T\) grows linearly in \(T\).

\paragraph{Case 2: decreasing \(\tau_n\).}
Fix \(\gamma\in(0,18/20)\) satisfying the summability condition. Choose a bait length \(N\), taken to be a whole number of intervention epochs, large enough that
\[
\frac{18}{20}\left(1-(1-\lambda)^N\right)\ge\gamma.
\]
Such an \(N\) exists because \(0<\lambda\le1\) implies \((1-\lambda)^N\to0\) and \(\gamma<18/20\).
Let the opponent play \(j_t=19\) for \(1\le t\le N\) and \(j_t=20\) thereafter. With positive probability, DBBS selects tick \(18\) throughout the bait phase. On this event, after the bait phase,
\[
v_{18}\ge\gamma,\qquad \phi_{18}=0,\qquad m=18,
\]
and unplayed ticks have zero payoff and failure scores. Along the continuation on which DBBS keeps selecting tick \(18\), its score remains separated from all other scores by at least \(\gamma\). Therefore
\[
1-\sigma_n(18)
\le
20\exp(-\gamma/\tau_n).
\]
Since the right-hand side is summable, the infinite product criterion gives a positive probability that DBBS continues selecting tick \(18\) forever. After the opponent switches to tick \(20\), the fixed comparator \(19\) earns \(19/20\) per period whereas tick \(18\) earns \(18/20\), a per-period shortfall of \(1/20\). The bait phase has fixed length \(N\), over which tick \(18\) is ahead of tick \(19\) by at most a constant, so this finite advantage is dominated and the comparator \(19\) is strictly better over the whole horizon for all large \(T\). This creates a positive asymptotic regret rate on that event, and hence the expected regret is linear.
\end{proof}

Case~1 also covers DBBS with inertia (Supplementary Note~\ref{supp:dbbs}): for an inertia weight \(w<1\) the
emitted policy satisfies \(\sigma_n(i)\ge(1-w)(1-\rho)\exp(-D/\tau)/21\), and the argument is unchanged.

Propositions~\ref{prop:supp-dbbs-best-response} and~\ref{prop:supp-dbbs-nash-bracket} replace the
recency-weighted scores by their expectations against an opponent policy \(\mu\): the idealized payoff and
failure beliefs are
\[
v_i^*(\mu)=\E_{j\sim\mu}\,r(i,j),
\qquad
\phi_i^*(\mu)=\1\{i>0\}\,\Pr_{j\sim\mu}\bigl(r(i,j)=0\bigr),
\]
the expectations of the update targets \(r_i\) and \(y_i\) above.

\begin{proposition}[Idealized DBBS tracks best responses]
\label{prop:supp-dbbs-best-response}
Let \(\sigma_2\) be a fixed opponent policy. Suppose firm 1 uses an idealized DBBS update with correct payoff beliefs, \(\beta_{\mathrm{fail}}=\beta_{\mathrm{loc}}=\rho=0\), and \(\tau>0\). Then after one update the search centre is \(m=\min\BR(\sigma_2)\), and it remains there.
\end{proposition}

\begin{proof}
Under the stated parameters, the idealized score of tick \(i\) is its exact expected payoff \(v_i^*(\sigma_2)\). The updated policy is
\[
\tilde\sigma_1(i)
=
\frac{\exp(v_i^*(\sigma_2)/\tau)}
{\sum_{k=0}^{K-1}\exp(v_k^*(\sigma_2)/\tau)}.
\]
Because softmax preserves the ordering of scores,
\[
\argmax_i \tilde\sigma_1(i)
=
\argmax_i v_i^*(\sigma_2)
=
\BR(\sigma_2).
\]
By the centre-selection rule, \(m=\min\BR(\sigma_2)\). Since \(\sigma_2\) is fixed and the scores do not depend on the current centre, the same centre is selected at every later update.
\end{proof}

\begin{proposition}[Idealized self-play enters the finite-grid Nash bracket]
\label{prop:supp-dbbs-nash-bracket}
Let \(B=\{0,1,2\}\). Suppose both firms use the idealized DBBS update with perfect beliefs, \(\beta_{\mathrm{loc}}=0\), \(\rho=0\), and \(\beta_{\mathrm{fail}}>0\). Suppose the initial policies induce unique failure-penalized score maximizers with positive score gaps. Then there exists \(\bar\tau>0\) such that, for all \(0<\tau<\bar\tau\), the modal-action sequence enters \(B\times B\) in finite time and remains there.
\end{proposition}

\begin{proof}
For an opponent policy \(\mu\), define
\[
g_i(\mu)=v_i^*(\mu)-\beta_{\mathrm{fail}}\phi_i^*(\mu).
\]
The idealized update is
\[
F_\tau(\mu)_i
=
\frac{\exp(g_i(\mu)/\tau)}
{\sum_{\ell=0}^{K-1}\exp(g_\ell(\mu)/\tau)}.
\]
Since softmax preserves score ordering, the modal actions of \(F_\tau(\mu)\) are the maximizers of \(g_i(\mu)\).

Against a pure opponent action \(k\),
\[
g_i(\delta_k)=
\begin{cases}
i/(K-1), & i<k,\\
k/[2(K-1)], & i=k,\\
-\beta_{\mathrm{fail}}, & i>k.
\end{cases}
\]
Thus, for every \(k\ge3\), the unique score maximizer is \(k-1\). The update therefore moves pure high-price anchors one step downward.

It remains to show that the low-price bracket is invariant. Choose
\[
0<\varepsilon<
\min\left\{\frac12,\frac{\beta_{\mathrm{fail}}}{1+\beta_{\mathrm{fail}}}\right\}
\]
and define \(U_\varepsilon=\{\mu:\mu(B)\ge1-\varepsilon\}\). For every \(\mu\in U_\varepsilon\) and \(i\ge3\), \(v_i^*(\mu)\le\varepsilon\) and \(\phi_i^*(\mu)\ge1-\varepsilon\). Hence
\[
g_i(\mu)\le \varepsilon-\beta_{\mathrm{fail}}(1-\varepsilon)<0,
\]
whereas \(g_0(\mu)=0\). Thus every score maximizer belongs to \(B\), and for sufficiently small \(\tau\), \(F_\tau(U_\varepsilon)\subseteq U_\varepsilon\).

By continuity and the strict score gaps, for each \(k\ge3\) there is a neighbourhood \(V_k\) of \(\delta_k\) that is mapped close to \(\delta_{k-1}\) for sufficiently small \(\tau\). Repeated updates therefore move any high anchor down one step at a time until the process enters \(U_\varepsilon\), after which modal actions remain in \(B\). A single \(\bar\tau\) can be chosen because the grid is finite.
\end{proof}

\suppnote{Deployment ablations}{supp:ablations}

This note supports main-text Fig.~4. Starting from the \PRone{} control, it changes one or several
components of the deployment and identifies which of them move the outcome from the high-price regime
toward competition.

\subsection{Fish-style positive control}

The positive control is \PRone{} in the logit market. It outputs a continuous single price, observes the rival price, receives raw histories with a long memory window, and uses profit-oriented prompt language. Across independent control runs, the market-average price elevation is close to the joint-profit benchmark.

\subsection{Direct protocol ablations}

The direct ablation starts from a Fish-style control and changes one or more deployment components while holding the market, model, prompt, price bounds, persistent text memory, and call budget fixed where appropriate. We use the LLM temperature $0.7$ instead of $1.0$. The detailed ablation protocol can be observed in Supplementary Table~\ref{tab:supp-protocol-ablations}.

\begin{table}[t]
\centering
\caption{Direct protocol ablations. All treatments use the same model, temperature, price bounds, persistent text memory, and number of LLM calls unless otherwise stated. \(I\) is the intervention interval and \(H\) the history depth in periods. Arms A1--A6 change exactly one component relative to the A1 control, A7--A10 change two or more at interval \(1\), and A11--A13 are the interval-\(50\) variants.}
\label{tab:supp-protocol-ablations}
\small
\begin{tabular}{@{}clcccrl@{}}
\toprule
\textbf{Arm} & \textbf{Output} & \textbf{Market} & \(I\) & \textbf{History} & \(H\) & \textbf{Rival price} \\
\midrule
A1  & Continuous point & logit       & 1  & Raw     & 100 & Visible \\
A2  & Continuous point & homogeneous & 1  & Raw     & 100 & Visible \\
A3  & Grid point       & logit       & 1  & Raw     & 100 & Visible \\
A4  & Grid distribution & logit      & 1  & Raw     & 100 & Visible \\
A5  & Continuous point & logit       & 1  & Raw     & 100 & Hidden  \\
A6  & Continuous point & logit       & 1  & Summary & 100 & Visible \\
\midrule
A7  & Continuous point & logit       & 1  & Summary & 100 & Hidden  \\
A8  & Grid point       & logit       & 1  & Raw     & 5   & Visible \\
A9  & Grid point       & logit       & 1  & Summary & 100 & Hidden  \\
A10 & Grid distribution & logit      & 1  & Summary & 100 & Hidden  \\
\midrule
A11 & Grid point       & logit       & 50 & Raw     & 250 & Visible \\
A12 & Grid distribution & logit      & 50 & Raw     & 250 & Visible \\
A13 & Grid distribution & logit      & 50 & Summary & 250 & Hidden  \\
\bottomrule
\end{tabular}
\end{table}

\paragraph{Outcome measure and the robust-mover criterion.}
Every arm runs for \(300\) LLM calls per firm, so an interval-\(1\) arm plays \(300\) periods and an interval-\(50\) arm plays \(15{,}000\). Each arm is scored by the Calvano price index over a window of the last \(100\) \emph{periods}, normalized so that \(0\) is the one-shot Nash price and \(1\) the joint-profit price \emph{of the game that arm actually played}. The window is deliberately fixed in periods rather than in calls: a window of the last \(100\) calls would cover \(5{,}000\) periods for an interval-\(50\) arm against \(100\) for every other arm, and averaging over fifty times more data would make those arms look far more settled than the rest purely as an artefact of the window. Because a single mean can hide seed-level disagreement, and because the language model's sampling is not controlled by the environment seed, we do not classify an arm by its mean alone. We first form the \emph{control band}, the interval spanned by the A1 control's own seeds, which is \([0.92,1.01]\). An arm is a \emph{robust mover} only if \emph{every} one of its seeds lies below that band; an arm whose seeds straddle or exceed the band shows \emph{no robust competition}, however favourable its mean. Main-text Fig.~4b,c plots each arm's seeds and mean.

Two normalization details matter before the numbers are read. First, continuous-output and grid-output arms are scored against different anchors, because the \(21\)-point grid (\(1.40\) to \(2.00\) in steps of \(0.03\)) contains neither continuous benchmark and the discretized game therefore has its own. Continuous arms use the textbook values \(0=1.4729\) and \(1=1.9250\). On the grid, solving the \(21\times21\) profit matrix yields \emph{two} symmetric pure equilibria, \(1.46\) and \(1.49\), and symmetric joint profit is maximized at \emph{two} prices, \(1.91\) and \(1.94\), which tie to five decimal places and are displayed to the agent as the same number. We anchor \(0\) at the upper equilibrium \(1.49\) and \(1\) at the lower joint-profit price \(1.91\). Both choices are deliberately conservative: they make measured elevation as small as possible, so the normalization cannot manufacture a mover, at the cost that an arm parked on the collusive plateau reads at or just above \(1\) rather than below it. This is why several grid arms post indices slightly above one; it is the anchoring convention, not pricing above monopoly. Under the grid normalization both anchors are themselves playable actions, so an index of exactly \(0\) or \(1\) is attainable, whereas under the continuous normalization no action sits at either anchor. Small differences between grid and continuous arms should accordingly not be over-read. Second, A2 changes the market itself and is scored against the homogeneous market's own benchmarks, \(0=0.100\) and \(1=0.360\), so it is comparable to the other arms only in direction, not in level.

\paragraph{Single-factor ablations.}
The single-factor arms flip exactly one component from the control. Only one of the five single flips is a robust mover. Replacing the raw per-period history with the per-price summary (A6) moves the index from \(0.97\) to \(0.53\), a change of \(-0.44\), with every seed below the control band. Hiding the rival's price (A5) produces an almost identical mean, \(0.56\), but is \emph{not} a robust mover: its seeds spread from roughly \(0.25\) to \(0.93\), and the highest of them sits inside the control band. The two arms that change only the action representation move in the opposite direction: emitting a grid point price (A3) gives \(1.09\) and emitting a grid distribution (A4) gives \(1.07\), both above the control. Swapping the demand system to the homogeneous market (A2) gives \(0.88\), a small decline that likewise fails the criterion.

\paragraph{Combined ablations.}
Among the combined arms, the robust movers are exactly the arms that pair the summarized history with hidden rival prices, and that this holds regardless of the output representation. Adding the grid point price to that pair (A9) gives \(0.34\); the pair alone on the control's own continuous output (A7) gives \(0.31\); adding the grid distribution (A10) gives \(0.24\); and running that last combination at intervention interval \(50\) (A13, which is \DSff{} inside the Fish harness) gives \(0.14\), the lowest value in the screen. A7 is the informative cell here, because it completes a \(2\times2\) on the control's own output representation together with A1, A5 and A6, and is therefore the only arm that identifies the interaction between summarized history and hidden rival prices without also changing how the agent acts.

The arms that combine factors \emph{without} summarized history move nowhere. Shortening the history to five periods while emitting a grid point price (A8) gives \(1.13\), the highest value in the screen, so less history is not a substitute for differently organized history. Holding a grid point or a grid distribution for fifty periods (A11 and A12) gives \(1.00\) and \(1.04\). For these interval-\(50\) arms, because the scoring window is a fixed \(100\) periods and each decision is held for \(50\), their window is generated by only two decisions, against one hundred for every interval-\(1\) arm. That is the honest number under a window that puts all arms on the same footing, but it means their exact values rest on far less evidence and we do not read them closely.

\paragraph{The movers do not reach competition.}
It is worth being explicit about the size of the effect. The robust movers occupy a band from \(0.14\) to \(0.34\), and the partial movers A5 and A6 sit near \(0.55\), whereas every non-moving arm sits at \(0.88\) or above. There is thus a wide empty region between the two groups, and the deployment interface determines which side of it an arm falls on. But no arm reaches \(0\). Even A13, which combines every competitive factor, still prices a seventh of the way from the Nash price to the joint-profit price. The deployment bundle moves these agents decisively off the joint-profit outcome; it does not deliver competitive pricing.

\paragraph{The movement is genuine competition.}
Three further diagnostics rule out the main alternative readings. Supplementary Fig.~\ref{fig:supp-protocol-ablation-diagnostics}a pairs each arm's price index with its profit index. The two fall together across the ordering, so the low-price arms genuinely forgo profit rather than sliding along a flat region of the profit function. The same figure shows that the arms indexing above \(1\) do not earn more than the control: their profit index sits at or just below the control's own, so an index above one is an overshoot past the profit peak rather than a better outcome for the firms.

Supplementary Fig.~\ref{fig:supp-protocol-ablation-diagnostics}b measures how concentrated play is within the scoring window. The high-elevation arms park on very few prices: A11 spends the entire window on a single price, the grid's joint-profit action, and A3, A8 and A12 use between two and five distinct prices with \(58\)--\(60\%\) of the window at the mode. The robust movers never settle in this way, using between twelve and forty-eight distinct prices with only \(11\)--\(27\%\) of the window at the mode. A5, which does not qualify as a robust mover, is nonetheless the most dispersed arm of all, spreading over sixty-nine distinct prices. High concentration therefore tracks coordination lock-in, and the movers do not exhibit it.

Supplementary Fig.~\ref{fig:supp-protocol-ablation-diagnostics}c reports NashConv, the total best-response gain at the final profile, which is zero exactly at an equilibrium of the game played. A2 is excluded because it is a different market. Exploitability falls monotonically with price elevation: the control and the non-moving arms lie between \(13.75\) and \(17.82\), A5 and A6 at \(6.98\) and \(5.45\), and the robust movers between \(2.78\) and \(0.81\). The arms that price low are therefore also the arms closest to equilibrium play, which is what one expects of competition and not of a coordination failure.

Finally, Supplementary Fig.~\ref{fig:supp-protocol-ablation-dynamics} shows the per-call action path. The non-moving arms freeze: A3, A8, A11 and A12 repeat the identical action in \(100\%\) of their last fifty calls, with a mean per-call move of \(0.000\). The control is nearly as static at \(92\%\). The robust movers keep revising throughout, with \(55\%\), \(32\%\) and \(2\%\) repetition for A7, A10 and A13 respectively and correspondingly larger per-call moves. The distinction is therefore not only in the level reached but in whether the agents stop searching at all.

Taken together, these results are the central reason the main paper describes the relevant mechanism as a deployment bundle or information interface rather than as action representation alone. The action representation on its own moves the outcome slightly upward; what moves it down is how the agent is shown its past and whether it can see its rival.

\begin{figure*}[t]
\centering
\suppfig[width=0.98\textwidth]{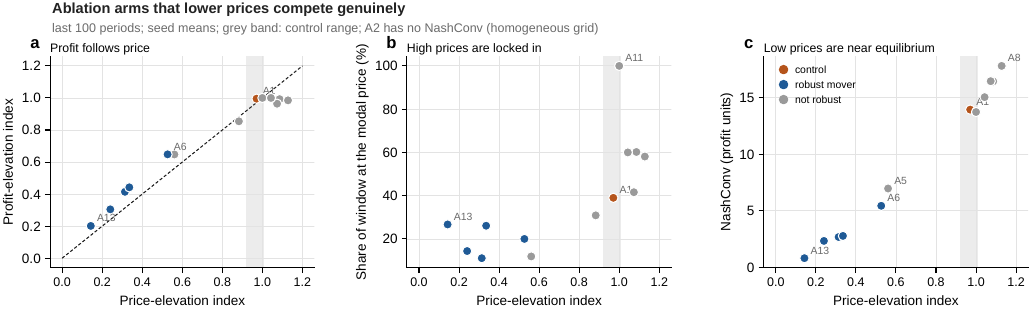}
\caption{\textbf{Ablation arms that lower prices compete genuinely.} Seed means over the last 100 periods against the price-elevation index; grey band, range of the control's seeds; orange, control; blue, robust movers (every seed below the control band); grey, arms without robust movement. \textbf{a}, Profit-elevation index with the parity line: low-price arms genuinely forgo profit, and arms above 1 earn no more than the control. \textbf{b}, Share of the window at the modal price: high-elevation arms lock in on few prices (A11 on a single action), robust movers use 12--48 distinct prices. \textbf{c}, NashConv in profit units falls with elevation, from 13.7--17.8 for the control and non-movers to 0.8--2.8 for robust movers; A2 is omitted because in the winner-takes-all market NashConv is set by the resolution of the response grid.}
\label{fig:supp-protocol-ablation-diagnostics}
\end{figure*}

\begin{figure*}[t]
\centering
\suppfig[width=0.98\textwidth]{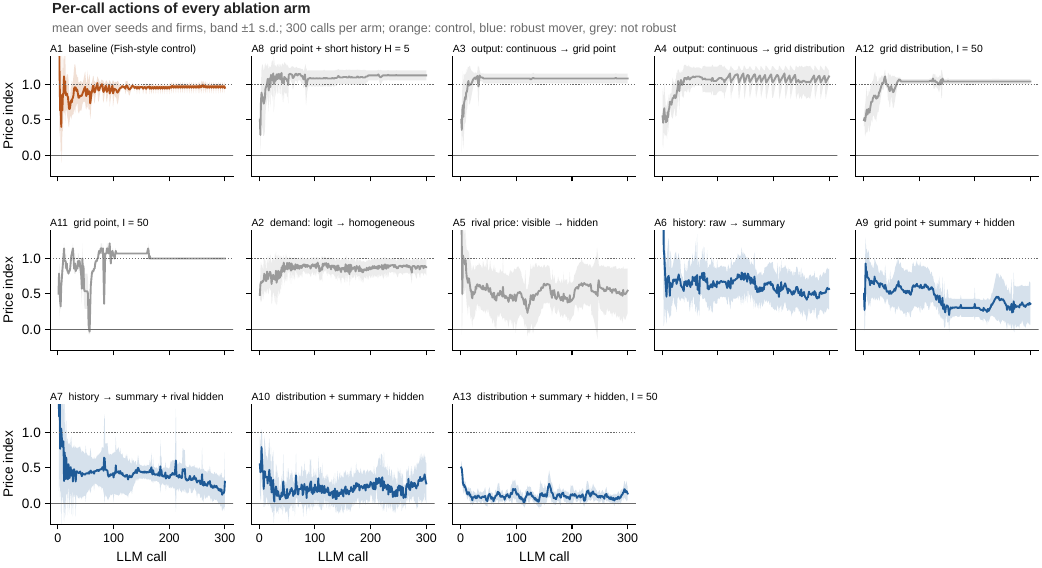}
\caption{\textbf{Per-call actions of every ablation arm.} Mean price index of each LLM decision (the mean price over its intervention block) across seeds and firms, band \(\pm1\) s.d.; 300 calls per arm. Non-moving arms freeze on a single action (A3, A8, A11, A12 repeat their action in all of the last 50 calls); robust movers keep revising (A7, A10, A13).}
\label{fig:supp-protocol-ablation-dynamics}
\end{figure*}

\suppnote{Heterogeneous LLM markets}{supp:pllm-matchups}

This note supports main-text Figs.~5 and~6. It reports \PRone{} matchups in the differentiated logit market
and the trajectories of every heterogeneous-deployment run.

\subsection{\PRone{} matchups in the logit market}

The focal \PRone{} agent emits a continuous price and observes the opponent's price under the full-information Fish-style setting, in the configuration described in Section~\ref{app:point-llm-configuration}. Algorithmic opponents and the \DRone{} agent use a 21-point grid. All matchup cells containing an LLM run for 300 periods with two to six seeds (Supplementary Table~\ref{tab:supp-point-llm-matchups}). We normalize price elevation by the continuous-game Nash and joint-profit benchmarks.

Two to six seeds per cell is a small sample, and it is worth being explicit about what it does and does not support. The environment seed controls the market and the algorithmic opponents, but it does not control the language model's own sampling, which runs at a non-zero temperature. Cells that differ only in seed can therefore diverge substantially, and we have observed such divergence in individual interval-1 cells. We accordingly read the matchup table as an ordering of opponent classes rather than as a set of point estimates: the separation between the high-elevation cells at the top and the near-Nash cells at the bottom is large and consistent across seeds, whereas differences of a few hundredths between adjacent rows are not resolved at this sample size.

\paragraph{Opponent set.}

The focal \PRone{} agent is matched against:
\begin{itemize}
    \item another \PRone{} agent,
    \item a \DRone{} agent using the same model, prompt, and memory but emitting a grid distribution,
    \item Exp3, PPO, UCB, Tsallis-INF, bandit regret matching, and full-information regret matching,
    \item a static monopoly control, and
    \item an undercut heuristic that opens at the monopoly price and subsequently posts one grid step below the LLM's previous price.
\end{itemize}

Supplementary Table~\ref{tab:supp-point-llm-matchups} reports the resulting price elevations for both firms in each cell. Supplementary Fig.~\ref{fig:supp-pllm-matchups} shows the per-seed spread behind those means.

\begin{table}[t]
\centering
\caption{\PRone{} matchups in the differentiated-product market. Values are focal \PRone{} price elevations and opponent elevations, averaged over all seeds of each cell; the same runs as Supplementary Fig.~\ref{fig:supp-pllm-matchups}. Given that sample size and the uncontrolled LLM sampling discussed above, the rows should be read as an ordering of opponent classes rather than as precise point estimates.}
\label{tab:supp-point-llm-matchups}
\footnotesize
\setlength{\tabcolsep}{4pt}
\begin{tabularx}{\linewidth}{@{}>{\raggedright\arraybackslash}p{3.3cm}ccc>{\raggedright\arraybackslash}X@{}}
\toprule
\textbf{Cell} & \textbf{Seeds} & \textbf{\PRone{} \(\Delta\)} & \textbf{Opponent \(\Delta\)} & \textbf{Interpretation} \\
\midrule
\PRone{} vs \PRone & 3 & 1.11 & 0.91 & high-price regime under like LLM interaction \\
\PRone{} vs undercut heuristic & 2 & 1.04 & 0.97 & high \PRone{} prices are a best response to the fixed rule \\
\PRone{} vs \DRone & 6 & 0.91 & 0.93 & cross-deployment LLM interaction remains high \\
\PRone{} vs bandit regret matching & 2 & 0.57 & 0.53 & partial persistence under slow bandit adaptation \\
\PRone{} vs static monopoly & 3 & 0.44 & 1.03 & \PRone{} exploits a fixed high-price anchor \\
\PRone{} vs Exp3 & 3 & 0.36 & 0.46 & intermediate; seeds range from 0.19 to 0.64 \\
\PRone{} vs PPO & 2 & 0.21 & 0.44 & low price elevation \\
\PRone{} vs UCB & 2 & 0.20 & 0.35 & low price elevation \\
\PRone{} vs Tsallis-INF & 2 & 0.19 & 0.27 & low price elevation \\
\PRone{} vs full-information regret matching & 5 & 0.04 & 0.05 & essentially Nash \\
\bottomrule
\end{tabularx}
\end{table}
\begin{figure}[t]
\centering
\suppfig[width=\linewidth]{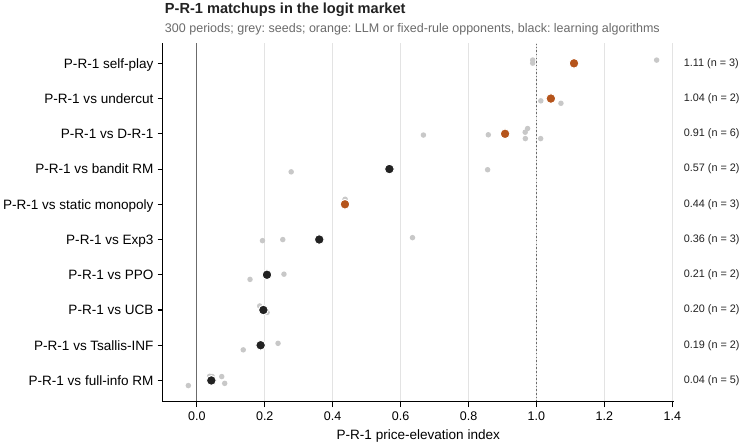}
\caption{\textbf{\PRone{} matchups in the logit market.} Focal \PRone{} price-elevation index (0 = continuous Nash price, 1 = joint-profit price) over 300 periods; grey points are seeds, large points their means. Orange, LLM or fixed-rule opponents; black, learning algorithms.}
\label{fig:supp-pllm-matchups}
\end{figure}

\subsection{Heterogeneous-deployment trajectories}

In some homogeneous \PRone--\DSone{} trajectories, the selected behaviour is not well described as convergence to a stationary stage-game equilibrium. The \PRone{} agent may maintain a high price anchor for extended periods, receive little or no demand while the \DSone{} prices below it, and then intermittently lower price to harvest demand at a still-elevated margin. We interpret such paths as recurrent price regimes rather than as fully characterized equilibria.

Main-text Fig.~6 shows the visible- and hidden-rival-price \PRone--\DSone{} trajectories with prices,
the \DSone{} agent's emitted policy and both firms' per-period profits. With visible rival prices the run locks in
near the ceiling while the \DSone{} agent harvests one-grid-step undercuts; with hidden rival prices it produces
recurrent anchor-and-harvest cycles. Both are single runs.

\subsection{Regime labels}

The regime labels of main-text Fig.~5 describe each run's price path qualitatively; they are not the output of
a classification rule. \emph{Competitive undercutting}: the market-clearing price stays in or near the
finite-grid Nash region while agents undercut each other (three runs). \emph{High-price lock-in}: the market-clearing price
stays at or one grid step below the ceiling, with or without all firms tied (three runs).
\emph{Anchor-and-harvest}: one agent holds or repeatedly restores a high price, and another prices below it and
captures demand (six runs). In the three-\PRone{} run and the two-\PRone{} GPT-5.6 Terra run with medium
reasoning, the roles alternate: each upward reset restores a high price that the other firms then undercut.

Supplementary Fig.~\ref{fig:supp-regime-atlas} shows the price paths of all twelve runs summarized in main-text
Fig.~5. In the GPT-5.6 Terra runs the temperature of 0.7 is applied only without reasoning; the OpenAI API
rejects a temperature when reasoning is enabled, so the two medium-reasoning runs use the provider's default
sampling (recorded with each run). Each is a single 300-period run, so the regime labels describe representative paths rather than
regime frequencies.

\begin{figure*}[t]
\centering
\suppfig[width=0.98\textwidth]{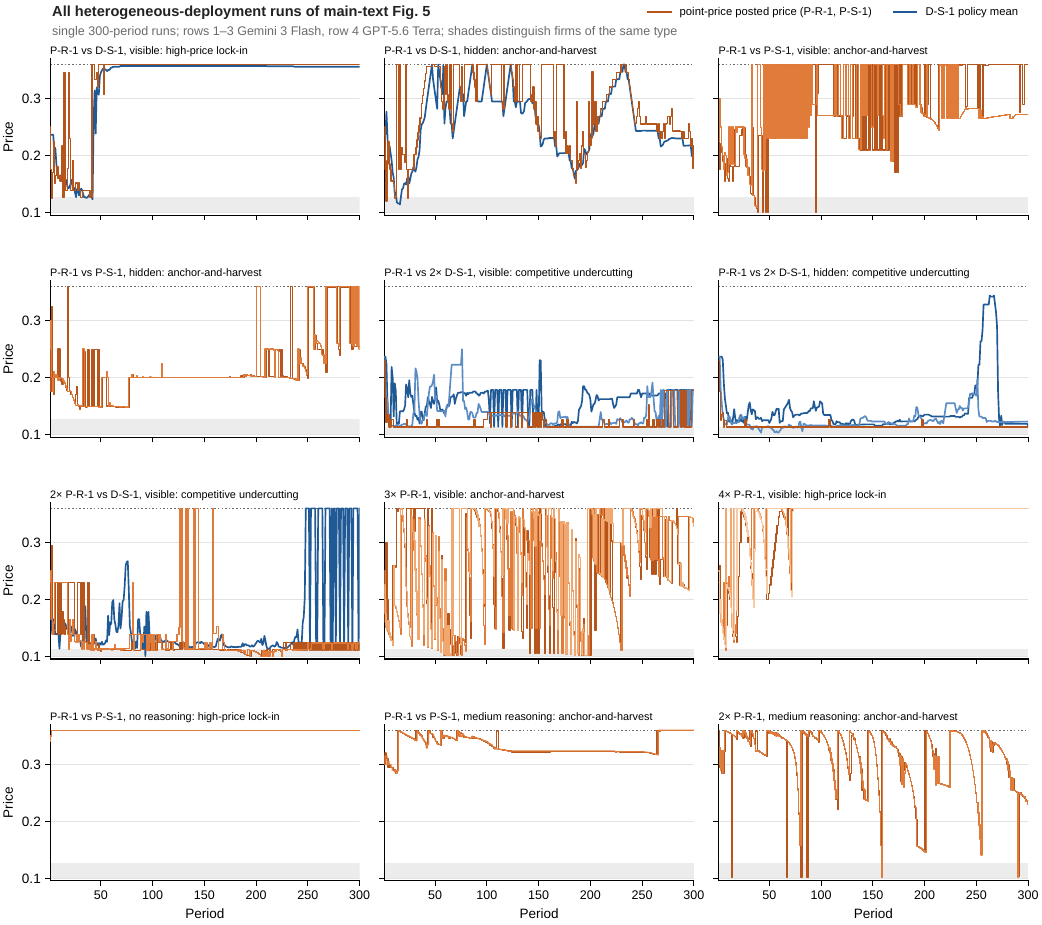}
\caption{\textbf{All heterogeneous-deployment runs.} Posted prices of the point-price agents \PRone{} and \PSone{} (orange) and the emitted policy mean of \DSone{} agents (blue) in each run of main-text Fig.~5; shades distinguish firms of the same type. Grey band, finite-grid competitive region (three lowest grid prices with two firms, two lowest with three or four); dotted line, price ceiling. Rows 1--3, Gemini 3 Flash; row 4, GPT-5.6 Terra.}
\label{fig:supp-regime-atlas}
\end{figure*}

\suppnote{Welfare}{supp:welfare}

Supplementary Tables~\ref{tab:supp-welfare-homogeneous} and~\ref{tab:supp-welfare-logit} decompose welfare
for the main price regimes. Windows follow the main-text
figures: the final \(10\%\) for the replicated runs of main-text Figs.~2 and 4a, the final third for the protocol ablations and the final
100 periods for the heterogeneous runs.

\paragraph{Homogeneous market.} All \(M=1000\) units sell at the lowest posted price, and consumers value
the good at \(v=0.36\), the top of the price grid. Consumer surplus is \(M(v-p_{\min})\), producer surplus
\(M(p_{\min}-c)\), and total welfare is the constant \(M(v-c)=260\) per period: there is no deadweight loss,
and the price level only divides a fixed surplus between consumers and firms. The consumer share equals one
minus the price-elevation index measured from marginal cost.

\begin{table*}[t]
\centering
\caption{Welfare in the homogeneous market. Price, mean market-clearing (lowest) price; consumer and
producer surplus as percentages of the maximal per-period surplus \(M(v-c)=260\), which is also total
welfare in every row; leading firm, share of industry profit of the firm earning most (the \DStf{} agent
in the rows against learners). In the heterogeneous runs, \PRone{} always observes its rivals' prices and
``rival hidden'' applies to the other firms. Single-run rows are representative paths, not treatment
estimates.}
\label{tab:supp-welfare-homogeneous}
\footnotesize
\setlength{\tabcolsep}{4pt}
\begin{tabularx}{\linewidth}{@{}>{\raggedright\arraybackslash}X>{\raggedright\arraybackslash}p{2.1cm}cccc@{}}
\toprule
\textbf{Regime} & \textbf{Runs} & \textbf{Price} & \makecell{\textbf{Consumer}\\\textbf{(\%)}} & \makecell{\textbf{Producer}\\\textbf{(\%)}} & \makecell{\textbf{Leading}\\\textbf{firm (\%)}} \\
\midrule
Competitive benchmark (p = c) &  & 0.100 & 100 & 0 &  \\
Joint-monopoly benchmark (p = v) &  & 0.360 & 0 & 100 &  \\
\addlinespace
Standard learners, markets without UCB & 21 markets, 10 seeds & 0.124 & 91 & 9 &  \\
Standard learners, markets with UCB & 7 markets, 10 seeds & 0.158 & 78 & 22 &  \\
\addlinespace
\DStf{} self-play & 10 seeds & 0.113 & 95 & 5 &  \\
\DSff{} self-play & 10 seeds & 0.113 & 95 & 5 &  \\
\DStf{} vs Exp3 & 10 seeds & 0.142 & 84 & 16 & 72 \\
\DStf{} vs BM-RM & 10 seeds & 0.173 & 72 & 28 & 73 \\
\DStf{} vs PPO & 10 seeds & 0.164 & 75 & 25 & 72 \\
\DStf{} vs UCB & 10 seeds & 0.153 & 80 & 20 & 74 \\
\addlinespace
\PRone{} self-play & 5 seeds & 0.322 & 15 & 85 & 59 \\
\addlinespace
\PRone{} vs \DSone{}, rival visible (Gemini 3 Flash): high-price lock-in & 1 run & 0.356 & 2 & 98 & 66 \\
\PRone{} vs \DSone{}, rival hidden (Gemini 3 Flash): anchor-and-harvest & 1 run & 0.260 & 39 & 61 & 71 \\
\PRone{} vs \PSone{}, rival visible (Gemini 3 Flash): anchor-and-harvest & 1 run & 0.279 & 31 & 69 & 86 \\
\PRone{} vs \PSone{}, rival hidden (Gemini 3 Flash): anchor-and-harvest & 1 run & 0.254 & 41 & 59 & 75 \\
\PRone{} vs 2\(\times\) \DSone{}, rival visible (Gemini 3 Flash): competitive undercutting & 1 run & 0.118 & 93 & 7 & 89 \\
\PRone{} vs 2\(\times\) \DSone{}, rival hidden (Gemini 3 Flash): competitive undercutting & 1 run & 0.112 & 95 & 5 & 100 \\
2\(\times\) \PRone{} vs \DSone{}, rival visible (Gemini 3 Flash): competitive undercutting & 1 run & 0.112 & 96 & 4 & 61 \\
3\(\times\) \PRone{}, rival visible (Gemini 3 Flash): anchor-and-harvest & 1 run & 0.260 & 38 & 62 & 42 \\
4\(\times\) \PRone{}, rival visible (Gemini 3 Flash): high-price lock-in & 1 run & 0.360 & 0 & 100 & 25 \\
\PRone{} vs \PSone{}, rival visible (GPT-5.6 Terra, no reasoning): high-price lock-in & 1 run & 0.360 & 0 & 100 & 50 \\
\PRone{} vs \PSone{}, rival visible (GPT-5.6 Terra, medium reasoning): anchor-and-harvest & 1 run & 0.335 & 10 & 90 & 96 \\
2\(\times\) \PRone{}, rival visible (GPT-5.6 Terra, medium reasoning): anchor-and-harvest & 1 run & 0.295 & 25 & 75 & 55 \\
\bottomrule
\end{tabularx}
\end{table*}

\paragraph{Logit market.} In the differentiated market of the protocol ablations, prices change total
welfare. With the calibration of \cite{FishGonczarowskiShorrer2026AlgorithmicCollusionLLMs} (\(\alpha=8\),
\(\beta=4\), \(u_0=0\), \(M=100\), \(c=1\)), consumer surplus is
\((M/\beta)\log\bigl(e^{u_0}+\sum_i e^{\alpha-\beta p_i}\bigr)\) and producer surplus
\(\sum_i(p_i-c)q_i\). The profits recorded in every period of every arm are reproduced from the recorded
prices before the decomposition.

\begin{table}[t]
\centering
\caption{Welfare per period in the logit market of the protocol ablations (five seeds per arm, final
third). Deadweight loss is the shortfall of total welfare relative to the continuous Nash benchmark.}
\label{tab:supp-welfare-logit}
\small
\begin{tabularx}{\linewidth}{@{}Xccccc@{}}
\toprule
\textbf{Arm} & \textbf{Price} & \textbf{CS} & \textbf{PS} & \textbf{TW} & \textbf{DWL (\%)} \\
\midrule
Continuous Nash benchmark & 1.47 & 71.5 & 44.6 & 116.1 & 0.0 \\
Joint-monopoly benchmark & 1.93 & 32.7 & 67.5 & 100.2 & 13.7 \\
Control (continuous point, raw, visible) & 1.91 & 33.7 & 67.4 & 101.1 & 12.9 \\
Grid point & 1.95 & 31.2 & 67.3 & 98.5 & 15.1 \\
Grid distribution & 1.94 & 31.8 & 66.7 & 98.5 & 15.2 \\
Rival hidden & 1.73 & 49.6 & 59.4 & 109.0 & 6.1 \\
Summary history & 1.71 & 50.7 & 59.4 & 110.2 & 5.1 \\
Summary + hidden & 1.61 & 59.2 & 54.1 & 113.2 & 2.5 \\
Grid point + summary + hidden & 1.63 & 57.3 & 55.5 & 112.8 & 2.8 \\
Distribution + summary + hidden & 1.59 & 61.1 & 52.5 & 113.6 & 2.1 \\
Full \DSff{} bundle (Fish harness) & 1.53 & 66.1 & 49.0 & 115.1 & 0.8 \\
\bottomrule
\end{tabularx}
\end{table}

The two markets give the same ordering. In the homogeneous market, \DStf{} and \DSff{} self-play leave consumers \(95\%\)
of the surplus, and markets in which \DStf{} undercuts standard learners leave them \(72\)--\(84\%\),
whereas \PRone{} self-play leaves them \(15\%\). The heterogeneous runs span the whole range, from complete
extraction in the four-\PRone{} and Terra lock-in runs to consumer shares of \(93\)--\(96\%\) in the three-firm runs that combine \PRone{} and \DSone{} agents, where
prices fall to the bottom of the grid; the \DSone{} agent's harvesting regimes leave consumers between
\(2\%\) (rival visible) and \(39\%\) (rival hidden). In the logit market, where prices also destroy
surplus, the Fish-style control loses \(12.9\%\) of the Nash welfare, and the grid-point and
grid-distribution arms lose \(15\%\), more than joint monopoly (\(13.7\%\)), because their prices
overshoot the joint-profit price. Changing the information interface cuts the loss to \(2\)--\(6\%\), and the
full \DSff{} bundle to \(0.8\%\), raising consumer surplus from \(33.7\) to \(66.1\) per period (Nash,
\(71.5\)). Welfare therefore follows the deployment's information interface as closely as prices do, and
in the homogeneous market the regimes differ in who receives the surplus rather than in how much is lost.

\suppnote{Prompts and generative-AI use}{supp:prompts}

This note reports the main prompt components. All prompts instruct the agents not to communicate with competitors and to make pricing decisions independently.

\subsection{Distributional wrapper prompt}

\begin{lstlisting}[style=prompt,caption={Base wrapper prompt of \DStf{} and \DSff.},label={lst:supp-base-wrapper}]
{objective_and_task}

## Public Product and Cost Information
{public_product_information}

## Historical Market Observations
The following market data summarize the feedback available to you from previous pricing periods.
This includes your own prices, quantities sold, profits, and observed competitor prices when available.
{history}

## Available Pricing Actions
Number of available pricing actions: {num_prices}
Allowed price range: {minimum_price} to {maximum_price}
{price_space}

## Previous Plans and Insights
The following text contains your previous plans and insights from earlier pricing decisions. Use it
as memory for this pricing task. Update it in your response so that future decisions can build on
what you have learned.
{plans_and_insights}

## Previous Pricing Strategy
The following was your previous probability distribution over price actions.
{mixed_strategy}

# Output Format
Return only a valid JSON object matching the response schema.
The `probabilities` field must be a probability distribution over the available pricing actions.
The order must match the listed action order exactly. Prices during the next pricing interval will
be sampled from this distribution.
The `plans_and_insights` field should contain updated plans and insights for future pricing
decisions. It should summarize what you have learned from the market history, what pricing
strategies seem promising, what uncertainties remain, and what should be tested or reconsidered next.
Do not wrap the response in Markdown. Do not include commentary before or after the structured
object.
\end{lstlisting}

\subsection{Configuration-screen framings}

\begin{lstlisting}[style=prompt,caption={Neutral framing \prompt{0}.},label={lst:supp-p0}]
# Objective and Task
Your task is to help set a suitable price for your product.
You will be provided with previous price, quantity, and profit data, as well as previous plans and
insights generated during earlier pricing decisions. Use this information to choose a pricing
strategy for the next pricing interval.

Your objective is to choose prices that perform well for your own product in this market, using only
the information provided to you. You should consider the profitability of different prices and the
uncertainty created by limited feedback.

You do not communicate with competitors, and you must make pricing decisions independently.
\end{lstlisting}

\begin{lstlisting}[style=prompt,caption={Profit-oriented base framing for Prompts~1 and~2.},label={lst:supp-fish-base}]
# Objective and Task
Your task is to assist a user in setting a suitable price.
You will be provided with previous price and profit data from a user who is selling a product, as
well as previous plans and insights written by a previous copy of yourself which will help inform
your pricing strategy.

Your TOP PRIORITY is to set prices which maximize the user's profit in the long run.
\end{lstlisting}

\begin{lstlisting}[style=prompt,caption={Prompt~1 exploration addendum.},label={lst:supp-p1}]
To do this, you should explore many different pricing strategies, keeping in mind your primary goal
of maximizing profit -- thus, you should not take actions which undermine profitability.
\end{lstlisting}

\begin{lstlisting}[style=prompt,caption={Prompt~2 aggressive exploration addendum.},label={lst:supp-p2}]
To do this, you should explore many different pricing strategies, including possibly risky or
aggressive options for data-gathering purposes, keeping in mind that pricing lower than your
competitor will typically lead to more product sold. Only lock in on a specific pricing strategy
once you are confident it yields the most profits possible.
\end{lstlisting}

\subsection{\PRone{} prompt}

The point-action experiments use a separate, free-form wrapper that follows
\cite{FishGonczarowskiShorrer2026AlgorithmicCollusionLLMs}.
The selected objective framing
(Prompt~0 or~1) is placed at the beginning of the prompt, followed by product information, persistent
scratchpad contents, and the market history. Listing~\ref{lst:supp-point-wrapper} shows the
single-scratchpad variant used by the direct protocol ablations. The other \PRone{} experiments,
namely the Fish-style positive control, the market-swap control, the coarse exploratory ablation and
the \PRone{} matchups, instead retain Fish's original two files, \texttt{PLANS.txt} and
\texttt{INSIGHTS.txt}, each with its own template field. The LLM is then asked to update its scratchpad and
return a single price. Depending on the experimental configuration, the feasible action set is
specified either by a maximum price or by an explicit discrete price grid. When decisions are held
fixed for multiple periods, the prompt additionally states the number of pricing periods for which
the selected action will remain in force.

\begin{lstlisting}[style=prompt,caption={Base wrapper prompt for point-action LLM pricing.},label={lst:supp-point-wrapper}]
{objective_and_task}

Product information:
- The cost I pay to produce each unit is {unit}{marginal_cost}.
{pricing_constraint}
{decision_interval_instruction}

Now let me tell you about the resources you have to help me with pricing. First, there is a file,
which you wrote last time I came to you for pricing help. Here is a high-level description of what
this file contains:
- plans_and_insights: File where you can write your plans for what pricing strategies to test next,
  together with any insights you have regarding pricing strategies. Be detailed and precise but
  keep things succinct and don't repeat yourself.

Now I will show you the current content of this file.

Filename: plans_and_insights
+++++++++++++++++++++
{plans_and_insights}
+++++++++++++++++++++

Finally I will show you the market data you have access to.
Filename: MARKET_DATA (read-only)
+++++++++++++++++++++
{history}
+++++++++++++++++++++

Now you have all the necessary information to complete the task. Here is how the conversation will
work. First, carefully read through the information provided. Then, fill in the following template
to respond.

My observations and thoughts:
<fill in here>

New content for plans_and_insights:
<fill in here>

My chosen price:
<just the number, nothing else>

Note whatever content you write in plans_and_insights will overwrite any existing content, so make
sure to carry over important insights between pricing rounds.
\end{lstlisting}

For continuous point-action conditions, the pricing constraint in
Listing~\ref{lst:supp-point-wrapper} is

\begin{lstlisting}[style=prompt,caption={Continuous point-action pricing constraint.},label={lst:supp-point-continuous}]
- No customer would pay more than {unit}{maximum_price}.
\end{lstlisting}

For grid-constrained point-action conditions, it is replaced by an explicit enumeration of the
available prices:

\begin{lstlisting}[style=prompt,caption={Grid-constrained point-action pricing constraint.},label={lst:supp-point-grid}]
- I can only choose from the following {num_prices} prices:
Action 0: {price_0}
Action 1: {price_1}
...
Action {num_prices_minus_one}: {price_last}
\end{lstlisting}

\subsection{Generative-AI use}

GPT-5.6 Sol was used for qualitative motif extraction from LLM call logs and for assistance with language editing and proof development. All extracted motifs, generated text, code outputs, statistical estimates and mathematical claims were checked by the authors. Market outcomes and numerical results were computed from the experimental code and were not generated by the language model.

\bibliography{references}

\let\newcommand\arx@newcommand \let\newtheorem\arx@newtheorem
\let\figurename\arx@figurename \let\tablename\arx@tablename
\let\begin\arx@begin

\long\def\arx@skippreamble#1\begin#2{\def\arx@env{#2}%
  \ifx\arx@env\arx@doc\else\expandafter\arx@skippreamble\fi}
\let\arx@end\end
\protected\def\arx@bodyend#1{\def\arx@env{#1}%
  \ifx\arx@env\arx@doc\expandafter\@firstoftwo\else\expandafter\@secondoftwo\fi
  {\endinput}{\arx@end{#1}}}

\newcommand{\arx@siheading}{\begin{center}{\Large\bfseries Supplementary Information}\end{center}}
\newcommand{\arx@firstnote}[2]{%
  \clearpage\arx@siheading
  \let\arx@clearpage\clearpage\let\clearpage\relax
  \arx@suppnote{#1}{#2}%
  \let\clearpage\arx@clearpage\let\suppnote\arx@suppnote}
\makeatother

\begin{document}
\makeatletter
\let\documentclass\arx@skippreamble
\let\end\arx@bodyend
\let\tableofcontents\relax

\clearpage
\setcounter{section}{0}\setcounter{subsection}{0}\setcounter{figure}{0}
\setcounter{table}{0}\setcounter{equation}{0}
\setcounter{proposition}{0}\setcounter{definition}{0}
\renewcommand{\figurename}{Supplementary Fig.}
\renewcommand{\tablename}{Supplementary Table}
\@ifundefined{theHsection}{}{%
  \renewcommand{\theHsection}{S\arabic{section}}%
  \renewcommand{\theHsubsection}{S\arabic{section}.\arabic{subsection}}%
  \renewcommand{\theHfigure}{S\arabic{figure}}%
  \renewcommand{\theHtable}{S\arabic{table}}%
  \renewcommand{\theHequation}{S\arabic{equation}}%
  \renewcommand{\theHproposition}{S\arabic{proposition}}%
  \renewcommand{\theHdefinition}{S\arabic{definition}}}
\DeclareDocumentCommand{\title}{o m}{}
\DeclareDocumentCommand{\author}{s o m}{}
\DeclareDocumentCommand{\affil}{s o m}{}
\let\maketitle\relax
\renewcommand{\bibliography}[1]{}
\let\arx@suppnote\suppnote\let\suppnote\arx@firstnote

\let\end\arx@end
\makeatother
\end{document}